\documentclass[a4paper,11pt]{article}
\pdfoutput=1
\usepackage{jcappub}
\usepackage{ulem}
\usepackage{bm}
\usepackage[dvipsnames]{xcolor}
\usepackage{float}
\usepackage{arydshln}
\usepackage{multirow}
\usepackage{comment}
\usepackage{varwidth}
\usepackage{cancel}

\allowdisplaybreaks

\renewcommand{\thesection}{\arabic{section}}

\def\laq{~\raise 0.4ex\hbox{$<$}\kern -0.8em\lower 0.62ex\hbox{$\sim$}~}
\def\gaq{~\raise 0.4ex\hbox{$>$}\kern -0.7em\lower 0.62ex\hbox{$\sim$}~}

\def\beq{\begin{equation}}
\def\eeq{\end{equation}}
\def\bea{\begin{eqnarray}}
\def\eea{\end{eqnarray}}

\def \pa {\partial}

\def \mc {\mathcal}
\def \ms {\mathscr}
\def \pr {\prime}
\def \t {\text}
\def \o {_\text{o}}
\def \oz {|^\text{o}_z}

\def \nex {\notag \\[1ex]}

\def \la {\lambda}

\def \ka {\epsilon}

\def \ga {\gamma}

\def \Hcal {\mathcal{H}}

\newcommand{\quotes}[1]{``#1''} 
\usepackage{amsmath, amssymb, amscd}
\usepackage{amsthm}
\usepackage[scr=boondox]{mathalfa}
\usepackage{tikz,tikz-cd}
\usepackage{booktabs}

    \title{A Light-Cone  Approach  to Higher-Order Cosmological Observables}

\author[a,b]{P. B\'echaz,}
\author[c]{G. Fanizza,}
\author[a,b]{G. Marozzi}
\author[b,d]{and M. R. Medeiros Silva}

\affiliation[a]{Dipartimento di Fisica, Universit\`a di Pisa, Largo B. Pontecorvo 3, 56127 Pisa, 
Italy}
\affiliation[b]{Istituto Nazionale di Fisica Nucleare, Sezione di Pisa, Largo B. Pontecorvo 3, 56127 Pisa, Italy}
\affiliation[c]{Dipartimento di Ingegneria, Universit\`a LUM, S.S. 100 km 18 - 70010 Casamassima (BA), Italy}
\affiliation[d]{Departamento de Física, Universidade Estadual de Londrina,
Rod. Celso Garcia Cid, Km 380, 86057-970, Londrina, Paraná, Brazil}

\emailAdd{pierre.bechaz@phd.unipi.it}
\emailAdd{fanizza@lum.it}
\emailAdd{giovanni.marozzi@unipi.it}
\emailAdd{matheusrmsilva@uel.br}

\abstract{
We develop a second-order cosmological perturbation theory on a background geometry expressed in terms of light-cone coordinates, extending the first-order analyses available in the literature. In particular, we investigate the gauge transformations of second-order perturbative quantities on the light-cone and establish their connection with standard perturbation theory. Through a consistent matching procedure, we identify the second-order gauge fixing that corresponds to the non-linear Geodesic Light-Cone gauge  within standard perturbation theory, known as the Observational Synchronous Gauge. We then emphasize its conceptual similarities and differences wrt the  standard Synchronous Gauge. Finally, within this new perturbative framework, and adopting a fully gauge-invariant approach, we compute the luminosity distance–redshift relation up to second order with anisotropic stress as seen by a free-falling observer. Remarkably, we show how divergences at the observer position can be eliminated in a completely model independent way.
These results validate our perturbative framework and establish it as a novel formalism for evaluating cosmological observables at second order.
}

\keywords{angular distance–redshift relation, cosmological perturbation theory,
geodesic light-cone gauge, observational synchronous gauge, relativistic cosmology
 
\vskip13pt plus8pt minus11pt

\noindent{\bfseries\large\sffamily{Preprints:}}}

\begin{document}

\maketitle

\section{Introduction}
\label{sec:intro}
Accurate measurements of the distance–redshift relation form the cornerstone of modern observational cosmology. In the simplest homogeneous and isotropic 
Friedmann-Lema\^itre-Robertson-Walker (FLRW) cosmology, the distance–redshift relation $d_\t{L}(z)$ is a direct probe of the expansion history of the Universe, underpinning the discovery of cosmic acceleration via Type Ia supernovae surveys. However, the presence of inhomogeneities -- arising from density fluctuations, gravitational potentials, and peculiar velocities -- modifies photon propagation and introduces corrections to the observed distance–redshift relation. Understanding these effects is essential for interpreting high-precision data from current and upcoming surveys such as the Vera C. Rubin Observatory’s LSST \cite{LSST:2008ijt}, Euclid \cite{Euclid:2024yrr}, and the Nancy Grace Roman Space Telescope \cite{akenson2019wfirst}.

At first order in cosmological perturbation theory, the impact of Large-Scale Structure (LSS) on the luminosity distance–redshift relation has been extensively studied. Pioneer works \cite{Sasaki1987,Pyne:2003bn,Bonvin:2005ps,Hui2006,Bonvin:2011bg,Challinor:2011bk}   derived expressions for the weak-lensing convergence, Shapiro time-delay, Sachs-Wolfe, and Doppler contributions. These results have been used to forecast the magnitude of cosmic variance in distance measurements and its correlation function \cite{Ben-Dayan:2014swa,Fanizza:2021tuh}. First-order calculations have proven adequate for many applications, but as observational uncertainties shrink towards the percent level and beyond, second-order effects become non-negligible.

Second-order perturbations generate new phenomena and couple different modes of the linear theory. For instance, lensing–lensing and lensing–velocity couplings produce corrections of second-order, while post-Born lensing gives rise to non-Gaussian contributions in the magnification distribution \cite{BenDayan:2013gc,Fleury:2016fda,Schiavone:2023olz}. Moreover, relativistic light-cone distortions and frame-dragging effects enter at second order, potentially biasing cosmological parameter estimation if left unaccounted for. In the literature, there are many possible approaches to compute such non-linear effects, as \textit{e.g.} the cosmic rulers approach\footnote{See also
\cite{Gervani:2025wmh} for a first attempt to go beyond second order for the galaxy number count.} \cite{Jeong:2011as,Schmidt:2012ne,Jeong:2014ufa,Villey:2025xfz}.
In general, many authors have tackled these challenges in \cite{BenDayan:2012wi,Umeh2014a,Umeh2014b,Marozzi:2014kua,Magi:2022nfy}, by computing  the second-order corrections to the luminosity distance–redshift relation, galaxy number counts and next-to-leading order lensing corrections \cite{Magi:2022nfy,Marozzi:2016uob,Marozzi:2016qxl,Marozzi:2016und,Pratten:2016dsm,Lewis:2016tuj}, but focusing on specific gauge fixings and/or not totally considering the adequate choice for terms at the observer position. 

Indeed, the distance–redshift relation depends on the class of observers by whom it is measured. This has been clearly pointed out for linear-order corrections in \cite{Biern:2016kys}, where it was shown that such a dependence also affects the well-posedness of the relation. At second order, several attempts have been made to consistently account for this effect in different types of observables, such as the redshift, the galaxy number counts, and the galaxy power spectrum \cite{Scaccabarozzi:2017ncm,Fanizza:2018qux,Scaccabarozzi:2018vux,Grimm:2020ays}. 

In these attempts, the observer is identified to be free-falling, which is well known to be the one associated with the definition of the proper time in the Synchronous Gauge (SG).
However,  as first shown in 
\cite{Ben-Dayan:2012uam}, this free-falling observer is also the one associated with the proper time of the Geodesic Light-Cone (GLC) gauge (first introduced in \cite{Gasperini:2011us}) or of the Observational Synchronous Gauge (OSG) (first introduced in \cite{Fanizza:2020xtv}). The latter is defined to be the standard, perturbative counterpart of the GLC gauge, and  consists in a gauge which preserves all the advantages of the GLC one, order by order in perturbation theory.
Therefore, the class of observers defined by the GLC gauge (or the OSG) is equivalent to the one associated with the SG, and it will then measure the same observables. On the other hand, far from the observer position, these two gauge fixings are not equivalent. For example, unlike  the SG, photons travel on constant angles in the GLC gauge, and this is one of the key ingredients allowing us to obtain non-perturbative expressions for cosmological observables within this gauge, as the redshift and the luminosity distance (see, for example, \cite{Gasperini:2011us,BenDayan:2012wi,Fanizza:2013doa}).

With the aim of solving the above-mentioned issues, here we report a new method based on the GLC gauge to compute relativistic corrections to cosmological observables and, using it, we develop a comprehensive derivation of the luminosity distance–redshift relation at second order in cosmological perturbation theory and as seen by a free-falling observer using a gauge-invariant and divergences-free approach.
Building upon the GLC formalism, we systematically expand the distance–redshift relation to second order and isolate the  observer dependent contributions,  as was done in \cite{Fanizza:2020xtv} at linear order.

Going into details, we  obtain the new results summarized below,  addressing also  three interrelated technical goals as listed in the last three points:
\begin{itemize}
\item A new method to obtain generic observables to non-linear order. By computing light-like observables on the past light-cone of a geodesic observer, the results are automatically expressed in term of the observed light-cone and observed angles. Therefore,  only an expansion around the observed redshift is needed to write  observables entirely in terms of observed quantities. This generalizes and extends to second order the first-order analysis described in \cite{Fanizza:2020xtv}.
\item General relativistic effects on the light-cone. The development of a higher-order gauge-invariant perturbation theory directly on the observed past light-cone allows for a better understanding of its intrinsic structure.  This is because, once rewritten in terms of gauge-invariant gravitational potentials, the  light-cone gauge modes (and their derivatives) encode well-defined   linear and non-linear relativistic corrections.
\item Identification of  the observer-position terms. We will systematically derive all second-order metric and geodesic perturbations evaluated at the observer’s space-time position. These include monopole and higher-multipole contributions to the redshift and luminosity distance, which have often been neglected or treated heuristically in previous works. 

\item Second-order gauge fixing at the observer position. By building upon the general gauge-transformation rules for cosmological perturbations, we implement a consistent gauge choice at the observer position to second order. This involves specifying residual gauge functions that uniquely fix time and spatial reparameterizations on the observer’s worldline, thereby eliminating unphysical degrees of freedom in the local expansion of the metric. 

\item Gauge invariance and cancellation of divergences. 
We will demonstrate that, once observer-position terms and gauge conditions are properly accounted for (see the last two points above), all would-be divergent contributions cancel in the final expression for the observed luminosity distance-redshift relation. In particular, we systematically show that all the terms at the observer position safely do not induce any divergences, ensuring a manifestly finite and gauge-invariant result. Notably, these cancellations occur without any assumption regarding the theory of gravity or the matter content of the Universe.
\end{itemize}

As mentioned, to achieve these aims we employ the GLC  gauge \cite{Gasperini:2011us,Fanizza:2013doa}, which facilitates the separation of light-propagation effects into contributions along and orthogonal to the null direction. We expand the photon wave-vector, affine parameter, and Sachs basis up to second order, carefully tracking observer and source evaluations. Gauge transformations are handled via Lie derivatives of the metric perturbations, with explicit computations of second-order residual generators on the observer’s worldline. In particular, we methodically compare and find a proper matching of our source terms  with the literature \cite{BenDayan:2012wi, Fanizza:2013doa, Marozzi:2014kua}, thus validating our light-cone approach to cosmological observables at second order.

This paper is organized as follows. In Sect.~\ref{ref:beyond-linear-order-perturb}, we describe two possible approaches to cosmological perturbation theory: the standard one, where space-time is foliated into  space-like hyper-surfaces, and the light-cone one, where space-time is foliated into light-like hyper-surfaces. After a proper splitting of the perturbations according to their symmetry properties, we compute the  gauge transformations in both the approaches, working up to the second perturbative order. We  also establish a fully non-linear dictionary between the standard and light-cone perturbations. In Sect.~\ref{sec:GLCgauge},  we start by defining the GLC gauge at a fully non-linear order. Then, we impose the GLC gauge-fixing conditions at first and second order within the light-cone perturbation theory, ending up with the forms that gauge modes needed to transform from a generic gauge  to the perturbative definition of the GLC one. The standard counterpart of such a gauge fixing, namely the OSG, is  then defined in Sect.~\ref{sec:OSG}, where we point out its similarities and differences  wrt the SG.  In Sect.~\ref{sec:obs-glc}, we recall how the GLC coordinates turn out to be a very powerful tool to compute cosmological observables to any order in perturbation theory. In particular, we provide exact, non-perturbative expressions for the redshift and the angular distance. Then, we compute a set of first- and second-order gauge-invariant variables associated with the GLC gauge and use them to define a new approach to evaluate higher-order cosmological observables. As an application of this new perturbative scheme, in Sect.~\ref{sec:dA} we focus on the computation of the  angular distance–redshift relation as seen by a free-falling observer. Thanks to a complete gauge-fixing procedure at the observer position, we show a definite way to solve the long-standing issue of divergences associated with the observer. To conclude, we provide the ultimate, gauge-invariant expression for the angular distance–redshift relation, whose source terms are systematically matched with the literature. Finally, the main conclusions and possible future directions of investigation are summarized in Sect.~\ref{sec:conclusions}. In Appendix \ref{app:A}, we prove that the observer defined by the OSG is free-falling, as well as the one of the  SG.
In Appendix \ref{app:table}, for the sake of clarity, we provide  a table summarizing the most relevant symbols and notations defined throughout the paper.

\section{Beyond Linear-Order Perturbation Theory on the Light-Cone}
\label{ref:beyond-linear-order-perturb}

In this section, basically following  the formalism first introduced in \cite{Fanizza:2020xtv} (see also  \cite{Frob:2021ore, Fanizza:2023ixk,Fanizza:2023fus} for further developments and applications of this perturbative framework), we will build a cosmological perturbation theory up to second order on top of the background light-cone geometry. Afterwards, we will also establish a rigorous connection between 
this new perturbation theory and the standard one.

\subsection{Definition of perturbations}
\label{subsec:perturb-glc-1-order}
In order to consistently fix the notation and provide the context in which we will introduce first- and second-order perturbations on a light-cone background, we just start by expressing the standard perturbation theory up to second order  using spherical coordinates.

\paragraph{Standard perturbation theory.} We consider the perturbed FLRW metric using the conformal time and spherical spatial coordinates,  $y^\mu = (\eta, r, \theta^a)$, as
\begin{align}
\text{d}s^2 & \equiv g_{\mu \nu} \text{d}y^\mu \text{d}y^\nu = (\bar{g}_{\mu \nu} +  g^{(1)}_{\mu \nu} +g^{(2)}_{\mu \nu} )\text{d}y^\mu \text{d}y^\nu \notag \\[1ex]
& = a^2(\eta) \Big [-\big (1+2(\phi^{(1)}+\phi^{(2)})\big )\text{d}\eta^2 -2(\mathcal{B}^{(1)}_r+\mathcal{B}^{(2)}_r) \text{d}\eta \text{d}r -2 (\mathcal{B}^{(1)}_a+ \mathcal{B}^{(2)}_a) \text{d}\eta \text{d} \theta^a  \notag \\[1ex]
& \quad + (1+ \mathcal{C}^{(1)}_{rr}+\mathcal{C}^{(2)}_{rr}) \text{d}r^2  + 2 (\mathcal{C}^{(1)}_{ra} + \mathcal{C}^{(2)}_{ra})\text{d} r \text{d} \theta^a+ (\bar{\gamma}^{\text{FLRW}}_{ab} + \mathcal{C}^{(1)}_{ab}+ \mathcal{C}^{(2)}_{ab}) \text{d}\theta^a \text{d}\theta^b \Big ] \, ,
\label{eq:FLRW-metric-perturb}
\end{align}
where $\bar{g}_{\mu \nu}$ is the homogeneous and isotropic background metric, while $g^{(1)}_{\mu \nu}$ and $g^{(2)}_{\mu \nu}$ respectively stand for first- and second-order fluctuations. In this notation, the 2-dimensional spatial background metric in polar coordinates is
\begin{equation}
\bar{\gamma}^{\text{FLRW}}_{ab} = \text{diag} \left(r^2, r^2 \sin^2 \theta\right) \, .
\label{eq:gamma-ab-flrw}
\end{equation}
Using an index $i=(r, \theta, \varphi)$ to label spatial coordinates and adopting from now on the superscript $n = 1,2$ to indicate first- and second-order perturbations, the Scalar-Vector-Tensor (SVT) decomposition splits  the various functions  according to their transformation properties under 3-dimensional rotations, that is,
\begin{align}
\mathcal{B}^{(n)}_i &= \partial_i B^{(n)} + B^{(n)}_i \, , \notag \\[1ex]
\mathcal{C}^{(n)}_{rr} &= -2\psi^{(n)} + 2 D_{rr} E^{(n)} + 2 \nabla_r F^{(n)}_r + 2 h^{(n)}_{rr} \, , \notag\\[1ex]
 \mathcal{C}^{(n)}_{ra} &=  2 D_{ra} E^{(n)} + 2 \nabla_{(r} F^{(n)}_{a)} + 2 h^{(n)}_{ra} \, , \notag\\[1ex]
 \mathcal{C}^{(n)}_{ab} &= -2\psi^{(n)} \, \bar{\gamma}^{\text{FLRW}}_{ab} + 2 D_{ab} E^{(n)} + 2 \nabla_{(a} F^{(n)}_{b)} + 2 h^{(n)}_{ab} \, , 
\label{eq:FLRW-metric-perturbations}
\end{align}
or, in a more compact form for the tensor $\mathcal{C}^{(n)}_{ij}$,
\begin{equation}
    \mathcal{C}^{(n)}_{ij} = -2\psi^{(n)} \bar{\gamma}_{ij} +2 D_{ij}E^{(n)} +2 \nabla_{(i}F^{(n)}_{j)}+2h^{(n)}_{ij}  \qquad  , \qquad \bar{\gamma}_{ij} = \text{diag}\left(1, \bar{\gamma}^{\text{FLRW}}_{ab}\right)\, .
    \label{eq:Cij-decomposed}
\end{equation}
Here we are indicating  with $\nabla_i$ the covariant derivative  wrt the spatial background metric, $D_{ij}$ is the traceless operator
\begin{equation}
    D_{ij} \equiv \nabla_{(i} \nabla_{j)} -\frac{1}{3} \bar{\gamma}_{ij} \Delta_3 \, ,
    \label{eq:Dij-def}
\end{equation}
and the symmetrization between the two lower indices of two vectors is defined as $X_{(\mu}Y_{\nu)} \equiv (1/2)(X_\mu Y_\nu + X_\nu Y_\mu)$.

In this decomposition, the vectors $B^{(n)}_i$ and $F^{(n)}_i$ are divergenceless and the tensor $h^{(n)}_{ij}$ is traceless and divergenceless, namely
\begin{equation}
\nabla^i B^{(n)}_i =0 \quad  , \quad \nabla^i F^{(n)}_i =0 \quad , \quad h^{(n)}_{ii}=0 \quad ,  \quad \nabla^i h^{(n)}_{ij}=0 \, .
\end{equation}
Next, from Eqs.~\eqref{eq:FLRW-metric-perturbations} we can compute the  SVT variables   in terms of the metric perturbations of Eq.~\eqref{eq:FLRW-metric-perturb}. Indeed, taking $\nabla^i \mathcal{B}^{(n)}_i$, together with the condition $\nabla^i B^{(n)}_i=0$, we obtain
\begin{equation}
    \Delta_3 B^{(n)} = \nabla^i \mathcal{B}^{(n)}_i \qquad  , \qquad  B^{(n)}_i = \mathcal{B}^{(n)}_i - \partial_i \frac{1}{\Delta_3} (\nabla^j \mathcal{B}^{(n)}_j) \, .
    \label{eq:extraction-B_i}
\end{equation}
Then, from the contractions $\bar{\gamma}^{ij} \mathcal{C}^{(n)}_{ij}$ and $D^{ij}\mathcal{C}^{(n)}_{ij}$, thanks to the transverse and traceless conditions recalled above, we get
\begin{equation}
    \psi^{(n)} = -\frac{1}{6}\bar{\gamma}^{ij} \mathcal{C}^{(n)}_{ij} \qquad  ,  \qquad (\Delta_3)^2 E^{(n)} = \frac{3}{4}D^{ij} \mathcal{C}^{(n)}_{ij} \, .
    \label{eq:extraction-psi-E}
\end{equation}
Furthermore, using these two equations to compute $\nabla^j \mathcal{C}^{(n)}_{ij}$ makes it possible to write 
\begin{equation}
  \Delta_3 F^{(n)}_i = \nabla^j \mathcal{C}^{(n)}_{ij} + 6 \nabla_i \psi^{(n)} -4 \nabla_i \bigg (\psi^{(n)} + \frac{1}{3} \Delta_3 E^{(n)} \bigg ) \, . 
  \label{eq:extraction-F_i}
\end{equation}
Finally, by means of the last two equations, we have that
\begin{equation}
    h^{(n)}_{ij} = \frac{1}{2} \mathcal{C}^{(n)}_{ij} - \frac{3 \nabla_i \nabla_j}{\Delta_3} \psi^{(n)} + \bigg (\bar{\gamma}_{ij} + \frac{\nabla_i \nabla_j}{\Delta_3} \bigg ) \bigg ( \psi^{(n)} + \frac{1}{3} \Delta_3 E^{(n)}\bigg )- \frac{1}{\Delta_3} \big [\nabla^k \nabla_{(i} \mathcal{C}^{(n)}_{j)k} \big ] \, .
    \label{eq:extraction-h_ij}
\end{equation}

\paragraph{Light-cone perturbation theory.} We can now move to the set of background light-cone coordinates indicated as $x^\mu = (\tau, w, \tilde{\theta}^a)$, where $\tau$ is the proper time of a time-like particle, $w$ defines the background light-cone of an observer and the angles $\tilde{\theta}^a$ coincide with the two polar angles $(\theta, \varphi)$ we observe in the sky. 
These coordinates are related to $y^\mu$ through the relations
\begin{equation}
\eta (\tau) = \int_{\tau_{\text{in}}}^\tau \frac{\text{d}\tau^\prime}{a(\tau^\prime)} \qquad  , \qquad r = w - \eta(\tau) \qquad  ,  \qquad \theta^a = \tilde{\theta}^a \, ,
\label{eq:spherical-to-glc}
\end{equation}
where $\tau_{\text{in}}$ is a very early initial time at which perturbations were negligible. Then,  the standard transformation rule under a diffeomorphism is given by
\begin{equation}
f_{\mu \nu} = \frac{\partial y^\rho}{\partial x^\mu} \frac{\partial y^\sigma}{\partial x^\nu} \, g_{\rho \sigma}  \, .
\label{eq:diff}
\end{equation}
Thus, the background light-cone metric  can be  directly computed using \eqref{eq:diff}, yielding 
\begin{equation}
\bar{f}_{\mu \nu}\t{d}x^\mu \t{d}x^\nu = a^2(\tau) \bigg [ -\frac{2}{a} \text{d}\tau \text{d}w + \text{d} w^2 + \bar{\gamma}_{ab} \text{d}\tilde{\theta}^a \text{d}\tilde{\theta}^b \bigg ]  \, , 
\label{eq:fbar}
\end{equation} 
where 
\begin{equation}
\bar{\gamma}_{ab} =  \big [ w- \eta(\tau)\big ]^2\text{diag} (1, \sin^2 \tilde{\theta}^1) \equiv \big [ w- \eta(\tau)\big ]^2 q_{ab}
\label{eq:gamma-ab-glc}
\end{equation}
and $q_{ab}$ is the metric on the 2-sphere. Then, by adding  perturbations on top of the metric $\bar{f}_{\mu \nu}$, we obtain
\begin{align}
\text{d}s^2 & \equiv f_{\mu \nu} \text{d}x^\mu \text{d}x^\nu = (\bar{f}_{\mu \nu} +  f^{(1)}_{\mu \nu}+ f^{(2)}_{\mu \nu}) \text{d}x^\mu \text{d}x^\nu  \notag \\[1ex]
& = a^2(\tau) \bigg [(L^{(1)}+L^{(2)}) \text{d}\tau^2 -\frac{2}{a}\big (1-a(M^{(1)}+M^{(2)}) \big ) \text{d}\tau \text{d}w\notag \\[1ex]
& \quad + 2 (V^{(1)}_a+V^{(2)}_a)  \text{d}\tau \text{d}\tilde{\theta}^a +(1+N^{(1)}+N^{(2)})\text{d}w^2  \notag \\[1ex]
& \quad + 2  (U^{(1)}_a+U^{(2)}_a) \text{d}w \text{d}\tilde{\theta}^a   +(\bar{\gamma}_{ab}+ \gamma^{(1)}_{ab}+\gamma^{(2)}_{ab}) \text{d}\tilde{\theta}^a \text{d}\tilde{\theta}^b \bigg ]  \, .
\label{eq:metricGLC}
\end{align}
For the sake of clarity,  in a more compact form, the background metric and the perturbed one read
\begin{equation}
\bar{f}_{\mu \nu} = a^2 \,
\begin{pmatrix}
0 & -1/a & \mathbf{0} \\
-1/a & 1 & \mathbf{0} \\
\mathbf{0}^\text{T} & \mathbf{0}^\text{T} & \bar{\gamma}_{ab}
\end{pmatrix} \qquad  ,  \qquad  f^{(n)}_{\mu \nu} = a^2 \, 
\begin{pmatrix}
L^{(n)} & M^{(n)} & V^{(n)}_b \\[1ex]
M^{(n)} & N^{(n)} & U^{(n)}_b \\[1ex]
V^{(n) \text{T}}_a & U^{(n) \text{T}}_a & \gamma^{(n)}_{ab}
\end{pmatrix} \, .
\label{eq:full-perturbed-glc-metric-matrixform}
\end{equation}
 We indicate with $D_a$ the covariant derivative on the sphere and with $\tilde{D}_a \equiv \epsilon^b_a D_b$ its dual, where
\begin{equation}
    \ka_{ab} \equiv \sqrt{\t{det}[q_{cd}]}  \, \varepsilon_{ab}  = \sin \tilde{\theta}^1 \, \varepsilon_{ab}
    \label{eq:volume-form}
\end{equation}
is the covariant volume form on the 2-dimensional sphere and $\varepsilon_{ab}$ is the usual totally anti-symmetric Levi-Civita symbol. Then, as also shown in \cite{Mitsou:2019nhj,Mitsou:2020czr},  we can decompose the vector/tensor-like perturbations of the metric as
\begin{equation}
\begin{split}
V^{(n)}_a & =r^2 \big [ D_a v^{(n)} + \tilde{D}_a \hat{v}^{(n)} \big ]  \, ,  \\[1ex]
U^{(n)}_a &=r^2 \big [ D_a u^{(n)}+ \tilde{D}_a \hat{u}^{(n)}  \big ]\, , \\[1ex]
 \gamma^{(n)}_{ab} &= 2r^2 \big [ q_{ab} \nu^{(n)} + D_{ab} \mu^{(n)} + \tilde{D}_{ab} \hat{\mu}^{(n)}  \big ] \, , 
 \label{eq:SPS-vec-tensor}
\end{split}
\end{equation}
where we have defined the traceless derivative operators
\begin{equation}
D_{ab} \equiv D_{(a}D_{b)} - \frac{1}{2}q_{ab}D^2 \qquad  , \qquad \qquad \tilde{D}_{ab} \equiv \tilde{D}_{(a}D_{b)}= D_{(a} \tilde{D}_{b)} \, 
\label{eq:D-ab-def}
\end{equation}
with $D^2$  the angular Laplacian. 

Note that $L^{(n)}$, $M^{(n)}$, $N^{(n)}$, $v^{(n)}$, $u^{(n)}$, $\nu^{(n)}$ and $\mu^{(n)}$ are scalars under rotations on the sphere, while $\hat{v}^{(n)}$, $\hat{u}^{(n)}$ and $\hat{\mu}^{(n)}$ behave as pseudo-scalars. For this reason, in \cite{Fanizza:2020xtv} this decomposition was dubbed \textit{Scalar-PseudoScalar} (SPS) decomposition. Because of their different behavior under rotations, the scalar modes may also be referred to as \textit{electric} and the pseudo-scalar ones as \textit{magnetic}.

\subsection{Standard and light-cone gauge transformations} 
\label{subsec:gauge-inv-first-order-light-cone}
Now we will proceed with a systematic computation of the first- and second-order gauge transformations for standard and light-cone perturbations.

\subsubsection{Gauge transformations at first order} 

\paragraph{Standard perturbation theory.} Considering the first-order map $y^\mu \rightarrow \tilde{y}^{\mu} = y^\mu + \ka_{(1)}^\mu$, where $\ka^\mu_{(1)}$ is the gauge field, the first-order metric fluctuation transforms as
\begin{equation}
\tilde{g}^{(1)}_{\mu \nu} (y) =  g^{(1)}_{\mu \nu}(y) - \mathcal{L}_{\ka_{(1)}} \bar{g}_{\mu \nu}(y) = g^{(1)}_{\mu \nu}(y) - \nabla_\mu \ka^{(1)}_\nu (y)- \nabla_\nu \ka^{(1)}_\mu (y) \, ,
\label{eq:lie-deriv-gauge-transf}
\end{equation}
where $\mathcal{L}_{\ka_{(1)}}$ is the Lie derivative along the 4-vector $\ka^\mu_{(1)}$ and we underline that both $g^{(1)}_{\mu \nu}$ and $\tilde{g}^{(1)}_{\mu \nu}$ are evaluated at the same coordinate $y$. Consequently, according to Eq.~\eqref{eq:lie-deriv-gauge-transf}, the FLRW metric perturbations introduced in \eqref{eq:FLRW-metric-perturb} transform as
\begin{align}
\tilde{\phi}^{(1)} &= \phi^{(1)} - \mathcal{H} \epsilon^\eta_{(1)}- \partial_\eta \epsilon^\eta_{(1)} \, , \notag\\[1ex]
\tilde{\mathcal{B}}^{(1)}_r &= \mathcal{B}^{(1)}_r - \partial_r \epsilon^\eta_{(1)} + \partial_\eta \epsilon^r_{(1)} \, , \notag\\[1ex]
\tilde{\mathcal{B}}^{(1)}_a &= \mathcal{B}^{(1)}_a - \partial_a \epsilon^\eta_{(1)} + \bar{\gamma}_{ab}\partial_\eta \epsilon^b_{(1)} \, , \notag\\[1ex]
\tilde{\mathcal{C}}^{(1)}_{rr} &= \mathcal{C}^{(1)}_{rr} - 2\mathcal{H}\epsilon^\eta_{(1)} - 2 \partial_r \epsilon^r_{(1)} \, , \notag\\[1ex]
\tilde{\mathcal{C}}^{(1)}_{ra} &= \mathcal{C}^{(1)}_{ra} - \partial_a \epsilon^r_{(1)} - \bar{\gamma}_{ab} \partial_r \epsilon^b_{(1)} \, , \notag\\[1ex]
\tilde{\mathcal{C}}^{(1)}_{ab} &= \mathcal{C}^{(1)}_{ab} - \frac{1}{a^2}\epsilon^\mu_{(1)} \partial_\mu \big (a^2 \bar{\gamma}_{ab} \big )- \big (\bar{\gamma}_{ac}\partial_b + \bar{\gamma}_{bc}\partial_a \big )\epsilon^c_{(1)} \, ,
\end{align}
with  $\mathcal{H} \equiv \partial_\eta a/a$ the conformal Hubble expansion rate. Then, following  \textit{e.g.} \cite{Mukhanov:1990me}, we can decompose $\epsilon^i_{(1)} = \partial^i \epsilon_{(1)} + \hat{\epsilon}^i_{(1)}$ (with $\nabla_i \hat{\epsilon}^i_{(1)}=0$) and, using the decomposition of Eqs.~\eqref{eq:FLRW-metric-perturbations} -- together with the fact that $\tilde{h}^{(1)}_{ij}= h^{(1)}_{ij}$ because no tensor d.o.f. is present in the gauge transformations -- we end up with the following transformation rules for scalars, 
\begin{align}
    \tilde{\phi}^{(1)} &= \phi^{(1)} - \partial_\eta \epsilon^\eta_{(1)}- \mathcal{H} \epsilon^\eta_{(1)}  \,  , \notag\\[1ex]
   \tilde{\psi}^{(1)} &= \psi^{(1)} + \mathcal{H} \epsilon^\eta_{(1)} + \frac{1}{3} \Delta_3 \epsilon_{(1)} \, , \notag \\[1ex]
   \tilde{E}^{(1)} &= E^{(1)}-\epsilon_{(1)} \, , \notag \\[1ex]
    \tilde{B}^{(1)} &= B^{(1)} - \epsilon^\eta_{(1)} + \partial_\eta \epsilon_{(1)} \, ,
\label{eq:gauge-transf-scalars-FLRW}
\end{align}
and for vectors, 
\begin{equation}
    \tilde{B}^{(1)}_i = B^{(1)}_i + \partial_\eta \bigg (\frac{\hat{\epsilon}^{(1)}_i}{a^2}\bigg ) \qquad  ,  \qquad \tilde{F}^{(1)}_i = F^{(1)}_i - \frac{\hat{\epsilon}^{(1)}_i}{a^2} \, .
\end{equation}
Using the above gauge transformations, the metric can be gauge-fixed eliminating two scalar and one vector perturbations. The most popular gauge choices (writing them at any order in perturbation theory) are the following: 
\begin{itemize}
    \item \textit{Poisson Gauge} (PG)\footnote{The PG is often referred to as \quotes{Newtonian/longitudinal gauge} at first order.}: $\,\,\,\,\,\,\,\,\,\,\,\,\mc{B}_i^{(n)} = 0 = E^{(n)}$;
    \item \textit{Synchronous Gauge} (SG): $\,\,\,\,\phi^{(n)} =0 =\mc{B}_i^{(n)}$;
    \item \textit{Uniform Curvature Gauge} (UCG): $\,\,\,\,\psi^{(n)} =0 = E^{(n)}$.
\end{itemize}
Since physics is independent of the gauge chosen to describe it,  one should write the particular physical quantity in which he is interested in a gauge-invariant way. A very important example is given by the gauge-invariant gravitational potentials $\Phi^{(1)}$ and $\Psi^{(1)}$, also known as \textit{Bardeen variables}. At first order, we define them as the value that the perturbations $\phi^{(1)}$ and $\psi^{(1)}$ acquire in the PG. In order to move to the aforementioned PG up to first order, we need the following conditions\footnote{up to arbitrary constants of integration} on the gauge modes,
    \begin{equation}
        \ka^\eta_{(1)} = B^{(1)}+\pa_\eta E^{(1)} \quad  , \quad \ka_{(1)} = E^{(1)} \quad   , \quad \hat{\ka}^{(1)}_i = - \int^\eta \text{d}\eta^\prime \, B^{(1)}_i (\eta^\prime) \, ,
        \label{eq:gauge-modes-NG-1st-standard}
    \end{equation}
obtained by imposing $\tilde{E}^{(1)}=0=\tilde{B}^{(1)}=\tilde{B}^{(1)}_i$.
Then, by performing a gauge transformation on $\phi^{(1)}$ and $\psi^{(1)}$  using the above gauge fields, we end up with 
\begin{align}
    \Phi^{(1)} & \equiv \phi^{(1)} - \partial_\eta (B^{(1)} + \partial_\eta E^{(1)}) - \mathcal{H}(B^{(1)}+\partial_\eta E^{(1)})  \, , \notag \\[1ex]
    \Psi^{(1)} &\equiv \psi^{(1)} + \mathcal{H} (B^{(1)}+ \partial_\eta E^{(1)}) + \frac{1}{3} \Delta_3 E^{(1)} \, .
\label{eq:bardeen-1st}
\end{align}

\paragraph{Light-cone perturbation theory.} Let us now consider a gauge transformation acting on the light-cone coordinates, namely
\begin{equation}
x^\mu \rightarrow \tilde{x}^\mu = x^\mu + \xi^\mu_{(1)} \qquad  , \qquad  \xi^\mu_{(1)} = \big (\xi^\tau_{(1)}, \xi^w_{(1)}, \xi^a_{(1)}\big ) \, .
\label{eq:gauge-modes-glc-def}
\end{equation}
Then, using the formula
\begin{equation}
\tilde{f}^{(1)}_{\mu \nu} (x) =   f^{(1)}_{\mu \nu}(x) - \nabla_\mu \xi^{(1)}_\nu (x)- \nabla_\nu \xi^{(1)}_\mu (x) \, ,\label{eq:LC-gaugetransf}
\end{equation}
as first computed in \cite{Fanizza:2020xtv}, we find that the first-order metric perturbations appearing in Eq.~\eqref{eq:metricGLC} transform as
\begin{align}
\tilde{L}^{(1)} &= L^{(1)} + \frac{2}{a} \dot{\xi}^w_{(1)}\, , \notag \\[1ex]
\tilde{M}^{(1)} &= M^{(1)} + \frac{H}{a}\xi^\tau_{(1)} - \dot{\xi}^w_{(1)} + \frac{1}{a} (\dot{\xi}^\tau_{(1)}+ \partial_w \xi^w_{(1)} ) \, , \notag \\[1ex]
\tilde{N}^{(1)} &= N^{(1)} -2H\xi^\tau_{(1)}+\frac{2}{a}\partial_w \xi^\tau_{(1)} - 2\partial_w \xi^w_{(1)} \, , \notag \\[1ex]
\tilde{V}^{(1)}_a &= V^{(1)}_a + \frac{\partial_a \xi^w_{(1)}}{a}- \bar{\gamma}_{ab}\dot{\xi}_{(1)}^b \, , \notag \\[1ex]
\tilde{U}^{(1)}_a &= U^{(1)}_a +\frac{\partial_a \xi^\tau_{(1)}}{a}-\partial_a \xi^w_{(1)} - \bar{\gamma}_{ab} \partial_w \xi^b_{(1)} \, , \notag \\[1ex]
\tilde{\gamma}^{(1)}_{ab} &= \gamma^{(1)}_{ab} - \frac{\xi^\tau_{(1)}}{a^2} \partial_\tau \big (a^2 \bar{\gamma}_{ab} \big )- \xi^w_{(1)} \partial_w \bar{\gamma}_{ab} - \big (\bar{\gamma}_{ac}D_b + \bar{\gamma}_{bc} D_a \big ) \xi^c_{(1)} \, .
\label{eq:gauge-transf-1st-order}
\end{align}
From now on, when using the light-cone coordinates, a \quotes{dot} denotes $\partial_\tau$, \textit{i.e.} a partial derivative wrt the time $\tau$. Then, analogously to what usually done with the SVT decomposition, we can further decompose the angular entries of the gauge field $\xi^\mu_{(1)}$ according to their transformation properties under 2-dimensional rotations, that is,
\begin{equation}
\xi^a_{(1)} = q^{ab}\, \big (D_b \chi_{(1)} + \tilde{D}_b \hat{\chi}_{(1)} \big ) \, , 
\label{eq:xi-a-decomposed}
\end{equation}
where $\chi_{(1)}$ is a pure scalar d.o.f. and $\hat{\chi}_{(1)}$ is a pure pseudo-scalar one. By doing so, we can rewrite the transformation laws of $V^{(1)}_a$, $U^{(1)}_a$ and $ \gamma^{(1)}_{ab}$ as transformation laws of the variables $v^{(1)}$, $\hat{v}^{(1)}$, $u^{(1)}$, $\hat{u}^{(1)}$, $\nu^{(1)}$, $\mu^{(1)}$ and $\hat{\mu}^{(1)}$ in the following way: 
\begin{align}
\tilde{\nu}^{(1)}& = \nu^{(1)} - \frac{1}{2}D^2 \chi_{(1)} - \xi^\tau_{(1)} \bigg (H-\frac{1}{ar} \bigg )- \frac{\xi^w_{(1)}}{r} \, , \notag \\[1ex]
\tilde{\mu}^{(1)} &= \mu^{(1)} - \chi_{(1)} \, ,  & \tilde{\hat{\mu}}^{(1)} &= \hat{\mu}^{(1)} - \hat{\chi}_{(1)} \, , \notag \\[1ex]
\tilde{v}^{(1)} &= v^{(1)} + \frac{\xi^w_{(1)}}{ar^2}- \dot{\chi}_{(1)} \, ,  & \tilde{\hat{v}}^{(1)} &= \hat{v}^{(1)}-\dot{\hat{\chi}}_{(1)}\, , \notag \\[1ex]
\tilde{u}^{(1)} &= u^{(1)} + \frac{\xi^\tau_{(1)}}{ar^2} - \frac{\xi^w_{(1)}}{r^2} -\partial_w \chi_{(1)} \, ,  & \tilde{\hat{u}}^{(1)} &= \hat{u}^{(1)} - \partial_w \hat{\chi}_{(1)} \, .
\label{eq:gauge-transf-1st-decom}
\end{align}

\subsubsection{Gauge transformations at second order}
\paragraph{Standard perturbation theory.}
Now, within the framework of standard perturbation theory, we consider an infinitesimal coordinate transformation,  identified by the 4-vectors $\ka^\mu_{(1)}$ and $\ka^\mu_{(2)}$ at first and second order respectively, 
\begin{equation}
y^\mu \rightarrow \tilde{y}^\mu \simeq y^\mu + \ka^\mu_{(1)}+ \frac{1}{2}  \left (\ka^\nu_{(1)} \partial_\nu \ka^\mu_{(1)}+ \ka^\mu_{(2)} \right )  \, ,
\label{eq:gauge-trans-coord}
\end{equation}
where we mean 
\begin{equation}
\ka^\mu_{(n)} = \big ( \ka^\eta_{(n)}, \ka^i_{(n)} \big ) = \big (\ka^\eta_{(n)}, \ka^r_{(n)}, \ka^a_{(n)} \big ) \, .
\end{equation}
Under the associated gauge transformation, the second-order perturbation of any tensor $X_{\mu \nu}=\bar{X}_{\mu \nu}+ X^{(1)}_{\mu \nu}+X^{(2)}_{\mu \nu}$ changes according to 
(see \cite{Bruni:1996im, Matarrese:1997ay})
\begin{equation}
X^{(2)}_{\mu \nu} \, \rightarrow \, \tilde{X}^{(2)}_{\mu \nu} = X^{(2)}_{\mu \nu}-\mathcal{L}_{\ka_{(1)}}X^{(1)}_{\mu \nu}+ \frac{1}{2} \left (\mathcal{L}^2_{\ka_{(1)}}\bar{X}_{\mu \nu}-\mathcal{L}_{\ka_{(2)}}\bar{X}_{\mu \nu}\right ) \,.
\label{eq:gauge-2nd-rule}
\end{equation} 
Thus, the gauge-transformed metric tensor $\tilde{g}^{(2)}_{\mu \nu}$, omitting the dependence on $y$, is 
\begin{align}
\tilde{g}^{(2)}_{\mu \nu}& =g^{(2)}_{\mu \nu} -\ka^\rho_{(1)} \partial_\rho g^{(1)}_{\mu \nu} -g^{(1)}_{\mu \rho} \partial_\nu \ka^\rho_{(1)}- g^{(1)}_{\nu \rho} \partial_\mu \ka^{\rho}_{(1)} \nex
& \quad -\frac{1}{2} \Big (\ka^\rho_{(2)}\partial_\rho \bar{g}_{\mu \nu}   + \bar{g}_{\mu \rho} \partial_\nu \ka^\rho_{(2)}+ \bar{g}_{\nu \rho} \partial_\mu \ka^\rho_{(2)} \Big ) + \ka^\rho_{(1)} \partial_\rho \bar{g}_{\nu \sigma} \partial_\mu \ka^\sigma_{(1)}+ \ka^\rho_{(1)} \partial_\rho \bar{g}_{ \mu \sigma}  \partial_\nu \ka^\sigma_{(1)}  \notag \\[1ex]
& \quad + \bar{g}_{\rho \sigma} \partial_\nu \ka^\rho_{(1)} \partial_\mu \ka^\sigma_{(1)}+\frac{1}{2} \Big ( \ka^\rho_{(1)} \ka^\sigma_{(1)} \partial_\rho \partial_\sigma \bar{g}_{\mu \nu} + \ka^\rho_{(1)} \partial_\sigma \bar{g}_{\mu \nu} \partial_\rho \ka^\sigma_{(1)} +  \bar{g}_{\mu \rho} \ka^\sigma_{(1)} \partial_\nu \partial_\sigma \ka^\rho_{(1)}   \notag\\[1ex]
& \quad +\bar{g}_{\mu \rho} \partial_\sigma \ka^\rho_{(1)}\partial_\nu \ka^\sigma_{(1)}+ \bar{g}_{\nu \rho} \ka^\sigma_{(1)} \partial_\mu \partial_\sigma \ka^\rho_{(1)}   + \bar{g}_{\nu \rho} \partial_\sigma \ka^\rho_{(1)}\partial_\mu \ka^\sigma_{(1)}\Big ) \, .
\label{eq:tildef2}
\end{align}
We can then compute the transformation laws for the second-order perturbations arising in Eq.~\eqref{eq:FLRW-metric-perturb}:
\begin{align}
 \tilde{\phi}^{(2)} &= \phi^{(2)}-\frac{\mathcal{H}}{2}\ka^\eta_{(2)}-\frac{1}{2}\pa_\eta\ka^\eta_{(2)}+\frac{1}{2}\bigg (\mathcal{H}^2+\frac{\pa^2_\eta a}{a} \bigg ) (\ka^\eta_{(1)})^2 -2 (\mathcal{H}\ka^\eta_{(1)}+\pa_\eta \ka^\eta_{(1)})\phi^{(1)}  \notag \\[1ex]
    & \quad  + \frac{5}{2}\mathcal{H}\ka^\eta_{(1)}\pa_\eta \ka^\eta_{(1)} + \frac{\mathcal{H}}{2}\ka^i_{(1)}\pa_i \ka^\eta_{(1)} + \frac{1}{2}\pa_\eta \ka^i_{(1)} (\pa_i \ka^\eta_{(1)}-2\mathcal{B}^{(1)}_i - \bar{\ga}_{ij}\pa_\eta \ka^j_{(1)})  \notag \\[1ex]
    & \quad + \frac{1}{2}\ka^\mu_{(1)}\pa_\mu (\pa_\eta \ka^\eta_{(1)}-2\phi^{(1)})\, , \\[1ex]
    \label{eq:phi-tilde-2}
\tilde{\mathcal{B}}^{(2)}_r &= \mathcal{B}^{(2)}_r - \frac{1}{2}\pa_r \ka^\eta_{(2)}+\frac{1}{2}\pa_\eta \ka^r_{(2)}-\frac{1}{2}(4\phi^{(1)}-3\pa_\eta \ka^\eta_{(1)}-\pa_r \ka^r_{(1)})\pa_r \ka^\eta_{(1)}  \notag \\[1ex]
    & \quad - \mathcal{B}^{(1)}_r (\pa_\eta \ka^\eta_{(1)}+\pa_r \ka^r_{(1)}) + \mathcal{C}^{(1)}_{rr}\pa_\eta \ka^r_{(1)} -\frac{1}{2}(\pa_\eta \ka^\eta_{(1)}+3\pa_r\ka^r_{(1)})\pa_\eta \ka^r_{(1)}\notag \\[1ex]
    & \quad -\frac{1}{2}(\pa_a \ka^r_{(1)}-2\mathcal{C}^{(1)}_{ra})\pa_\eta \ka^a_{(1)}-\frac{1}{2} (2\mathcal{B}^{(1)}_a- \pa_a \ka^\eta_{(1)}+2\bar{\ga}_{ab}\pa_\eta \ka^b_{(1)})\pa_r \ka^a_{(1)}  \notag \\[1ex]
    & \quad -\frac{1}{2}\ka^i_{(1)} (2\pa_i \mathcal{B}^{(1)}_r - \pa_r \pa_i \ka^\eta_{(1)}+ \pa_\eta \pa_i \ka^r_{(1)})-2\mathcal{H}\ka^\eta_{(1)} (\mathcal{B}^{(1)}_r-\pa_r \ka^\eta_{(1)}+\pa_\eta \ka^r_{(1)} )\notag \\[1ex]
    & \quad - \frac{1}{2}\ka^\eta_{(1)} (2\pa_\eta \mathcal{B}^{(1)}_r - \pa_\eta \pa_r \ka^\eta_{(1)}+ \pa^2_\eta \ka^r_{(1)}) \, ,\\[1ex]
 \tilde{\mathcal{B}}^{(2)}_a &= \mathcal{B}^{(2)}_a -\frac{1}{2}\pa_a \ka^\eta_{(2)}+ \frac{1}{2}\bar{\ga}_{ab}\pa_\eta \ka^b_{(2)}-2\mathcal{H}\ka^\eta_{(1)}(\mathcal{B}^{(1)}_a +\bar{\ga}_{ab}\pa_\eta \ka^b_{(1)}-\pa_a \ka^\eta_{(1)})-\ka^\eta_{(1)}\pa_\eta \mathcal{B}^{(1)}_a\notag \\[1ex]
    & \quad -\ka^i_{(1)}\pa_i \mathcal{B}^{(1)}_a - \mathcal{B}^{(1)}_i \pa_a \ka^i_{(1)}-\mathcal{B}^{(1)}_a \pa_\eta \ka^\eta_{(1)}-2\phi^{(1)}\pa_a \ka^\eta_{(1)}+\mathcal{C}^{(1)}_{ai}\pa_\eta \ka^i_{(1)}\notag \\[1ex]
    & \quad -\ka^i_{(1)}\pa_i \bar{\ga}_{ab}\pa_\eta \ka^b_{(1)}-\bar{\ga}_{bc}\pa_a \ka^b_{(1)}\pa_\eta \ka^c_{(1)} + \frac{3}{2}\pa_\eta \ka^\eta_{(1)} \pa_a \ka^\eta_{(1)} - \pa_\eta \ka^r_{(1)}\pa_a \ka^r_{(1)}  \notag \\[1ex]
    & \quad + \frac{1}{2}\big [ \ka^\eta_{(1)}\pa_\eta \pa_a \ka^\eta_{(1)}+ \ka^i_{(1)}\pa_a \pa_i \ka^\eta_{(1)} + \pa_i \ka^\eta_{(1)}\pa_a \ka^i_{(1)}\nex
    & \quad -\bar{\ga}_{ab}(\ka^\eta_{(1)}\pa^2_\eta \ka^b_{(1)} -\ka^i_{(1)}\pa_\eta \pa_i \ka^b_{(1)}-\pa_\eta \ka^\eta_{(1)}\pa_\eta \ka^b_{(1)}- \pa_i \ka^b_{(1)}\pa_\eta \ka^i_{(1)}) \big ]\, , \\[1ex]
\tilde{\mathcal{C}}^{(2)}_{rr} &= \mathcal{C}^{(2)}_{rr} - \mathcal{H}\ka^\eta_{(2)} - \pa_r \ka^r_{(2)}+ \bigg (\mathcal{H}^2+ \frac{\pa^2_\eta a}{a}\bigg )(\ka^\eta_{(1)})^2 + 2(\pa_r \ka^r_{(1)}-\mathcal{C}^{(1)}_{rr})\pa_r \ka^r_{(1)} \notag \\[1ex]
    & \quad + (\pa_a \ka^r_{(1)}+\bar{\ga}_{ab}\pa_r \ka^b_{(1)}-2\mathcal{C}^{(1)}_{ra})\pa_r \ka^a_{(1)} + \ka^i_{(1)}(2\pa_r \pa_i \ka^r_{(1)}+ \pa_i \mathcal{C}^{(1)}_{rr})  \notag \\[1ex]
    & \quad + \mathcal{H} \big [\ka^i_{(1)} \pa_i \ka^\eta_{(1)}+ \ka^\eta_{(1)} (\pa_\eta \ka^\eta_{(1)}+4\pa_r \ka^r_{(1)}-2\mathcal{C}^{(1)}_{rr}) \big ] + \ka^\eta_{(1)}(2\pa_\eta \pa_r \ka^r_{(1)}-\pa_\eta \mathcal{C}^{(1)}_{rr}) \notag \\[1ex]
    & \quad + (2\mathcal{B}^{(1)}_r - \pa_r \ka^\tau_{(1)}+ \pa_\eta \ka^r_{(1)})\pa_r \ka^\eta_{(1)} \, , \\[1ex]
\tilde{\mathcal{C}}^{(2)}_{ra} &= \mathcal{C}^{(2)}_{ra} -\frac{1}{2}\pa_a \ka^r_{(2)}-\frac{1}{2}\bar{\ga}_{ab} \pa_r \ka^b_{(2)}+2\mathcal{H}\ka^\eta_{(1)}(\pa_a \ka^r_{(1)}+\bar{\ga}_{ab}\pa_r \ka^b_{(1)}-\mathcal{C}^{(1)}_{ra})- \ka^\eta_{(1)}\pa_\eta \mathcal{C}^{(1)}_{ra}\notag \\[1ex]
    & \quad - \ka^i_{(1)}\pa_i \mathcal{C}^{(1)}_{ra} + 2\mathcal{B}^{(1)}_{(r} \pa_{a)} \ka^\eta_{(1)} -\mc{C}^{(1)}_{i(r}\pa_{a)}\ka^i_{(1)} + \ka^i_{(1)}\pa_r \ka^b_{(1)}\pa_i \bar{\ga}_{ab} - \pa_a \ka^\eta_{(1)}\pa_r \ka^\eta_{(1)} \notag \\[1ex]
    & \quad + \pa_a \ka^r_{(1)}\pa_r \ka^r_{(1)}+\bar{\ga}_{bc}\pa_a \ka^b_{(1)}\pa_r \ka^c_{(1)}+ \frac{1}{2}\big [\pa_a (\ka^\eta_{(1)}\pa_\eta \ka^r_{(1)}+\ka^i_{(1)}\pa_i \ka^r_{(1)}) \notag \\[1ex]
    & \quad  +\bar{\ga}_{ab} \pa_r (\ka^\eta_{(1)}\pa_\eta \ka^b_{(1)}+\ka^i_{(1)}\pa_i \ka^b_{(1)}) \big ] \, , \\[1ex]
\tilde{\mathcal{C}}^{(2)}_{ab} &= \mc{C}^{(2)}_{ab} - \frac{1}{2a^2} \ka^\mu_{(2)}\pa_\mu (a^2 \bar{\ga}_{ab})-\bar{\ga}_{c(a}\pa_{b)}\ka^c_{(2)}-2\Hcal \ka^\eta_{(1)}\mc{C}^{(1)}_{ab}-\ka^\eta_{(1)}\pa_\eta \mc{C}^{(1)}_{ab} - \ka^i_{(1)}\pa_i \mc{C}^{(1)}_{ab}\notag \\[1ex]
    & \quad + 2\mc{B}^{(1)}_{(a}\pa_{b)}\ka^\eta_{(1)}-2 \mc{C}^{(1)}_{i(a}\pa_{b)}\ka^i_{(1)}+\frac{1}{a^2} \ka^\mu_{(1)}\pa_\mu [a^2 \bar{\ga}_{c(b}]\pa_{a)}\ka^c_{(1)}-\pa_a \ka^\eta_{(1)}\pa_b \ka^\eta_{(1)}-\pa_a \ka^r_{(1)}\pa_b \ka^r_{(1)}\notag \\[1ex]
    & \quad + \bar{\ga}_{cd}\pa_a \ka^c_{(1)}\pa_b \ka^d_{(1)}+\frac{1}{2a^2}\ka^\mu_{(1)}\pa_\mu \big [\ka^\nu_{(1)}\pa_\nu (a^2\bar{\ga}_{ab}) \big ] + \bar{\ga}_{c(a} \pa_{b)} (\ka^\eta_{(1)}\pa_\eta \ka^c_{(1)}+\ka^i_{(1)}\pa_i \ka^c_{(1)}) \, .
\end{align}
Next, in order to work out the gauge transformations of the SVT perturbations, we make use of the above formulae together with the decompositions of Eqs.~\eqref{eq:FLRW-metric-perturbations}. Moreover, as usual, we write $\ka^i_{(n)} = \pa^i \ka_{(n)}+ \hat{\ka}^i_{(n)}$ with $\nabla_i \hat{\ka}^i_{(n)}=0$ and we exploit  Eqs.~\eqref{eq:extraction-B_i}-\eqref{eq:extraction-h_ij}. For the scalars we then get
\begin{align}
    \tilde{\phi}^{(2)} &= \phi^{(2)}-\frac{\mathcal{H}}{2}\ka^\eta_{(2)}-\frac{1}{2}\pa_\eta\ka^\eta_{(2)}+\frac{1}{2}\bigg (\mathcal{H}^2+\frac{\pa^2_\eta a}{a} \bigg ) (\ka^\eta_{(1)})^2 -2 (\mathcal{H}\ka^\eta_{(1)}+\pa_\eta \ka^\eta_{(1)})\phi^{(1)}  \notag \\[1ex]
    & \quad + \frac{5}{2}\mathcal{H}\ka^\eta_{(1)}\pa_\eta \ka^\eta_{(1)} + \frac{\mathcal{H}}{2}(\pa^i \ka_{(1)}+\hat{\ka}^i_{(1)})\pa_i \ka^\eta_{(1)} \notag \\[1ex]
    & \quad + \frac{1}{2}\pa_\eta (\pa^i \ka_{(1)}+\hat{\ka}^i_{(1)}) \big [\pa_i \ka^\eta_{(1)}  -2(\pa_i B^{(1)}+B^{(1)}_i) - \bar{\ga}_{ij}\pa_\eta (\pa^j \ka_{(1)}+\hat{\ka}^j_{(1)}) \big ] \nex
    & \quad + \frac{1}{2}
    \big [\ka^\eta_{(1)}\pa_\eta + (\pa^i \ka_{(1)}+\hat{\ka}^i_{(1)}) \pa_i \big ]  \big [\pa_\eta \ka^\eta_{(1)}-2\phi^{(1)} \big ] \, , \notag \\[1ex]
    \tilde{\psi}^{(2)} &= \psi^{(2)}+ \frac{\Hcal}{2}\ka^\eta_{(2)}+\frac{1}{6}\Delta_3 \ka_{(2)}-2\Hcal \ka^\eta_{(1)}\psi^{(1)}-\ka^\eta_{(1)} \pa_\eta \psi^{(1)}- (\pa^i \ka_{(1)}+\hat{\ka}^i_{(1)})\pa_i \psi^{(1)}  \notag \\[1ex]
    & \quad - \frac{1}{2}\bigg [ \bigg (\Hcal^2 + \frac{\pa^2_\eta a}{a} \bigg )(\ka^\eta_{(1)})^2+\Hcal\ka^\eta_{(1)}\pa_\eta \ka^\eta_{(1)} + \Hcal (\pa^i\ka_{(1)}+ \hat{\ka}^i_{(1)})\pa_i \ka^\eta_{(1)}\bigg ] - \frac{1}{6}\bar{\ga}^{ij}\Pi_{ij}\, , \notag \\[1ex]
    \tilde{E}^{(2)} &= E^{(2)}-\frac{1}{2}\ka_{(2)}+\frac{3}{4(\Delta_3)^2} (D^{ij}\Pi_{ij})\, , \notag \\[1ex]
    \tilde{B}^{(2)} &= B^{(2)}-\frac{1}{2}\ka^\eta_{(2)}+\frac{1}{2}\pa_\eta \ka_{(2)}-\pa_i B^{(1)}\pa^i \ka_{(1)}- \hat{\ka}^i_{(1)}\pa_i B^{(1)}+\frac{1}{2}\ka^\eta_{(1)}\pa_\eta \ka^\eta_{(1)}+\frac{1}{2}\pa^i \ka_{(1)}\pa_i \ka^\eta_{(1)}  \notag \\[1ex]
    & \quad + \frac{1}{2}\hat{\ka}^i_{(1)}\pa_i \ka^\eta_{(1)} + \frac{\nabla^i}{\Delta_3} \Pi_i \, , 
    \label{eq:spt-gauge-transf-2nd}
\end{align}
where we have defined
\begin{align}
    \Pi_i &= -\big [ \ka^\eta_{(1)} ( 2\Hcal+\pa_\eta  ) + \pa_\eta \ka^\eta_{(1)} \big ] \big [\pa_i B^{(1)}+B^{(1)}_i\big ]- (\pa^j \ka_{(1)}+\hat{\ka}^j_{(1)})\pa_j B^{(1)}_i - 2\phi^{(1)}\pa_i \ka^\eta_{(1)} \notag \\[1ex]
    & \quad - B^{(1)}_j \pa_i (\pa^j \ka_{(1)}+\hat{\ka}^j_{(1)}) +      \big \{[-2\psi^{(1)}\bar{\ga}_{ij}+2D_{ij}E^{(1)}+2\nabla_{(i}F^{(1)}_{j)}+2h^{(1)}_{ij}] \pa_\eta \notag \\[1ex]
    & \quad -2\Hcal \ka^\eta_{(1)}\bar{\ga}_{ij} \big \} \big [\pa^j \ka_{(1)}+\hat{\ka}^j_{(1)} \big ] + \big [\pa_\eta(\pa^k \ka_{(1)}+\hat{\ka}^k_{(1)}) \big ]\big [\bar{\ga}_{jk}\pa_i -\pa_j \bar{\ga}_{ik}\big ] \big [\pa^j \ka_{(1)}+\hat{\ka}^j_{(1)}\big ]  \notag \\[1ex]
    & \quad + \pa_i \ka^\eta_{(1)}(2\Hcal + \pa_\eta) \ka^\eta_{(1)} -\frac{1}{2}\bar{\ga}_{ij} \big [\pa_\eta \ka^\eta_{(1)}+ (\pa^k \ka_{(1)}+\hat{\ka}^k_{(1)})\pa_k \big ] \big [\pa_\eta (\pa^j \ka_{(1)}+\hat{\ka}^j_{(1)}) \big ]  \notag \\[1ex]
    & \quad -\frac{1}{2}\ka^\eta_{(1)}\bar{\ga}_{ij}\pa^2_\eta (\pa^j \ka_{(1)}+\hat{\ka}^j_{(1)})  -\frac{1}{2}\bar{\ga}_{ij}\big [\pa_k (\pa^j \ka_{(1)}+\hat{\ka}^j_{(1)}) \big ] \big [\pa_\eta (\pa^k \ka_{(1)}+\hat{\ka}^k_{(1)})\big ] \, , \notag \\[1ex]
    \Pi_{ij} &= -2\big [ \ka^\eta_{(1)} (2\Hcal + \pa_\eta) + (\pa^k \ka_{(1)}+\hat{\ka}^{k}_{(1)})\pa_k \big ]\big [D_{ij}E^{(1)}+\nabla_{(i}F^{(1)}_{j)}+h^{(1)}_{ij} \big ] \notag \\[1ex]
    & \quad - 4\big \{ [-\psi^{(1)}\bar{\ga}_{k(i} + D_{k(i}E^{(1)}+ \nabla_{((i}F^{(1)}_{k)}+ h^{(1)}_{k(i}]\big \} \pa_{j)} (\pa^k \ka_{(1)}+\hat{\ka}^k_{(1)}) \notag \\[1ex]
    & \quad + 4\Hcal \ka^\eta_{(1)}\bar{\ga}_{k(i}\pa_{j)} (\pa^k \ka_{(1)}+\hat{\ka}^k_{(1)}) + 2(\pa^\ell \ka_{(1)}+\hat{\ka}^\ell_{(1)})\pa_\ell \bar{\ga}_{k(j}\pa_{i)} (\pa^k \ka_{(1)}+\hat{\ka}^k_{(1)})  \notag \\[1ex]
    & \quad + \bar{\ga}_{k\ell} \pa_i (\pa^k \ka_{(1)}+\hat{\ka}^k_{(1)})\pa_j (\pa^\ell \ka_{(1)}+\hat{\ka}^\ell_{(1)})+2\Hcal \ka^\eta_{(1)}(\pa^k \ka_{(1)}+\hat{\ka}^k_{(1)})\pa_k \bar{\ga}_{ij}\notag \\[1ex]
    & \quad + \ka^\eta_{(1)}\pa_k \bar{\ga}_{ij} (\pa^k \ka_{(1)}+\hat{\ka}^k_{(1)})+(\pa^k \ka_{(1)}+\hat{\ka}^k_{(1)})\pa_\ell \bar{\ga}_{ij} \pa_k (\pa^\ell \ka_{(1)}+\hat{\ka}^\ell_{(1)})\notag \\[1ex]
    & \quad +\bar{\ga}_{k(i} \big [\ka^\eta_{(1)}\pa_{j)} \pa_\eta(\pa^k \ka_{(1)}+\hat{\ka}^k_{(1)}) + (\pa^\ell \ka_{(1)}+\hat{\ka}^\ell_{(1)})\pa_{j)}\pa_\ell (\pa^k \ka_{(1)}+\hat{\ka}^k_{(1)}) \notag \\[1ex]
    & \quad + \pa_\eta (\pa^k \ka_{(1)}+\hat{\ka}^k_{(1)})\pa_{j)}\ka^\eta_{(1)}+\pa_\ell (\pa^k \ka_{(1)}+\hat{\ka}^k_{(1)})\pa_{j)}(\pa^\ell \ka_{(1)}+\hat{\ka}^\ell_{(1)})\big ] \, .
    \label{eq:Pi-i&Pi-ij-2nd}
\end{align}
For the vectors we get
\begin{align}
\tilde{B}^{(2)}_i &=B^{(2)}_i + \frac{1}{2}\pa_\eta \hat{\ka}^{(2)}_i + \Pi_i - \pa_i \frac{1}{\Delta_3}(\nabla^j \Pi_j) \, , \notag \\[1ex]
\tilde{F}^{(2)}_i &= F^{(2)}_i -\frac{1}{2}\hat{\ka}^{(2)}_i  +\frac{1}{\Delta_3} \bigg [\nabla^j \Pi_{ij} -\nabla_i \Pi^j_j + \nabla_i \bigg (\frac{2}{3}\Pi^j_j - \frac{1}{\Delta_3} (D^{jk}\Pi_{jk}) \bigg ) \bigg ]
\end{align}
and finally the gauge-transformed of the tensor $h^{(2)}_{ij}$ is
\begin{align}
    \tilde{h}^{(2)}_{ij} &=h^{(2)}_{ij} -\ka^\eta_{(1)}\pa_\eta h^{(1)}_{ij} +[\pa_{(i}B^{(1)}+B^{(1)}_{(i}]\pa_{j)}\ka^\eta_{(1)}-\frac{1}{2}\pa_i \ka^\eta_{(1)}\pa_j \ka^\eta_{(1)} + \frac{1}{2}\Pi_{ij}+\frac{1}{2}\frac{\nabla_i \nabla_j}{\Delta_3} \Pi^k_k\notag \\[1ex]
    & \quad +\frac{1}{2}\bigg [\bar{\ga}_{ij}+\frac{\nabla_i \nabla_j}{\Delta_3}\bigg ] \bigg [ -\frac{\Pi^k_k}{3}+\frac{1}{2\Delta_3} (D^{k\ell}\Pi_{k\ell})\bigg ] -\frac{1}{\Delta_3}[\nabla^k \nabla_{(i} \Pi_{j)k}]\, .
\end{align}
As an interesting application of the above formulae, we compute the gravitational gauge-invariant potentials at the second perturbative order. We need to make use of  Eqs.~\eqref{eq:gauge-modes-NG-1st-standard}, providing the form of the gauge mode $\ka^\mu_{(1)}$ to move to the PG, as well as 
\begin{align}
    \ka^\eta_{(2)} &= 2 \bigg [B^{(2)}+\pa_\eta E^{(2)} + \frac{3}{4(\Delta_3)^2}(D^{ij}\Pi^{(\text{PG})}_{ij}) - \pa_i B^{(1)}\pa^i E^{(1)} +\pa_i B^{(1)}\int^\eta \text{d}\eta^\prime \, B^i_{(1)}(\eta^\prime)\notag \\[1ex]
    & \quad + \frac{1}{2}(B^{(1)}+\pa_\eta E^{(1)}) \pa_\eta (B^{(1)}+\pa_\eta E^{(1)})+\frac{1}{2}\pa^i E^{(1)}\pa_i (B^{(1)}+\pa_\eta E^{(1)})\notag \\[1ex]
    & \quad - \frac{1}{2}\pa_i (B^{(1)}+\pa_\eta E^{(1)})\int^\eta \text{d}\eta^\prime \, B^i_{(1)}(\eta^\prime)+ \frac{\nabla^i}{\Delta_3}\Pi^{(\text{PG})}_i\bigg ]\, , \notag \\[1ex]
    \ka_{(2)} &= 2 \bigg [E^{(2)}+\frac{3}{4(\Delta_3)^2}(D^{ij}\Pi^{(\text{PG})}_{ij}) \bigg ] \, ,
    \label{eq:gauge-modes-NG-2nd-standard}
\end{align}
respectively stemming from the conditions $\tilde{B}^{(2)}=0=\tilde{E}^{(2)}$ 
applied to Eqs.~\eqref{eq:spt-gauge-transf-2nd}. In the above equations, $\Pi^{(\text{PG})}_i$ and $\Pi^{(\text{PG})}_{ij}$ are the quantities defined in Eqs.~\eqref{eq:Pi-i&Pi-ij-2nd} where the first-order gauge fields have been replaced by the forms they must have to move to the PG, namely  Eqs.~\eqref{eq:gauge-modes-NG-1st-standard}. Therefore, by performing a gauge transformation using the first two of Eqs.~\eqref{eq:spt-gauge-transf-2nd} and  Eqs.~\eqref{eq:gauge-modes-NG-2nd-standard},  we obtain
\begin{align}
    \Phi^{(2)} &= \phi^{(2)} - \pa_\eta (B^{(2)}+\pa_\eta E^{(2)})-\Hcal (B^{(2)}+\pa_\eta E^{(2)}) - \Hcal \bigg [ \frac{3}{4(\Delta_3)^2}(D^{ij}\Pi^{(\text{PG})}_{ij}) \notag \\[1ex]
    & \quad - \pa_i B^{(1)}\pa^i E^{(1)} +\pa_i B^{(1)}\int^\eta \text{d}\eta^\prime \, B^i_{(1)}(\eta^\prime)+\frac{1}{2}(B^{(1)}+\pa_\eta E^{(1)}) \pa_\eta (B^{(1)}+\pa_\eta E^{(1)})\notag \\[1ex]
    & \quad  + \frac{1}{2}\pa^i E^{(1)}\pa_i (B^{(1)}+\pa_\eta E^{(1)}) - \frac{1}{2}\pa_i (B^{(1)}+\pa_\eta E^{(1)})\int^\eta \text{d}\eta^\prime \, B^i_{(1)}(\eta^\prime)+ \frac{\nabla^i}{\Delta_3}\Pi^{(\text{PG})}_i\bigg ] \notag \\[1ex]
    & \quad - \pa_\eta \bigg [ \frac{3}{4(\Delta_3)^2}(D^{ij}\Pi^{(\text{PG})}_{ij}) - \pa_i B^{(1)}\pa^i E^{(1)} +\pa_i B^{(1)}\int^\eta \text{d}\eta^\prime \, B^i_{(1)}(\eta^\prime)\notag \\[1ex]
    & \quad + \frac{1}{2}(B^{(1)}+\pa_\eta E^{(1)}) \pa_\eta (B^{(1)}+\pa_\eta E^{(1)})+\frac{1}{2}\pa^i E^{(1)}\pa_i (B^{(1)}+\pa_\eta E^{(1)})\notag \\[1ex]
    & \quad - \frac{1}{2}\pa_i (B^{(1)}+\pa_\eta E^{(1)})\int^\eta \text{d}\eta^\prime \, B^i_{(1)}(\eta^\prime)+ \frac{\nabla^i}{\Delta_3}\Pi^{(\text{PG})}_i\bigg ] +\frac{1}{2}\bigg ( \Hcal^2 + \frac{\pa^2_\eta a}{a}\bigg ) (B^{(1)}+\pa_\eta E^{(1)})^2   \notag \\[1ex]
    & \quad -2 \big [\Hcal (B^{(1)}+\pa_\eta E^{(1)}) + \pa_\eta (B^{(1)}+\pa_\eta E^{(1)}) \big ]\phi^{(1)} + \frac{5}{2}\Hcal (B^{(1)}+\pa_\eta E^{(1)}) \pa_\eta (B^{(1)}+\pa_\eta E^{(1)}) \notag \\[1ex]
    & \quad + \frac{\Hcal}{2}\bigg (\pa^i E^{(1)}-\int^\eta \text{d}\eta^\prime \, B^i_{(1)}(\eta^\prime) \bigg ) \pa_i (B^{(1)}_i + \pa_\eta E^{(1)}) \notag \\[1ex]
    & \quad + \frac{1}{2} \big [\pa_\eta\pa^i E^{(1)}+B^i_{(1)} \big ] \big [\pa_i (B^{(1)}+\pa_\eta E^{(1)})-2 (\pa_i B^{(1)}+B^{(1)}_i) - \bar{\ga}_{ij}( \pa_\eta \pa^i E^{(1)}-B^{(1)}) \big ] \notag \\[1ex]
    & \quad +\frac{1}{2}\bigg [(B^{(1)}+\pa_\eta E^{(1)})\pa_\eta + \bigg (\pa^i E^{(1)}-\int^\eta \text{d}\eta^\prime \, B^i_{(1)}(\eta^\prime) \bigg ) \pa_i  \bigg ] \big [\pa_\eta (B^{(1)}+\pa_\eta E^{(1)}) -2\phi^{(1)} \big ]\, , \notag \\[1ex]
    \Psi^{(2)} &= \psi^{(2)}+\Hcal (B^{(2)}+\pa_\eta E^{(2)})+\frac{1}{3}\Delta_3 E^{(2)}+\Hcal \bigg [ \frac{3}{4(\Delta_3)^2}(D^{ij}\Pi^{(\text{PG})}_{ij}) - \pa_i B^{(1)}\pa^i E^{(1)} \notag \\[1ex]
    & \quad  + \pa_i B^{(1)}\int^\eta \text{d}\eta^\prime \, B^i_{(1)}(\eta^\prime)
   + \frac{1}{2}(B^{(1)}+\pa_\eta E^{(1)}) \pa_\eta (B^{(1)}+\pa_\eta E^{(1)})\notag \\[1ex]
   & \quad +\frac{1}{2}\pa^i E^{(1)}\pa_i (B^{(1)}+\pa_\eta E^{(1)}) - \frac{1}{2}\pa_i (B^{(1)}+\pa_\eta E^{(1)})\int^\eta \text{d}\eta^\prime \, B^i_{(1)}(\eta^\prime)+ \frac{\nabla^i}{\Delta_3}\Pi^{(\text{PG})}_i\bigg ]  \notag \\[1ex]
   & \quad + \frac{1}{4\Delta_3}(D^{ij}\Pi^{(\text{NG})}_{ij}) - 2\Hcal (B^{(1)}+\pa_\eta E^{(1)})\psi^{(1)} - (B^{(1)}+\pa_\eta E^{(1)})\pa_\eta \psi^{(1)}\notag \\[1ex]
    & \quad - \pa_i \psi^{(1)} \bigg (\pa^i E^{(1)}-\int^\eta \text{d}\eta^\prime \, B^i_{(1)}(\eta^\prime) \bigg ) -\frac{1}{2}\bigg [\bigg (\Hcal^2 + \frac{\pa^2_\eta a}{a} \bigg )(B^{(1)}+\pa_\eta E^{(1)})^2 \notag \\[1ex]
    & \quad + \Hcal (B^{(1)}+\pa_\eta E^{(1)})\pa_\eta (B^{(1)}+\pa_\eta E^{(1)}) \notag \\[1ex]
    & \quad + \Hcal \pa_i (B^{(1)}+\pa_\eta E^{(1)}) \bigg (\pa^i E^{(1)}-\int^\eta \text{d}\eta^\prime \, B^i_{(1)}(\eta^\prime) \bigg ) - \frac{1}{6}\bar{\ga}^{ij}\Pi^{(\text{PG})}_{ij} \bigg ]\, .
    \label{eq:bardeen-2nd}
\end{align}


\paragraph{Light-cone perturbation theory.}
By applying the general gauge transformation rule \eqref{eq:tildef2} to the metric tensor $f^{(2)}_{\mu \nu}$, as written in the line element \eqref{eq:metricGLC}, and by considering two gauge fields $\xi^\mu_{(1)}$ and $\xi^\mu_{(2)}$, as first done in \cite{MasterThesis}, we obtain that the light-cone perturbations transform as
\begin{align}
\tilde{L}^{(2)} & = L^{(2)}+\frac{1}{a} \dot{\xi}^w_{(2)} -\xi^\mu_{(1)} \partial_\mu L^{(1)} -2 \big ( L^{(1)}\dot{\xi}^\tau_{(1)} +M^{(1)} \dot{\xi}^w_{(1)}+ V^{(1)}_a \dot{\xi}^a_{(1)}\big ) \nex
& \quad - \frac{2H}{a}\xi^\tau_{(1)} \dot{\xi}^w_{(1)} + \big (\dot{\xi}^w_{(1)} \big )^2 + \bar{\gamma}_{ab} \dot{\xi}^a_{(1)} \dot{\xi}^b_{(1)}\notag \\[1ex]
& \quad - \frac{1}{a} \big [\partial_\tau \big ( \xi^\mu_{(1)} \partial_\mu \xi^w_{(1)} \big )+ 2\dot{\xi}^\tau_{(1)}\dot{\xi}^w_{(1)}+ 2\dot{a} L^{(1)}\xi^\tau_{(1)} \big ] \label{eq:tildeL} \, , \\[1ex]
\tilde{M}^{(2)} & = M^{(2)}+\frac{H}{2a} \xi^\tau_{(2)}-\frac{1}{2}\dot{\xi}^w_{(2)}+\frac{1}{2a} \big (\partial_w \xi^w_{(2)}+ \dot{\xi}^\tau_{(2)} \big ) \nex
& \quad + \frac{1}{2a^2} \bigg \{ -\ddot{a}\big (\xi^\tau_{(1)} \big )^2  - \dot{a} \big (\xi^\mu_{(1)} \partial_\mu \xi^\tau_{(1)}+ 2 \xi^\tau_{(1)}\dot{\xi}^\tau_{(1)}+2\xi^\tau_{(1)}\partial_w \xi^w_{(1)}\big ) \bigg \}\nex
& \quad - \frac{1}{2a} \bigg \{ \dot{\xi}^a_{(1)} \partial_a \xi^\tau_{(1)}  + \partial_a \xi^w_{(1)}\partial_w \xi^a_{(1)}+ \big (\partial_w \xi^w_{(1)}+ \dot{\xi}^\tau_{(1)} \big )^2 + 4 \dot{\xi}^w_{(1)} \partial_w \xi^\tau_{(1)}\notag \\[1ex]
& \quad + \xi^\mu_{(1)} \partial_\mu \big (\dot{\xi}^\tau_{(1)}+ \partial_w \xi^w_{(1)} \big )+ 4\dot{a} \xi^\tau_{(1)}(M^{(1)}-\dot{\xi}^w_{(1)}) \bigg \}\notag \\[1ex]
& \quad + \frac{1}{2}  \bigg \{-2 \big (L^{(1)}\partial_w \xi^\tau_{(1)}+ M^{(1)}\partial_w \xi^w_{(1)}+V^{(1)}_a \partial_w \xi^a_{(1)} \big ) \notag \\[1ex]
& \quad -2 \big ( 
M^{(1)} \dot{\xi}^\tau_{(1)}+ N^{(1)} \dot{\xi}^w_{(1)} + U^{(1)}_a \dot{\xi}^a_{(1)} \big )+\dot{\xi}^w_{(1)} \big (\dot{\xi}^\tau_{(1)}+3\partial_w \xi^w_{(1)})\notag \\[1ex]
& \quad + \dot{\xi}^a_{(1)} \partial_a \xi^w_{(1)}+2\bar{\gamma}_{ab} \partial_w \xi^a_{(1)}\dot{\xi}^b_{(1)}+ \xi^\mu_{(1)} \partial_\mu (-2M^{(1)}+\dot{\xi}^w_{(1)} \big )
\bigg \} \label{eq:tildeM}\, , \\[1ex]
\tilde{N}^{(2)}&= N^{(2)}-H\xi^\tau_{(2)}+\frac{1}{a}\partial_w \xi^\tau_{(2)}-\partial_w \xi^w_{(2)}+\frac{\dot{a}}{a^2}\xi^\tau_{(1)} \big (\dot{a} \xi^\tau_{(1)}-2\partial_w  \xi^\tau_{(1)} \big )\nex
& \quad -2 \big (M^{(1)}\partial_w \xi^\tau_{(1)}  + N^{(1)} \partial_w \xi^w_{(1)}+U^{(1)}_a \partial_w \xi^a_{(1)} \big ) + \big (\partial_w \xi^w_{(1)} \big )^2 + \partial_w \big (\xi^\mu_{(1)}\partial_\mu \xi^w_{(1)} \big )\nex
& \quad + \bar{\gamma}_{ab} \partial_w \xi^a_{(1)} \partial_w \xi^b_{(1)} - \xi^\mu_{(1)}\partial_\mu N^{(1)}-\frac{1}{a} \bigg \{-\ddot{a} (\xi^\tau_{(1)})^2-\dot{a} \xi^\mu_{(1)}\partial_\mu \xi^\tau_{(1)}+2 \partial_w \xi^\tau_{(1)} \partial_w \xi^w_{(1)}\nex
& \quad +\partial_w \big (\xi^\mu_{(1)} \partial_\mu \xi^\tau_{(1)} \big ) + \dot{a}\xi^\tau_{(1)} \big (2N^{(1)}-4 \partial_w \xi^w_{(1)} \big ) \bigg \}\label{eq:tildeN}  \, ,\\[1ex]
\tilde{V}^{(2)}_a & = V^{(2)}_a + \frac{1}{2a}\partial_a \xi^w_{(2)}-\frac{1}{2}\bar{\gamma}_{ab} \dot{\xi}^b_{(2)}-\frac{H}{a}\xi^\tau_{(1)} \partial_a \xi^w_{(1)}\nex
& \quad -\frac{1}{2a} \bigg \{\partial_a \big ( \xi^\mu_{(1)}\partial_\mu \xi^w_{(1)} \big )   + 2 \big (\dot{\xi}^\tau_{(1)} \partial_a \xi^w_{(1)}+ \dot{\xi}^w_{(1)} \partial_a \xi^\tau_{(1)} \big ) \bigg \}  - 2 H \xi^\tau_{(1)} \big ( V^{(1)}_a-\bar{\gamma}_{ab} \dot{\xi}^b_{(1)}\big ) \nex
& \quad - \big (L^{(1)} \partial_a \xi^\tau_{(1)} + M^{(1)} \partial_a \xi^w_{(1)}+V^{(1)}_b \partial_a \xi^b_{(1)}+ V^{(1)}_a \dot{\xi}^\tau_{(1)}+U^{(1)}_a \dot{\xi}^w_{(1)}+ \gamma^{(1)}_{ab} \dot{\xi}^b_{(1)}  \big ) \notag \\[1ex]
& \quad + \frac{1}{2} \bar{\gamma}_{ab} \partial_\tau \big ( \xi^\mu_{(1)} \partial_\mu \xi^b_{(1)}\big )- \xi^\mu_{(1)} \partial_\mu V^{(1)}_a + \dot{\xi}^w_{(1)}\partial_a \xi^w_{(1)}+ \xi^\mu_{(1)} \partial_\mu \bar{\gamma}_{ab} \dot{\xi}^b_{(1)}+ \bar{\gamma}_{bc} \partial_a \xi^b_{(1)} \dot{\xi}^c_{(1)} \label{eq:tildeV} \, ,\\[1ex]
\tilde{U}^{(2)}_a & = U^{(2)}_a +\frac{1}{2a} \partial_a \xi^\tau_{(2)}-\frac{1}{2}\partial_a \xi^w_{(2)}-\frac{1}{2}\bar{\gamma}_{ab} \partial_w \xi^b_{(2)}-\frac{H}{a} \xi^\tau_{(1)} \partial_a \xi^\tau_{(1)} \notag \\[1ex]
& \quad - \frac{1}{2a} \bigg \{ \partial_a \big ( \xi^\mu_{(1)}\partial_\mu \xi^\tau_{(1)} \big )  + 2 \big ( \partial_a \xi^w_{(1)} \partial_w \xi^\tau_{(1)}+ \partial_a \xi^\tau_{(1)} \partial_w \xi^w_{(1)}\big ) \bigg \} \notag \\[1ex]
& \quad  -2H \xi^\tau_{(1)} \big ( U^{(1)}_a -\partial_a \xi^w_{(1)}-\bar{\gamma}_{ab} \partial_w \xi^b_{(1)}\big ) - \big (M^{(1)} \partial_a \xi^\tau_{(1)}+ N^{(1)} \partial_a \xi^w_{(1)}\notag \\[1ex]
& \quad + U^{(1)}_b \partial_a \xi^b_{(1)} + V^{(1)}_a \partial_w \xi^\tau_{(1)}+U^{(1)}_a \partial_w \xi^w_{(1)}+ \gamma^{(1)}_{ab} \partial_w \xi^b_{(1)} \big ) \notag  \\[1ex]
& \quad + \frac{1}{2} \bar{\gamma}_{ab} \partial_w \big (\xi^\mu_{(1)} \partial_\mu \xi^b_{(1)} \big )- \xi^\mu_{(1)} \partial_\mu U^{(1)}_a + \partial_w \xi^w_{(1)}\partial_a \xi^w_{(1)}+ \xi^\mu_{(1)} \partial_\mu \bar{\gamma}_{ab} \partial_w \xi^b_{(1)}  \notag \\[1ex]
& \quad + \bar{\gamma}_{bc} \partial_a \xi^b_{(1)} \partial_w \xi^c_{(1)} +\frac{1}{2}\pa_a (\xi^\mu_{(1)}\pa_\mu \xi^w_{(1)}) \label{eq:tildeU} \, ,\\[1ex]
\tilde{\gamma}^{(2)}_{ab} &= \gamma^{(2)}_{ab}-\frac{1}{2a^2} \xi^\tau_{(2)} \partial_\tau (a^2 \bar{\gamma}_{ab}) -\frac{1}{2}\xi^w_{(2)} \partial_w \bar{\gamma}_{ab} - \bar{\gamma}_{a(c} D_{b)} \xi^c_{(2)} -2\gamma^{(1)}_{c(a} \pa_{b)} \xi^c_{(1)}-\xi^{c}_{(1)}\partial_{c}\gamma_{ab}^{(1)} \notag \\[1ex]
& \quad -\frac{1}{a^2} \xi^\tau_{(1)} \partial_\tau \big (a^2 \gamma^{(1)}_{ab} \big )- \xi^w_{(1)} \partial_w \gamma^{(1)}_{ab} + (\dot{H}+2H^2) \big (\xi^\tau_{(1)} \big )^2 \bar{\gamma}_{ab}+   \partial_a \xi^w_{(1)} \partial_b \xi^w_{(1)}  +   \notag \\[1ex]
& \quad  -2\big [ V^{(1)}_{(a}\partial_{b)}\xi^\tau_{(1)}+ U^{(1)}_{(a}\partial_{b)}\xi^w_{(1)} \big ] -\frac{2}{a} \partial_{(a} \xi^\tau_{(1)} \partial_{b)} \xi^w_{(1)} + H \bar{\gamma}_{ab} \xi^\mu_{(1)} \partial_\mu \xi^\tau_{(1)}  \notag\\[1ex]
& \quad  +4H \xi^\tau_{(1)} \bar{\gamma}_{c(a} D_{b)}\xi^c_{(1)}+ 2H\xi^\tau_{(1)} \big ( \xi^\tau_{(1)}  \dot{\bar{\gamma}}_{ab} + \xi^w_{(1)} \partial_w \bar{\gamma}_{ab} \big ) + 2\xi^\mu_{(1)} \partial_\mu [\bar{\gamma}_{c(a} \partial_{b)}  ] \xi^c_{(1)}\notag \\[1ex]
& \quad  + \partial_{(a} \xi^c_{(1)} \partial_{b)} \xi^d_{(1)} \bar{\gamma}_{cd}  + \bar{\gamma}_{c(a} \partial_{b)}  \big ( \xi^\mu_{(1)} \partial_\mu \xi^c_{(1)}\big ) +  \frac{1}{2}\mathbb{G}^{(2)}_{ab}\, ,
\label{eq:tilde-delta-gamma}
\end{align}
where we have defined 
\begin{align}
    \mathbb{G}^{(2)}_{ab} &\equiv \xi^\rho_{(1)}\xi^\sigma_{(1)}\pa_\rho \pa_\sigma \bar{\ga}_{ab}+ \xi^\rho_{(1)}\pa_\rho \xi^\sigma_{(1)}\pa_\sigma \bar{\ga}_{ab}\notag \\ 
    & =(\xi^\tau_{(1)})^2 \ddot{\bar{\ga}}_{ab} + 2\xi^\tau_{(1)}\xi^w_{(1)} \pa_w \dot{\bar{\ga}}_{ab} + 2\xi^\tau_{(1)}\xi^c_{(1)}\pa_c \dot{\bar{\ga}}_{ab} + (\xi^w_{(1)})^2 \pa^2_w \bar{\ga}_{ab}  + 2\xi^w_{(1)}\xi^c_{(1)}\pa_w \pa_c \bar{\ga}_{ab}\notag \\[1ex]
    & \quad + \xi^c_{(1)}\xi^d_{(1)}\pa_c \pa_d \bar{\ga}_{ab} + \xi^\tau_{(1)}\dot{\xi}^\tau_{(1)}\dot{\bar{\ga}}_{ab} + \xi^w_{(1)}\pa_w \xi^\tau_{(1)}\dot{\bar{\ga}}_{ab} + \xi^c_{(1)}\pa_c \xi^\tau_{(1)}\dot{\bar{\ga}}_{ab} + \xi^\tau_{(1)}\dot{\xi}^w_{(1)}\pa_w \bar{\ga}_{ab} \notag \\[1ex]
    & \quad + \xi^w_{(1)}\pa_w \xi^w_{(1)}\pa_w \bar{\ga}_{ab} + \xi^c_{(1)}\pa_c \xi^w_{(1)}\pa_w \bar{\ga}_{ab} 
    + \xi^\tau_{(1)}\dot{\xi}^{c}_{(1)} \pa_c \bar{\ga}_{ab} + \xi^w_{(1)}\pa_w \xi^c_{(1)} \pa_c \bar{\ga}_{ab} + \xi^d_{(1)}\pa_d \xi^c_{(1)}\pa_c \bar{\ga}_{ab} \, .
    \label{eq:G2-ab}
\end{align}
At this stage, we can make use of the SPS decomposition of Eqs.~\eqref{eq:SPS-vec-tensor}  to obtain the transformation properties of the SPS variables. 
In order to do so, we note that all the r.h.s. of Eqs.~\eqref{eq:tildeV}-\eqref{eq:tilde-delta-gamma} can always be written according to this decomposition. In particular, for the vectors, calling $\Pi_a$ each term appearing in Eqs.~\eqref{eq:tildeV} and \eqref{eq:tildeU}, this \textit{must} have the form\footnote{We extract the factor $r^2$ from the decomposition of $\Pi_a$ and $\Pi_{ab}$ (see below) bearing in mind Eqs.~\eqref{eq:SPS-vec-tensor}.}
\begin{equation}
\Pi_a = r^2 \, \big (D_a \Pi + \tilde{D}_a \hat{\Pi}\big ) \, ,
\label{eq:trick-gauge-vector}
\end{equation}
where $\Pi$ is a pure scalar and $\hat{\Pi}$ is a pure pseudo-scalar. We can now use the fact that, regardless of the coordinates we are using to parameterize the 2-sphere, it holds (see \cite{Fanizza:2020xtv}) 
\begin{equation}
q^{ab} D_{(a}\tilde{D}_{b)}\equiv 0 \qquad  ,  \qquad q^{ab} D_{(a}D_{b)} = D^2 = q^{ab} \tilde{D}_{(a} \tilde{D}_{b)} \, .
\label{eq:properties-derivatives}
\end{equation}
Then,  by contracting Eq.~\eqref{eq:trick-gauge-vector} with $q^{ab}D_b$ and $q^{ab}\tilde{D}_b$ and by symmetrizing in the bottom indices $a,b$, we can express both $\Pi$ and $\hat{\Pi}$ in terms of $\Pi_a$ as follows:
\begin{equation}
\Pi = \frac{1}{r^2} \frac{1}{D^2}q^{ab} D_{(a} \Pi_{b)} \qquad  ,  \qquad \hat{\Pi} = \frac{1}{r^2}\frac{1}{D^2} q^{ab} \tilde{D}_{(a} \Pi_{b)} \, .
\label{eq:scalar-pseudoscalar-extraction}
\end{equation}
A similar procedure can be adopted for tensor perturbations, by calling $\Pi_{ab}$ each term of Eq.~\eqref{eq:tilde-delta-gamma} and by decomposing it as
\begin{equation}
\Pi_{ab} = 2r^2 \, \big (q_{ab} \rho + D_{ab} \sigma + \tilde{D}_{ab} \hat{\sigma} \big ) \, ,
\end{equation}
where $\rho$ contains  the trace of $\Pi_{ab}$, $\sigma$ is pure scalar and $\hat{\sigma}$ is pure pseudo-scalar. Then, we need to compute the contractions $q^{ab} \Pi_{ab}, D^{ab} \Pi_{ab}, \tilde{D}^{ab} \Pi_{ab}$ remembering that  (see again \cite{Fanizza:2020xtv})
\begin{equation}
q^{ab} D_{ab}=0=q^{ab} \tilde{D}_{ab} = D^{ab} \tilde{D}_{ab} \qquad ,  \qquad D^{ab} D_{ab} = \frac{1}{2}(D^2)^2 = \tilde{D}^{ab} \tilde{D}_{ab} \, .
\end{equation}
For $\rho$ we immediately get that
\begin{equation}
\rho = \frac{1}{4r^2} q^{ab} \Pi_{ab} \, .
\label{eq:rho-extraction}
\end{equation}
For $\sigma$ and $\hat{\sigma}$ we must pay attention to the fact that, using the definition of the derivative operator $D_{ab}$, we can write
\begin{equation}
D^{ab}\Pi_{ab} = D^{ab}D_{ab} \sigma = D^aD^bD_{ab} \sigma = \bigg ( D^aD^bD_aD_b  -\frac{1}{2}(D^2)^2 \bigg ) \sigma \, ,
\end{equation}
where 
\begin{equation}
D^aD^bD_aD_b \sigma = \Big ( D^aD_aD^bD_b + D^a \big ([D_b,D_a]D^b \big )\Big ) \sigma \, .
\end{equation}
Exploiting the usual definition of the Riemann tensor in terms of the commutator of covariant derivatives, the second contribution to the above expression is
\begin{equation}
D^a \big ([D_b,D_a]D^b \sigma \big )= D^a (R^b_{cba}D^c \sigma) = D^a q_{ac} D^c \sigma = D^2 \sigma \, ,
\end{equation}
where, as provided \textit{e.g.} in \cite{Weinberg:1972kfs}, we have used the general fact that, for a $d$-sphere of fixed radius $r$ and metric $\bar{\gamma}_{ab}$, the Ricci tensor is given by
\begin{equation}
R^b_{cab} = \frac{d-1}{r^2} \bar{\gamma}_{ac} = (d-1)q_{ac} \, .
\end{equation}
It follows that, summing the two contributions, we have
\begin{equation}
D^{ab}\Pi_{ab} = \frac{1}{2} \bigg ((D^2)^2 + 2D^2 \bigg )\sigma \equiv \frac{1}{2}\mathcal{D} \sigma \qquad \Rightarrow \qquad \sigma = \frac{1}{r^2} \mathcal{D}^{-1} D^{ab}\Pi_{ab}
\label{eq:mu-extraction}
\end{equation}
and similarly 
\begin{equation}
\hat{\sigma} = \frac{1}{r^2} \mathcal{D}^{-1} \tilde{D}^{ab}\Pi_{ab} \, .
\label{eq:hat-mu-extractions}
\end{equation}
Moreover, in order to work out the gauge transformations of $v^{(2)}$, $\hat{v}^{(2)}$, $u^{(2)}$, $\hat{u}^{(2)}$, $\nu^{(2)}$, $\mu^{(2)}$ and $\hat{\mu}^{(2)}$, we need to decompose the angular components of $\xi^\mu_{(n)}$ as
\begin{equation}
\xi^a_{(n)} = q^{ab} \big ( D_b \chi_{(n)}+ \tilde{D}_b \hat{\chi}_{(n)}\big ) \, ,
\label{eq:xi-a-decomposed-general}
\end{equation}
 where $\chi_{(n)}$ are pure scalar gauge d.o.f. while $\hat{\chi}_{(n)}$ are pure pseudo-scalar ones.
 
In this way, for $\tilde{v}^{(2)}$ and $\tilde{\hat{v}}^{(2)}$ we get 
\begin{align}
D^2 \tilde{v}^{(2)} &= D^2 \bigg [ v^{(2)} + \frac{1}{2ar^2} \xi^w_{(2)}-\frac{1}{2}\dot{\chi}_{(2)}-\frac{1}{2ar^2}\xi^\mu_{(1)} \partial_\mu \xi^w_{(1)} + \frac{1}{r^2}\mc{V}^{(2)}\bigg ] \, , \notag \\[1ex]
D^2 \tilde{\hat{v}}^{(2)} &=D^2 \bigg [  \hat{v}^{(2)}-\frac{1}{2}\dot{\hat{\chi}}_{(2)}+\frac{1}{r^2}\hat{\mc{V}}^{(2)}\bigg ]  \, ,
\label{eq:v-2-tilde}
\end{align}
where we have defined\footnote{To get $\hat{\mc{V}}^{(2)}$ (as well as $\hat{\mc{U}}^{(2)}$, see later), we have used the properties \eqref{eq:properties-derivatives} and the fact that, for a vector $X^a$ it holds
\begin{align*}
[\tilde{D}_a, D_c]X^a &=\varepsilon^b_a R^a_{dbc}X^d = \frac{d}{2}(\varepsilon^b_c q_{bd} - \varepsilon^b_a q_{bc})X^d = -d\varepsilon_{cd}X^d \, , 
\end{align*}
where $d=2$ for the unit 2-sphere.}
\begin{align}
    \mc{V}^{(2)}&\equiv -2Hr^2 q^{ab}D_a \xi^\tau_{(1)} (D_b v^{(1)} + \tilde{D}_b \hat{v}^{(1)}) -2Hr^2 \xi^\tau_{(1)}D^2 v^{(1)} \nex
    & \quad -D_a \xi^\mu_{(1)} \pa_\mu \big [r^2 q^{ab}(D_b v^{(1)} + \tilde{D}_b \hat{v}^{(1)}) \big ]-\xi^\mu_{(1)}\pa_\mu (r^2 D^2 v^{(1)})\nex
    & \quad -r^2 q^{ab}(D_a \chi_{(1)} + \tilde{D}_a \hat{\chi}_{(1)})(D_b v^{(1)} + \tilde{D}_b \hat{v}^{(1)})-q^{ab}D_a L^{(1)} D_b \xi^\tau_{(1)} -q^{ab}D_a M^{(1)}D_b \xi^w_{(1)} \nex
    & \quad - r^2 q^{ab}q^{cd} D_a (D_c v^{(1)} + \tilde{D}_c \hat{v}^{(1)})D_b (D_d \chi_{(1)} + \tilde{D}_d \hat{\chi}_{(1)}) - L^{(1)}D^2 \xi^\tau_{(1)}-M^{(1)} D^2 \xi^w_{(1)}\nex
    & \quad -r^2q^{ab} (D_a v^{(1)} + \tilde{D}_a \hat{v}^{(1)})D^2(D_b \chi_{(1)} + \tilde{D}_b \hat{\chi}_{(1)})-r^2\dot{\xi}^\tau_{(1)}D^2 v^{(1)}-r^2\dot{\xi}^w_{(1)}D^2 u^{(1)}\nex
    & \quad -2r^2q^{ab}q^{cd}D_a (q_{bc}\nu^{(1)}+D_{bc}\mu^{(1)}+\tilde{D}_{bc}\hat{\mu}^{(1)})(D_d \dot{\chi}_{(1)}+\tilde{D}_d 
    \dot{\hat{\chi}}_{(1)})\nex
    & \quad -r^2q^{ab}(D_a v^{(1)} + \tilde{D}_a \hat{v}^{(1)})D_b \dot{\xi}^\tau_{(1)}-r^2q^{ab}(D_a u^{(1)} + \tilde{D}_a \hat{u}^{(1)})D_b \dot{\xi}^w_{(1)}\nex
    & \quad -2r^2 q^{ab}q^{cd}(q_{ac}\nu^{(1)}+D_{ac}\mu^{(1)}+\tilde{D}_{ac}\hat{\mu}^{(1)})D_b (D_d \dot{\chi}_{(1)}+\tilde{D}_d 
    \dot{\hat{\chi}}_{(1)})\nex
    & \quad +2Hr^2 q^{ab}D_a \xi^\tau_{(1)} (D_b \dot{\chi}_{(1)}+\tilde{D}_b 
    \dot{\hat{\chi}}_{(1)}) +2Hr^2 \xi^\tau_{(1)}D^2 \dot{\chi}_{(1)}\nex
    & \quad +2q^{ab}\bigg (-\frac{r}{a}D_a \xi^\tau_{(1)}+rD_a \xi^w_{(1)} \bigg )(D_b \dot{\chi}_{(1)} + \tilde{D}_b \dot{\hat{\chi}}_{(1)})+2\bigg (-\frac{r}{a}\xi^\tau_{(1)}+r\xi^w_{(1)} \bigg )D^2 \dot{\chi}_{(1)}\nex
    & \quad -\frac{H}{a}q^{ab}D_a \xi^\tau_{(1)}D_b \xi^w_{(1)}-\frac{H}{a}\xi^\tau_{(1)}D^2\xi^w_{(1)}-\frac{1}{a}(q^{ab}D_a \dot{\xi}^\tau_{(1)}D_b \xi^w_{(1)}+\dot{\xi}^\tau_{(1)}D^2 \xi^w_{(1)})\nex
    & \quad -\frac{1}{a}(q^{ab}D_a \dot{\xi}^w_{(1)}D_b \xi^\tau_{(1)}+\dot{\xi}^w_{(1)}D^2 \xi^\tau_{(1)})+q^{ab}D_a \dot{\xi}^w_{(1)}D_b \xi^w_{(1)}+\dot{\xi}^w_{(1)}D^2 \xi^w_{(1)}\nex
    & \quad +r^2 q^{ab}D_a (D_c \dot{\chi}_{(1)}+\tilde{D}_c \dot{\hat{\chi}}_{(1)})\pa_b \big [q^{cd}(D_d \chi_{(1)}+\tilde{D}_d \hat{\chi}_{(1)}) \big ]  \nex
    & \quad +r^2q^{ab}(D_a \dot{\chi}_{(1)}+\tilde{D}_a \dot{\hat{\chi}}_{(1)})D^2(D_b \chi_{(1)}+\tilde{D}_b\hat{\chi}_{(1)})\nex
    & \quad +\frac{a^2r^2}{2} \pa_\tau \Big [D_a \xi^\mu_{(1)} \pa_\mu \big [q^{ab}(D_b \chi_{(1)}+\tilde{D}_b \hat{\chi}_{(1)}) \big ]+\xi^\mu_{(1)} \pa_\mu (D^2 \chi_{(1)}) \Big ]\nex
    & \quad +a^2r^2 q^{ab}(D_a \chi_{(1)}+\tilde{D}_a \hat{\chi})(D_b \dot{\chi}_{(1)}+\tilde{D}_b \dot{\hat{\chi}}) \, , \nex
    \hat{\mc{V}}^{(2)}&\equiv -2Hr^2 \xi^\tau_{(1)}D^2 \hat{v}^{(1)}- r^2 \tilde{D}_a \big [q^{bc}(D_c \chi_{(1)}+\tilde{D}_c \hat{\chi}_{(1)}) \big ]\pa_b \big [q^{ad} (D_d v^{(1)}+\tilde{D}_d \hat{v}^{(1)})  \big ]\nex
    & \quad - \xi^\mu_{(1)}\pa_\mu (r^2 D^2 \hat{v}^{(1)}) +2r^2 \varepsilon^{ab}(D_a \chi_{(1)}+\tilde{D}_a \hat{\chi}_{(1)}) (D_b v_{(1)}+\tilde{D}_b \hat{v}_{(1)})\nex
    & \quad -r^2 \dot{\xi}^\tau_{(1)}D^2 \hat{v}_{(1)} -r^2 \dot{\xi}^w_{(1)}D^2 \hat{u}_{(1)}\nex
    & \quad -2r^2 q^{ab}q^{cd}\tilde{D}_a (q_{bc}\nu^{(1)}+D_{bc}\mu^{(1)}+\tilde{D}_{bc}\hat{\mu}^{(1)})(D_d \dot{\chi}_{(1)}+\tilde{D}_d \dot{\hat{\chi}}_{(1)})\nex
    & \quad -2r^2 q^{ab}q^{cd}(q_{ac}\nu^{(1)}+D_{ac}\mu^{(1)}+\tilde{D}_{ac}\hat{\mu}^{(1)})\tilde{D}_b (D_d \chi_{(1)}+\tilde{D}_d \hat{\chi}_{(1)}) + 2Hr^2 \xi^\tau_{(1)} D^2 \dot{\hat{\chi}}_{(1)}\nex
    & \quad +2 \bigg ( -\frac{r}{a}\xi^\tau_{(1)}+r\xi^w_{(1)}\bigg ) D^2 \dot{\hat{\chi}}_{(1)}+r^2 q^{ab}\tilde{D}_a (D_c \dot{\chi}_{(1)}+\tilde{D}_c \dot{\hat{\chi}}_{(1)})\pa_b \big [q^{cd}(D_d \chi_{(1)}+\tilde{D}_d \hat{\chi}_{(1)}) \big ]\nex
    & \quad + a^2 r^2  \pa_\tau \Big [q^{cd}\tilde{D}_a (D_d \chi_{(1)}+\tilde{D}_d \hat{\chi}_{(1)}) \pa_c \big [q^{ab}(D_b \chi_{(1)}+\tilde{D}_b \hat{\chi}_{(1)}) \big ]+\xi^\mu_{(1)}\pa_\mu (D^2 \hat{\chi}_{(1)}) \Big ]\,  .
\end{align}
Regarding $\tilde{u}^{(2)}$ and $\tilde{\hat{u}}^{(2)}$, we obtain
\begin{align}
D^2\tilde{u}^{(2)}&= D^2 \bigg [u^{(2)}+ \frac{1}{2ar^2}\xi^\tau_{(2)}-\frac{1}{2r^2} \xi^w_{(2)}-\frac{1}{2} \partial_w \chi_{(2)}-\frac{1}{2ar^2}\xi^\mu_{(1)} \partial_\mu \xi^\tau_{(1)}\nex
& 
\quad +\frac{1}{2r^2}\xi^\mu_{(1)}\pa_\mu \xi^w_{(1)}+ \frac{1}{r^2}\mc{U}^{(2)} \bigg ]\, , \notag \\[1ex]
D^2\tilde{\hat{u}}^{(2)} &=D^2 \bigg [ \hat{u}^{(2)}-\frac{1}{2}\pa_w \hat{\chi}_{(2)} + \frac{1}{r^2} \hat{\mc{U}}^{(2)}\bigg ]\, , 
\label{eq:u-2-tilde}
\end{align}
where 
\begin{align}
    \mc{U}^{(2)} & \equiv -2Hr^2 q^{ab}D_a \xi^\tau_{(1)} (D_b u^{(1)}+\tilde{D}_b \hat{u}^{(1)})-2Hr^2 \xi^\tau_{(1)}D^2 u^{(1)}\nex
    & \quad - D_a \xi^\mu_{(1)}\pa_\mu \big [r^2 q^{ab}(D_b u^{(1)}+\tilde{D}_b \hat{u}^{(1)}) \big ]-\xi^\mu_{(1)}\pa_\mu (r^2 D^2 u^{(1)})\nex
    & \quad -r^2 q^{ab}(D_a \chi_{(1)}+\tilde{D}_a \hat{\chi}_{(1)}) (D_b u^{(1)}+\tilde{D}_b \hat{u}^{(1)}) - q^{ab}D_a M^{(1)}D_b \xi^\tau_{(1)}-q^{ab}D_a N^{(1)}D_b \xi^w_{(1)}\nex
    & \quad - r^2 q^{ab}q^{cd}D_a (D_c u^{(1)}+\tilde{D}_c \hat{u}^{(1)})D_b (D_d \chi_{(1)}+\tilde{D}_d \hat{\chi}_{(1)})-M^{(1)}D^2 \xi^\tau_{(1)}-N^{(1)}D^2 \xi^w_{(1)}\nex
    & \quad -r^2 q^{ab}(D_a u^{(1)}+\tilde{D}_a \hat{u}^{(1)})D^2 (D_b \chi_{(1)}+\tilde{D}_b \hat{\chi}_{(1)})-r^2 \pa_w \xi^\tau_{(1)}D^2 u^{(1)}-r^2 \pa_w \xi^w_{(1)} D^2 v^{(1)} \nex
    & \quad -2r^2 q^{ab}q^{cd}D_a (q_{bc}\nu^{(1)}+D_{bc}\mu^{(1)}+\tilde{D}_{bc}\hat{\mu}^{(1)})\pa_w (D_d \chi_{(1)}+\tilde{D}_d \hat{\chi}_{(1)})\nex
    & \quad -r^2 q^{ab}(D_a v^{(1)}+\tilde{D}_a \hat{v}^{(1)})\pa_w (D_b \xi^\tau_{(1)})-r^2 q^{ab}(D_a u^{(1)}+\tilde{D}_a \hat{u}^{(1)}) \pa_w (D_b \xi^w_{(1)})\nex
    & \quad -2r^2q^{ab}q^{cd}(q_{ac}\nu^{(1)}+D_{ac}\mu^{(1)}+\tilde{D}_{ac}\hat{\mu}^{(1)}) \pa_w (D_b D_d \chi_{(1)}+D_b \tilde{D}_d \hat{\chi}_{(1)})\nex
    & \quad + 2Hr^2q^{ab}D_a \xi^\tau_{(1)}\pa_w (D_b  \chi_{(1)}+\tilde{D}_b \hat{\chi}_{(1)})+2Hr^2 \xi^\tau_{(1)}\pa_w (D^2 \chi_{(1)})\nex
    & \quad +2^{ab}\bigg (-\frac{r}{a}D_a \xi^\tau_{(1)}+rD_a \xi^w_{(1)} \bigg )\pa_w (D_b  \chi_{(1)}+\tilde{D}_b \hat{\chi}_{(1)})+2 \bigg (-\frac{r}{a}\xi^\tau_{(1)}+r\xi^w_{(1)} \bigg )\pa_w (D^2 \chi_{(1)})\nex
    & \quad -\frac{H}{a}q^{ab}(D_a \xi^\tau_{(1)}D_b \xi^\tau_{(1)}+\xi^\tau_{(1)}D^2 \xi^\tau_{(1)})+2Hq^{ab}(D_a \xi^\tau_{(1)}D_b \xi^w_{(1)}+\xi^\tau_{(1)}D^2 \xi^w_{(1)})\nex
    & \quad -\frac{1}{a}\big [ q^{ab}\pa_w (D_a \xi^\tau_{(1)})D_b \xi^w+\pa_w \xi^\tau_{(1)}D^2 \xi^w_{(1)} \big ]-\frac{1}{a}\big [q^{ab}\pa_w (D_a \xi^w_{(1)})D_b \xi^\tau_{(1)} + \pa_w \xi^w_{(1)}D^2 \xi^\tau_{(1)} \big ]\nex
    & \quad +q^{ab}\pa_w (D_a \xi^w_{(1)})D_b \xi^w_{(1)}+\pa_w \xi^w_{(1)}D^2 \xi^w_{(1)}\nex
    & \quad +r^2 q^{ab}\pa_w (D_a D_c\chi_{(1)}+D_a\tilde{D}_c \hat{\chi}_{(1)}) \pa_b \big [q^{cd}(D_d  \chi_{(1)}+\tilde{D}_d \hat{\chi}_{(1)}) \big ]\nex
    & \quad +r^2 q^{ab}\pa_w (D_a  \chi_{(1)}+\tilde{D}_a \hat{\chi}_{(1)}) D^2 (D_b  \chi_{(1)}+\tilde{D}_b \hat{\chi}_{(1)}) \nex
    & \quad +\frac{a^2r^2}{2}\pa_w \Big [D_a \xi^\mu_{(1)}\pa_\mu  \big [q^{ab} (D_b \chi_{(1)}+\tilde{D}_b \hat{\chi}_{(1)})\big ] +\xi^\mu_{(1)}\pa_\mu (D^2 \chi_{(1)})\Big ]\nex
    & \quad +a^2r^2 q^{ab}(D_a  \chi_{(1)}+\tilde{D}_a \hat{\chi}_{(1)})\pa_w (D_b  \chi_{(1)}+\tilde{D}_b \hat{\chi}_{(1)})\, , \nex
     \hat{\mc{U}}^{(2)} & \equiv   -2Hr^2 \xi^\tau_{(1)}D^2 \hat{u}^{(1)} - r^2 \tilde{D}_a \big [q^{bc}(D_c \chi_{(1)}+\tilde{D}_c \hat{\chi}_{(1)}) \big ]\pa_b \big [q^{ad} (D_d u^{(1)}+\tilde{D}_d \hat{u}^{(1)})  \big ]\nex
     & \quad - \xi^\mu_{(1)}\pa_\mu (r^2 D^2 \hat{u}^{(1)}) +2r^2 \varepsilon^{ab}(D_a \chi_{(1)}+\tilde{D}_a \hat{\chi}_{(1)}) (D_b u^{(1)}+\tilde{D}_b \hat{u}^{(1)})\nex
     & \quad -r^2 \pa_w \xi^\tau_{(1)}D^2 \hat{u}^{(1)}-r^2 \pa_w \xi^w_{(1)}D^2 \hat{u}^{(1)}\nex
     & \quad -2r^2q^{ab}q^{cd}\tilde{D}_a (q_{bc}\nu^{(1)}+D_{bc}\mu^{(1)}+\tilde{D}_{bc}\hat{\mu}^{(1)})\pa_w (D_d \chi_{(1)}+\tilde{D}_d \hat{\chi}_{(1)})\nex
     & \quad - 2r^2 q^{ab}q^{cd}(q_{ac}\nu^{(1)}+D_{ac}\mu^{(1)}+\tilde{D}_{ac}\hat{\mu}^{(1)})\pa_w (\tilde{D}_b D_d \chi_{(1)}+\tilde{D}_b D_d \hat{\chi}_{(1)})\nex
     & \quad +2Hr^2 \xi^\tau_{(1)}\pa_w (D^2 \hat{\chi}_{(1)})+2\bigg (-\frac{r}{a}\xi^\tau_{(1)}+r\xi^w_{(1)} \bigg ) \pa_w (D^2 \hat{\chi}_{(1)})\nex
     & \quad +r^2 q^{ab}\pa_w \tilde{D}_a(D_c \chi_{(1)}+\tilde{D}_c \hat{\chi}_{(1)})\pa_b \big [q^{cd}(D_d \chi_{(1)}+\tilde{D}_d \hat{\chi}_{(1)}) \big ]\nex
     & \quad +\frac{a^2r^2}{2}\pa_w \Big [\tilde{D}_a (D_d \chi_{(1)}+\tilde{D}_d \hat{\chi}_{(1)})\pa_c  \big [q^{ab} (D_b \chi_{(1)}+\tilde{D}_b \hat{\chi}_{(1)})\big ] +\xi^\mu_{(1)}\pa_\mu (D^2 \hat{\chi}_{(1)})\Big ]\, .
\label{eq:mathbU-2-defined}
\end{align}
Finally, let us focus on the transformation properties of $\nu^{(2)}, \mu^{(2)}$ and $\hat{\mu}^{(2)}$. 
We find that
\begin{align}
\tilde{\nu}^{(2)} & = \nu^{(2)}-\frac{1}{4}D^2 \chi_{(2)}-\frac{1}{2}\xi^\tau_{(2)} \bigg (H-\frac{1}{ra} \bigg )-\frac{1}{2}\frac{\xi^w_{(2)}}{r}-\frac{1}{a^2r^2}\xi^\tau_{(1)} \partial_\tau (a^2r^2\nu^{(1)}) \notag\\[1ex]
& \quad  -\frac{1}{r^2}\xi^w_{(1)} \partial_w (r^2 \nu^{(1)})+ \frac{1}{2}(\dot{H}+2H^2)\big (\xi^\tau_{(1)} \big )^2  + \frac{H}{2}\xi^\mu_{(1)}\partial_\mu \xi^\tau_{(1)}\notag\\[1ex]
& \quad + 2 H\xi^\tau_{(1)} \bigg (-\frac{\xi^\tau_{(1)}}{ar}+\frac{\xi^w_{(1)}}{r} \bigg ) + \frac{1}{4r^2} q^{ab} \mc{V}^{(2)}_{ab}  \, ,
\label{eq:nu-2-tilde}
\end{align}
where 
\begin{align}
\mc{V}^{(2)}_{ab} & \equiv  \Big \{ -4r^2 [q_{c(a}\nu^{(1)}+D_{c(a}\mu^{(1)}+\tilde{D}_{c(a}\hat{\mu}^{(1)}]\pa_{b)} -2r^2 \big [(\pa_c(q_{ab}\nu^{(1)})+\pa_c (D_{ab}\mu^{(1)})\nex
& \quad + \pa_c (\tilde{D}_{ab}\hat{\mu}^{(1)})\big ] + 4Hr^2\xi^\tau_{(1)}q_{c(a}D_{b)} +2\xi^\mu_{(1)}\partial_\mu [r^2 q_{c(a}\partial_{b)}]\notag \\[1ex]
& \quad + r^2 q_{c(a}\partial_{b)}  (\xi^\mu_{(1)}\partial_\mu) \Big \} \big [ q^{cd} (D_d \chi_{(1)}+\tilde{D}_d \hat{\chi}_{(1)})\big ] + \frac{1}{2}\mathbb{G}^{(2)}_{ab}-\frac{2}{a} \partial_{(a}\xi^\tau_{(1)}\partial_{b)}\xi^w_{(1)}+ \pa_a \xi^w_{(1)}\pa_b\xi^w_{(1)}\notag \\[1ex]
& \quad  + r^2q_{cd}\Big \{ \partial_{(a} \big [q^{ce}(D_e \chi_{(1)}+\tilde{D}_e\hat{\chi}_{(1)}) \big ] \partial_{b)} \big [q^{df}(D_f \chi_{(1)}+\tilde{D}_f\hat{\chi}_{(1)}) \big ] \Big \} \notag \\[1ex]
& \quad -2r^2 \Big \{[D_{(a}v^{(1)}+\tilde{D}_{(a}\hat{v}^{(1)}]\partial_{b)} \xi^\tau_{(1)} +[D_{(a}u^{(1)}+\tilde{D}_{(a}\hat{u}^{(1)}]\partial_{b)} \xi^w_{(1)} \Big \} \, .
\label{eq:mathscrN}
\end{align}
Moreover
\begin{align}
\tilde{\mu}^{(2)}&= \mu^{(2)}-\frac{1}{2}\chi_{(2)} +\frac{1}{r^2} \mathcal{D}^{-1} D^{ab} \mc{M}^{(2)}_{ab} \, , \notag \\[1ex]
\tilde{\hat{\mu}}^{(2)}&= \hat{\mu}^{(2)}-\frac{1}{2}\hat{\chi}_{(2)} +\frac{1}{r^2} \mathcal{D}^{-1} \tilde{D}^{ab} \mc{M}^{(2)}_{ab} \, ,
\label{eq:mu-2-tilde}
\end{align}
where
\begin{align}
 \mc{M}^{(2)}_{ab} & \equiv \Big \{ -4r^2 [q_{c(a}\nu^{(1)}+D_{c(a}\mu^{(1)}+\tilde{D}_{c(a}\hat{\mu}^{(1)}]D_{b)}   + Hr^2 \xi^\tau_{(1)}q_{c(a}D_{b)} +\xi^\mu_{(1)}\partial_\mu  [r^2 q_{c(a}\partial_{b)}] \notag \\[1ex]
& \quad + r^2 q_{c(a}\partial_{b)} (\xi^\mu_{(1)}\partial_\mu) \Big \} \big [ q^{cd} (D_b \chi_{(1)}+\tilde{D}_b \hat{\chi}_{(1)})\big ] + \mathbb{G}^{(2)}_{ab}- \frac{4}{a}\partial_{(a}\xi^\tau_{(1)}\partial_{b)}\xi^w_{(1)}  \notag \\[1ex]
& \quad  + r^2 \Big \{ \partial_{(a} \big [q^{cd}(D_d \chi_{(1)}+\tilde{D}_d\hat{\chi}_{(1)}) \big ] \partial_{b)} \big [q^e_c(D_e \chi_{(1)}+\tilde{D}_e\hat{\chi}_{(1)}) \big ] \Big \} \notag \\[1ex]
& \quad -4r^2 \Big \{[D_{(a}v^{(1)}+\tilde{D}_{(a}\hat{v}^{(1)}]\partial_{b)} \xi^\tau_{(1)} +[D_{(a}u^{(1)}+\tilde{D}_{(a}\hat{u}^{(1)}]\partial_{b)} \xi^w_{(1)} \Big \}  \notag \\[1ex]
& \quad +  2\partial_{(a} \xi^w_{(1)}\partial_{b)}\xi^w_{(1)}-\frac{1}{a^2}\partial_\tau \big [2a^2r^2 (q_{ab}\nu^{(2)}+D_{ab}\mu^{(2)}+\tilde{D}_{ab}\hat{\mu}^{(2)}) \big ] \notag \\[1ex]
& \quad - \partial_w \big [\xi^w_{(1)}(q_{ab}\nu^{(1)}+D_{ab}\mu^{(1)}+\tilde{D}_{ab}\hat{\mu}^{(1)}) \big ] \, .
\label{eq:Mab-2-defined}
\end{align}
The results of this subsection are the technical pillars on which our light-cone perturbation theory at second order is based. They will also  form the basis of the gauge-invariant formalism that we will develop in the next section.

\subsection{Map between standard and light-cone perturbation theory}
\label{subsec:map-FLRW-GLC}

To conclude this section, we establish a dictionary between the standard metric perturbations $g^{(n)}_{\mu \nu}$ and the light-cone ones $f^{(n)}_{\mu \nu}$. In order to do so, exploiting the usual diffeomorphism law 
\begin{equation}
f^{(n)}_{\mu \nu} = \frac{\partial y^\rho}{\partial x^\mu} \frac{\partial y^\sigma}{\partial x^\nu} \, g^{(n)}_{\rho \sigma} \, ,
\end{equation}
we obtain the following relations and their inverse expressions:
\begin{equation}
\begin{cases}
a^2L = -2 \big ( \phi - \frac{1}{2} \mathcal{C}_{rr} - \mathcal{B}_r\big )\\
aM = - \mathcal{B}_{r}- \mathcal{C}_{rr}\\
N = \mathcal{C}_{rr}\\
aV_a = -\mathcal{B}_a - \mathcal{C}_{ra}\\
U_a = \mathcal{C}_{ra}\\
\delta \gamma_{ab} = \mathcal{C}_{ab} 
\end{cases}
\qquad \Rightarrow \qquad 
\begin{cases}
\phi =-\frac{1}{2} ( a^2 L + N + 2aM )\\
\mathcal{B}_r =  -N-aM\\
\mathcal{C}_{rr} = N\\
\mathcal{B}_a = -U_a - aV_a\\
\mathcal{C}_{ra} = U_a\\
\mathcal{C}_{ab} =\delta \gamma_{ab}
\end{cases} \,\,\,\,\,\,\,\, .
\label{eq:dictionary}
\end{equation}
 We have omitted the superscript $n = 1,2$ in the above equations  because they have a fully non-linear behavior, meaning that they are valid at any perturbative order. 

As a consequence, using Eqs.~\eqref{eq:dictionary}, we have the possibility to express the SPS variables in terms of the SVT ones at any order in perturbation theory. In particular, to obtain $v$, $\hat{v}$, $u$ and $\hat{u}$ from $V_a$ and $U_a$ we use Eqs.~\eqref{eq:scalar-pseudoscalar-extraction}, while to obtain $\nu$, $\mu$ and $\hat{\mu}$ from $\delta \gamma_{ab}$ we use Eqs.~\eqref{eq:rho-extraction}, \eqref{eq:mu-extraction} and \eqref{eq:hat-mu-extractions}: 
\begin{align}
    L &= -\frac{2}{a^2} \bigg \{\phi + \psi - \bigg (\nabla_r \nabla_r - \frac{1}{3}\Delta_3 \bigg ) E - \partial_r B - B_r - \nabla_r F_r - h_{rr}\bigg \} \, , \notag \\[1ex]
    M &= -\frac{1}{a} \bigg \{-2\psi + 2 \bigg (\nabla_r \nabla_r - \frac{1}{3}\Delta_3 \bigg ) E + \partial_r B + B_r + 2 \nabla_r F_r + 2 h_{rr}  \bigg \} \, , \notag \\[1ex]
    N &= -2\psi + 2 \bigg (\nabla_r \nabla_r - \frac{1}{3}\Delta_3  \bigg ) E + 2 \nabla_r F_r + 2 h_{rr} \, , \notag \\[1ex]
    v &=-\frac{1}{a}\frac{1}{D^2}\bigg \{ \bar{\ga}^{ab} D_{a} \big [ 2 \nabla_{(r}\nabla_{b)}E + \partial_{b}B + B_{b} + 2 \nabla_{(r} F_{b)} + 2h_{rb}\big ] \bigg \}\, \, , \notag \\[1ex]
    \hat{v} &=-\frac{1}{a}\frac{1}{D^2}\bigg \{ \bar{\ga}^{ab} \tilde{D}_{a} \big [2 \nabla_{(r}\nabla_{b)}E + \partial_{b}B + B_{b} + 2 \nabla_{(r} F_{b)} + 2h_{rb}\big ] \bigg \}\, , \notag \\[1ex]
    u &= \frac{2}{D^2}\bigg \{ \bar{\ga}^{ab} D_{a} \big [\nabla_{(r}\nabla_{b)}E + \nabla_{(r}F_{b)}+ h_{rb}\big ] \bigg \}\, , \notag \\[1ex] 
    \hat{u} &= \frac{2}{D^2}\bigg \{ \bar{\ga}^{ab} \tilde{D}_{a} \big [\nabla_{(r}\nabla_{b)}E + \nabla_{(r}F_{b)}+ h_{rb}\big ] \bigg \}\, , \notag \\[1ex]
    \nu &= - \bigg ( \psi + \frac{1}{3}\Delta_3 E \bigg ) + \frac{\bar{\ga}^{ab}}{2} \bigg \{\nabla_{a} \big (\nabla_{b}E+ F_{b} \big )+ h_{ab} \bigg \}\, , \notag \\[1ex]
    \mu &= \frac{2}{r^2} \frac{1}{D^2 (D^2+2)} \bigg \{D^{ab} \bigg [\nabla_{a} \big (\nabla_{b}E + F_{b} \big ) + h_{ab} \bigg ] \bigg \}\, , \notag \\[1ex]
    \hat{\mu} &= \frac{2}{r^2} \frac{1}{D^2 (D^2+2)} \bigg \{\tilde{D}^{ab} \bigg [\nabla_{a} \big (\nabla_{b}E + F_{b} \big ) + h_{ab} \bigg ] \bigg \}\, .
    \label{eq:SPS-in-terms-SVT}
\end{align}
These new relations will be useful when we compute the angular distance in terms of the PG gravitational potentials $\Phi$ and $\Psi$  defined in Sect.~\ref{subsec:gauge-inv-first-order-light-cone}. We emphasize that, as previously shown at first order (see \cite{Fanizza:2020xtv,Fanizza:2023ixk}), the gauge transformations are properly mapped from the standard perturbations to the light-cone ones.

\section{The Geodesic Light-Cone Gauge}
\label{sec:GLCgauge}
In this section, we will define the so-called \textit{Geodesic Light-Cone} gauge at a fully non-linear level. Then,  we will show the realizations of the linearized and quadratic versions of such a gauge fixing within the light-cone perturbation theory. In particular, we will provide the explicit expressions of the gauge modes needed to fix the GLC gauge at first and second order.

\subsection{Fully non-linear framework}
\label{subsec:GLC-anyorder}
In order to describe light signals  propagating along the past light-cone of a  free-falling observer, it is convenient to resort to the  so-called \textit{Geodesic Light-Cone} (GLC) coordinates $x^\mu = (\tau, w, \tilde{\theta}^a)$, first introduced in \cite{Gasperini:2011us}. These coordinates,  defined by the line element\footnote{See \cite{Fleury:2016htl} for a \quotes{bottom-up} derivation of this metric.}
\begin{equation}
\text{d}s^2 =  - 2 \Upsilon \text{d} \tau \text{d}w + \Upsilon^2 \text{d}w^2 + g_{ab} (\text{d}\tilde{\theta}^a-\mathcal{U}^a \text{d}w)(\text{d}\tilde{\theta}^b - \mathcal{U}^b \text{d}w) \, ,
\label{eq:GLCmetric-start}
\end{equation}
are such that $\tau$ is the proper time of a free-falling observer and the level sets of $w$ identify its past light-cone. Then, the intersection between  constant $\tau$ and constant $w$ hyper-surfaces defines a space diffeomorphic to a 2-sphere parameterized by the angles $\tilde{\theta}^a$.

In \eqref{eq:GLCmetric-start}, we have introduced the six arbitrary functions $\Upsilon$, $\mathcal{U}^a$ and $g_{ab} = g_{ba}$, where the symmetric metric $g_{ab}$ corresponds to the metric induced on the 2-sphere generated by the vectors $\partial_{\tilde{\theta}^a}$. In matrix form, the metric and its inverse read
\begin{equation}
g_{\mu \nu} = 
\begin{pmatrix}
0 & -\Upsilon & \mathbf{0}\\
-\Upsilon & \Upsilon^2 + \mathcal{U}^2 & -\mathcal{U}_b\\
\mathbf{0}^\t{T} & - \mathcal{U}^{\text{T}}_a & g_{ab}
\end{pmatrix} \qquad , \qquad 
g^{\mu \nu} = 
\begin{pmatrix}
- 1 & -1/\Upsilon & -\mathcal{U}^b/\Upsilon\\
-1/\Upsilon & 0  & \mathbf{0}\\
- (\mathcal{U}^a)^\text{T} / \Upsilon & \mathbf{0}^\t{T}  & g^{ab}
\end{pmatrix}
\end{equation}
with $\mathcal{U}^a \equiv g^{ab}\mathcal{U}_b$, $\mathcal{U}^2 \equiv \mathcal{U}^a \mathcal{U}_a$ and $g^{ab} \equiv \big (g_{ab} \big )^{-1}$. Exploiting the above formulae, it is easy to check that
\begin{equation}
    \sqrt{-g_{(4)}} = \Upsilon \sqrt{|g|} \quad , \quad \partial^\mu w \partial_\mu w =0 \quad , \quad \partial^\nu \tau \nabla_\nu \big (\partial_\mu \tau \big )=0 \, , 
    \label{eq:properties-glc}
\end{equation}
where $g_{(4)}=\t{det}[g_{\mu \nu}]$ and $g \equiv \text{det}[g_{ab}]$. The second equality confirms that $w$ is indeed a null coordinate. The third equality, directly stemming from $g^{\tau \tau}=-1$, shows that $\pa_\mu \tau$ defines a geodesic flow, \textit{i.e.}  $u_\mu = - \partial_\mu \tau = -\delta^\tau_\mu$  is the 4-velocity of a free-falling observer, solution to the time-like geodesic equations. Note also that 
\begin{equation}
    u^\mu = \bigg ( 1, \frac{1}{\Upsilon}, \frac{\mc{U}^a}{\Upsilon} \bigg ) \, ,
    \label{eq:4-velocity-GLC}
\end{equation}
so, since  light-rays propagate tangentially to the past light-cone of a free-falling observer sitting on the tip of its light-cone, future time-like motion (\textit{i.e.} photons incoming to the observer) requires $u^w >0$, hence $\Upsilon >0$.
Furthermore, the normal vector to the $w = \text{const.}$ hyper-surfaces can be identified with 
\begin{equation}
    k^\mu \equiv -\omega_{\text{phys}} g^{\mu \nu} \partial_\nu w = -\omega_{\text{phys}} g^{\mu w} = \frac{\omega_{\text{phys}}}{\Upsilon} \delta^\mu_\tau \, .
    \label{eq:normal-vector-glc-light-cone}
\end{equation}
Using the metric \eqref{eq:GLCmetric-start}, we see that this vector is null ($k^\mu k_\mu = 0$), geodesic ($k^\nu \nabla_\nu k^\mu=0$) and orthogonal to the 2-sphere parameterized by the angles $\tilde{\theta}^a$ ($k^\mu \partial_\mu \tilde{\theta}^a \propto g^{w a}=0$). Therefore, it can be correctly interpreted as the wave-vector of photons incoming to an observer, being  the solution to the light-like geodesics $\tilde{\theta}^a = \text{const.}$ So, $\omega_{\text{phys}}$ is the physical frequency of the signal. The fact that, in these coordinates, photons propagate with constant values of $w$ and $\tilde{\theta}^a$ makes it possible to  obtain
fully non-linear expressions for light-like observables, as shown in \cite{Gasperini:2011us, Fanizza:2013doa}.

From a geometrical point of view,   the standard FLRW metric (in spherical coordinates) is recovered starting from the GLC one once we set
\begin{equation}
\begin{split}
\tau = t \quad  , \quad w = r+\eta (\tau) \quad  ,  \quad  \Upsilon = a(t) \quad , \quad \mathcal{U}^a = 0 \quad , \quad g_{ab} \text{d}\tilde{\theta}^a \text{d}  \tilde{\theta}^b = a^2(t) r^2 \text{d} \Omega^2 \, .
\end{split}
\label{eq:GLC-background-FLRW}
\end{equation}
In particular, as we will use later on, $\mc{U}^a$ vanishes in the FLRW limit. 

To conclude, the metric \eqref{eq:GLCmetric-start} does not completely specify the gauge, because the following  residual gauge transformations, \textit{i.e.}
\begin{equation}
    w \rightarrow w^\prime = w^\prime (w) \qquad  , \qquad
      \tilde{\theta}^a \rightarrow \tilde{\theta}^{a \prime} = \tilde{\theta}^{a \prime} (w, \tilde{\theta}^a) \, , 
     \label{eq:glc-residual-freedom}
\end{equation}
preserve the form of the metric. As first pointed out in \cite{Fanizza:2013doa} (see also \cite{Fleury:2016htl}),  the first transformation amounts to relabelling light-cones, while the second one to relabelling light-rays with the angles $\tilde{\theta}^a$ and to  how
such a  labelling is transferred from one light-cone to another.

\subsection{Gauge modes at first order}
Following \cite{Fanizza:2020xtv}, now we recall which conditions must be imposed on the first-order perturbations appearing in the metric \eqref{eq:metricGLC} such that the GLC gauge of Eq.~\eqref{eq:GLCmetric-start} is recovered at a perturbative level. 

Fixing the GLC gauge in Eq.~\eqref{eq:metricGLC}, such that its forms coincides with Eq.~\eqref{eq:GLCmetric-start}, requires that the $\tau\tau$ and $\tau a$ metric entries  must vanish. This leads to the following conditions:
\begin{equation}
L^{(1)}=0 \qquad  ,\qquad V^{(1)}_a =0 \,.
\label{eq:GLC-linear-1}
\end{equation}
Moreover, from Eq.~\eqref{eq:metricGLC} we also see that
the $\tau w$ and $w w$ metric entries are related and lead to the following additional condition:
\begin{equation}
N^{(1)}+2aM^{(1)}=0\,.
\label{eq:GLC-linear-2}
\end{equation}
The condition $V^{(1)}_a =0$ can be then safely obtained by
\begin{equation}
{v}^{(1)}={\hat{v}}^{(1)}=0 \,.
\label{eq:GLC-linearized}
\end{equation}
Next, using Eqs.~\eqref{eq:gauge-transf-1st-order} and \eqref{eq:gauge-transf-1st-decom}, we can  determine the dependence of the gauge field $\xi^\mu_{(1)}$ on the metric perturbations $f^{(1)}_{\mu \nu}$ to move from a generic gauge to the GLC one. In particular, we obtain the following differential equations for the gauge fields, 
\begin{align}
\dot{\xi}^\tau_{(1)} +\frac{\partial_w \xi^\tau_{(1)}}{a}&=-\frac{1}{2} (a^2L^{(1)} + N^{(1)}+2aM^{(1)}) \, , \notag \\[1ex]
\dot{\xi}_{(1)}^w &= -\frac{a}{2}L^{(1)} \, , \notag \\[1ex]
\dot{\chi}_{(1)}&= v^{(1)}+\frac{\xi^w_{(1)}}{ar^2} \, , \notag \\[1ex]
\dot{\hat{\chi}}_{(1)} &= \hat{v}^{(1)} \, ,
\label{eq:gauge-inv-1st-order}
\end{align}
whose solutions are
\begin{align}
\xi^\tau_{(1)} &= -\frac{1}{2}\int_{\tau_{\text{in}}}^\tau \text{d}\tau^\prime \, \big ( a^2L^{(1)} + N^{(1)}+2aM^{(1)}\big )\big (\tau^\prime, w-\eta (\tau)+ \eta (\tau^\prime) \big ) \ ,  \notag \\[1ex]
\xi^w_{(1)} &= \frac{1}{2} \int_\tau^{\tau_\t{o}} \text{d}\tau^\prime aL^{(1)} + w^{(1)}_0 \, , \notag\\[1ex]
\chi_{(1)} &= - \int_\tau^{\tau_\t{o}} \text{d}\tau^\prime \bigg (v^{(1)}+\frac{1}{2ar^2} \int_{\tau^\prime}^{\tau_\t{o}} \text{d}\tau^{\prime \prime} aL^{(1)} + \frac{w^{(1)}_0}{ar^2} \bigg )+ \chi^{(1)}_0 \, , \notag \\[1ex]
\hat{\chi}_{(1)} &= -\int_\tau^{\tau_\t{o}} \text{d}\tau^\prime \, \hat{v}^{(1)} + \hat{\chi}^{(1)}_0 \, , 
\label{eq:xi-1st-order}
\end{align}
where $\tau\o$ is the present time,  $w^{(1)}_0 = w^{(1)}_0 (w)$, $\chi^{(1)}_0= \chi^{(1)}_0(w, \tilde{\theta}^a)$ and $\hat{\chi}^{(1)}_0= \hat{\chi}^{(1)}_0(w, \tilde{\theta}^a)$. As stressed in \cite{Fanizza:2020xtv}, these last three functions reflect the fact that the GLC gauge is defined up to the residual transformations of Eqs.~\eqref{eq:glc-residual-freedom}.

\subsection{Gauge modes at second order}
\label{subsec:GLC-2nd-order}
Proceeding as above, the GLC gauge-fixing conditions at second order are
\begin{align}
& L^{(2)}=0 \, , \notag \\[1ex]
& V^{(2)}_a =0 \, , \notag \\[1ex]
& 4N^{(2)}- (N^{(1)})^2 +8aM^{(2)}- 4 U^2_{(1)} =0 \, .
\label{eq:fixing-GLC-gauge}
\end{align}
Using these equations and Eqs.~\eqref{eq:gauge-inv-1st-order} and \eqref{eq:xi-1st-order}, we can then compute the expressions for the correspondent second-order gauge modes.

The constraint $\tilde{L}^{(2)}=0$, imposed in Eq.~\eqref{eq:tildeL}, gives the following differential equation for $\xi^w_{(2)}$,
\begin{align}
\partial_\tau \xi^w_{(2)} &= -aL^{(2)} + \partial_\tau \big (aL^{(1)}\xi^\tau_{(1)} \big ) + a \xi^i_{(1)}\partial_i L^{(1)} - a^2 L^{(1)} \bigg ( M^{(1)}+\frac{a}{4}L^{(1)}\bigg )\notag \\[1ex]
& \quad +  \partial_\tau \big (\xi^\mu_{(1)} \partial_\mu \xi^w_{(1)} \big )+ ar^2 q^{ab} \mathscr{L}^{(2)}_{ab} \, ,
\end{align}
where we have introduced
\begin{align}
    \mathscr{L}^{(2)}_{ab} & \equiv \bigg ( D_a v^{(1)}+ \tilde{D}_a \hat{v}^{(1)}+\frac{1}{ar^2} D_a\xi^w_{(1)}\bigg )\bigg ( D_b v^{(1)}+ \tilde{D}_b \hat{v}^{(1)}-\frac{1}{ar^2}D_b \xi^w_{(1)}\bigg ) \, .  
\end{align}
The solution can then be written as
\begin{align}
\xi^w_{(2)} & = \int_\tau^{\tau_\t{o}} \text{d}\tau^\prime \, a \bigg [ L^{(2)} -\xi^i_{(1)} \partial_i L^{(1)}+ aL^{(1)} \bigg (M^{(1)}+ \frac{a}{4}L^{(1)} \bigg )-r^2 q^{ab} \mathscr{L}^{(2)}_{ab} \bigg ] \notag\\[1ex]
& \quad + a L^{(1)}\xi^\tau_{(1)}+\xi^\mu_{(1)}\pa_\mu \xi^w_{(1)} + w^{(2)}_0 (w, \tilde{\theta}^a) \, .
\label{eq:xi-w-2nd-order}
\end{align}
Using Eq.~\eqref{eq:xi-w-2nd-order}, we can now impose the constraints $\tilde{v}^{(2)}=\hat{\tilde{v}}^{(2)}=0$ in Eqs.~\eqref{eq:v-2-tilde}. From this, we obtain the following expressions for $\chi_{(2)}$ and $\hat{\chi}_{(2)}$:
\begin{align}
\chi_{(2)} &= - \int_\tau^{\tau_\t{o}} \text{d}\tau^\prime \, \bigg [2v^{(2)}+\frac{\xi^w_{(2)}}{ar^2}-\frac{\xi^\mu_{(1)}\partial_\mu \xi^w_{(1)}}{ar^2}+\frac{2}{r^2D^2} \mathcal{V}^{(2)} \bigg ] + \chi^{(2)}_0(w, \tilde{\theta}^a) \notag \\[1ex]
& = - \int_\tau^{\tau_\t{o}} \text{d}\tau^\prime \bigg \{
2v^{(2)}+\frac{1}{ar^2} \bigg [\int_{\tau^\prime}^{\tau_\t{o}} \text{d}\tau^{\prime \prime} \, a \bigg (L^{(2)}-\xi^i_{(1)}\partial_i L^{(1)}+ aL^{(1)}\bigg (M^{(1)}+\frac{a}{4}L^{(1)} \bigg ) \notag \\[1ex]
& \quad - r^2 q^{ab} \mathscr{L}^{(2)}_{ab}
\bigg ) +aL^{(1)}\xi^\tau_{(1)}+w^{(2)}_0 (w, \tilde{\theta}^a)\bigg ] + \frac{2}{r^2D^2}\mc{V}^{(2)}
\bigg \}  + \chi^{(2)}_0 (w, \tilde{\theta}^a)\, , \notag \\[1ex]
\hat{\chi}_{(2)} &=-2 \int_\tau^{\tau_\t{o}} \text{d}\tau^\prime \, \bigg [\hat{v}^{(2)}+ \frac{1}{r^2D^2}\hat{\mc{V}}^{(2)} \bigg ] + \hat{\chi}^{(2)}_0(w, \tilde{\theta}^a) \, .
\label{eq:chi-2nd-order}
\end{align}
Finally, the condition $4\tilde{N}^{(2)}- \big ( \tilde{N}^{(1)}\big )^2+ 8a\tilde{M}^{(2)}-4\tilde{U}^2_{(1)}=0$ gives the following partial differential equation for $\xi^\tau_{(2)}$,
\begin{equation}
\begin{split}
\dot{\xi}^\tau_{(2)}+\frac{\partial_w \xi^\tau_{(2)}}{a} = -(a^2L^{(2)}+2aM^{(2)}+ N^{(2)}) +\frac{1}{4}(N^{(1)})^2+ F^{(2)} \, ,
\end{split}
\label{eq:eqdiff-xi0-2}
\end{equation}
where we have defined 
\begin{align}
F^{(2)} & = a \bigg [  \partial_\tau \big (aL^{(1)}\xi^\tau_{(1)} \big ) + a \xi^i_{(1)}\partial_i L^{(1)} - a^2 L^{(1)} \bigg ( M^{(1)}+\frac{a}{4}L^{(1)}\bigg )+   \partial_\tau \big (\xi^\mu_{(1)} \partial_\mu \xi^w_{(1)} \big )\notag \\[1ex]
& \quad + ar^2 q^{ab} \mathscr{L}^{(2)}_{ab} \bigg ]  +  2aL^{(1)}\partial_w \xi^\tau_{(1)}+2M^{(1)}\Big [a\dot{\xi}^\tau_{(1)}+ \partial_w \xi^\tau_{(1)}+2\dot{a}\xi^\tau_{(1)}\notag \\[1ex]
& \quad + a \partial_w \xi^w_{(1)} \Big ] + N^{(1)} \bigg (H\xi^\tau_{(1)}+ 2a \dot{\xi}^w_{(1)}+\frac{1}{a}\partial_w \xi^\tau_{(1)}+\partial_w \xi^w_{(1)} \bigg ) \notag \\[1ex]
& \quad +  \Big \{ 2ar^2(D_au^{(1)}+\tilde{D}_a\hat{u}^{(1)}) - 2ar^2 \partial_w (D_a \chi_{(1)}+\tilde{D}_a \hat{\chi}_{(1)})  \notag \\[1ex]
& \quad +\partial_a (\xi^\tau_{(1)}- a\xi^w_{(1)}) \Big \} \big [q^{ab}(D_b \dot{\chi}_{(1)}+\tilde{D}_b \dot{\hat{\chi}}_{(1)}) \big ] \notag \\[1ex]
& \quad + \Big \{ 2r^2 \big [ a (D_a v^{(1)}+\tilde{D}_a \hat{v}^{(1)}) +D_a u^{(1)}+\tilde{D}_a \hat{u}^{(1)}\big ] - r^2 \partial_w (D_a \chi_{(1)}+\tilde{D}_a \hat{\chi}_{(1)}) \notag \\[1ex]
& \quad + \partial_a \xi^w_{(1)}\Big \} \big [q^{ab}\partial_w (D_b \chi_{(1)}+\tilde{D}_b \hat{\chi}_{(1)}) \big ] +\frac{\big (\partial_w \xi^\tau_{(1)} \big )^2}{a^2}+2H\xi^\tau_{(1)}\dot{\xi}^\tau_{(1)}\notag \\[1ex]
& \quad + \xi^\mu_{(1)}\partial_\mu N^{(1)} -a \xi^\mu_{(1)} \partial_\mu \big (\dot{\xi}^w_{(1)}-2M^{(1)} \big ) + \frac{1}{a}\partial_w \big (\xi^\mu_{(1)}\partial_\mu \xi^\tau_{(1)} \big ) -\pa_w (\xi^\mu_{(1)} \partial_\mu \xi^w_{(1)}) \notag \\[1ex] 
& \quad + \big (\dot{\xi}^\tau_{(1)}+ \partial_w \xi^w_{(1)} \big )^2 + 4 \dot{\xi}^w_{(1)}\partial_w \xi^\tau_{(1)}-4\dot{a}\xi^\tau_{(1)}\dot{\xi}^w_{(1)}\notag \\[1ex]
& \quad + \xi^\mu_{(1)}\partial_\mu \big ( \dot{\xi}^\tau_{(1)} + \partial_w \xi^w \big )  - a\big (\dot{\xi}^\tau_{(1)}\dot{\xi}^w_{(1)}+ 3 \dot{\xi}^w_{(1)}\partial_w \xi^w_{(1)}\big ) \notag \\[1ex]
& \quad + \bigg [r^2 (D_a u^{(1)}+\tilde{D}_a \hat{u}^{(1)})+\frac{1}{a}\pa_a \xi^\tau_{(1)}-\pa_a \xi^w_{(1)}-r^2 \pa_w(D_a \chi_{(1)}+\tilde{D}_a \hat{\chi}_{(1)}) \bigg ]^2\, .
\label{eq:F2-def}
\end{align}
 The solution can then be written as
\begin{equation}
\begin{split}
\xi^\tau_{(2)} & =  \int_{\tau_{\text{in}}}^\tau \text{d}\tau^{\prime} \bigg (-(a^2L^{(2)}+2aM^{(2)}+N^{(2)})+\frac{1}{4}(N^{(1)})^2+F^{(2)} \bigg )\big (\tau^{\prime}, w-\eta (\tau) + \eta (\tau^{\prime}) \big )
\end{split} \, .
\label{eq:xi-0-2nd-order}
\end{equation}


\section{The Observational Synchronous Gauge}
\label{sec:OSG}
In this section, we will show how to impose the analogous of the GLC gauge in standard perturbation theory at second order.
This fixing is called the  \textit{Observational Synchronous Gauge}, first introduced in  \cite{Fanizza:2020xtv} at first order in perturbation theory. We will then examine the conceptual similarities and differences between this gauge  and the Synchronous Gauge (SG).

\subsection{First order}
As first pointed out in \cite{Fanizza:2020xtv}, using Eqs.~\eqref{eq:dictionary}, it is easy to see that the first-order gauge fixing in standard perturbation theory, equivalent to the GLC one described in Eqs.~\eqref{eq:GLC-linear-1}-\eqref{eq:GLC-linearized}, is given by
\begin{equation}
\phi^{(1)} =0 \qquad  ,  \qquad \mathcal{B}^{(1)}_r = -\frac{1}{2}\mathcal{C}^{(1)}_{rr} \qquad  ,  \qquad \mathcal{B}^{(1)}_a = - \mathcal{C}^{(1)}_{ra} \, ,
\label{eq:OSG}
\end{equation}
 where we note that we still have the condition $\phi^{(1)}=0$, as in the SG. 
Furthermore, let us recall that within the GLC gauge  not only the time $\tau$  is the one of a free-falling observer, but we also have that the angles $\tilde{\theta}^a$, representing the directions of observation in the sky, are constant along the light-cone.
Consequently, in \cite{Fanizza:2020xtv} the standard counterpart of the GLC gauge, described in \eqref{eq:OSG}, was dubbed \textit{Observational Synchronous Gauge} (OSG). 

As a conclusive remark, note also that the naive to all-orders extension of the SG consists in imposing the conditions $\phi=0$  and $B=0$ at a fully non-linear level, whereas here an order-by-order fixing of the OSG is performed.

\subsection{Second order}
\label{subsec:OSG-2nd-order}

Let us now show how to fix the OSG at second order and explain why this choice must be exploited order by order. 
Starting from Eqs.~\eqref{eq:dictionary}, which 
are valid at any perturbative order,
 the conditions $L^{(2)}=0$ and $V^{(2)}_a=0$ immediately give 
\begin{equation}
\mathcal{B}^{(2)}_a = - \mathcal{C}^{(2)}_{ra} \qquad , \qquad 
\phi^{(2)} = -\frac{1}{2} (N^{(2)}+2aM^{(2)})\, .
\end{equation}
By considering the remaining constraint
of Eqs.~\eqref{eq:fixing-GLC-gauge}, i.e. $4N^{(2)}+8aM^{(2)}-\big ( N^{(1)}\big)^2 - 4U^2_{(1)}=0$, we can then note that
\begin{equation}
\phi^{(2)} = -\frac{1}{8} \Big [ \big (N^{(1)} \big )^2 + 4 U^2_{(1)} \Big ]\, .
\end{equation}
Furthermore, using the above results and Eqs.~\eqref{eq:dictionary}
we can easily obtain the relation
\begin{equation}
\mathcal{B}^{(2)}_r =\phi^{(2)}  -\frac{1}{2}\mathcal{C}^{(2)}_{rr}\,.
\end{equation}
Summarizing, we have the 
following relations that can be used to define the OSG at the second perturbative order:
\begin{equation}
\phi^{(2)} =  -\frac{1}{8} \Big [ \big (N^{(1)} \big )^2 + 4U^2_{(1)} \Big ] \quad  , \quad \mathcal{B}^{(2)}_r =\phi^{(2)}  -\frac{1}{2}\mathcal{C}^{(2)}_{rr}\quad  ,  \quad \mathcal{B}^{(2)}_a = - \mathcal{C}^{(2)}_{ra} \, .
\label{eq:OSG-second-order}
\end{equation}
Let us note that $\phi^{(2)} \neq 0$, whereas this is not the case at first order. However, as first pointed out in \cite{Ben-Dayan:2012uam}, the proper time $\tau$ of the GLC gauge is non-perturbatively equal to the one of the SG.
This implies that, at any perturbative order, the observers of the SG and OSG  are both free-falling and then physically equivalent.
Therefore, they measure the same physical observables. For the sake of completeness, and as a non-trivial check of our new perturbation theory, we will prove the above point that both the SG and OSG observers are free-falling  in Appendix \ref{app:A} by working at the perturbative level within the light-cone perturbation theory.

Instead, we now present a detailed comparison between the two gauges, focusing on their similarities and differences through the GLC gauge modes derived in the previous section. In particular, by expressing these gauge modes starting from the SG formulation defined on the light-cone background with perturbations, we demonstrate that the GLC/OSG and SG gauges are fully equivalent at the observer position, but differ elsewhere. More specifically, in the GLC/OSG gauge the hyper-surfaces $w = \text{const.}$ non-perturbatively define the past light-cone of the observer, and the angular coordinates $\tilde{\theta}^a$ remain constant along this light-cone ($\tilde{\theta}^a_{\t{source}}=
\tilde{\theta}^a_{\t{observer}}$). These properties do not hold in the SG.

In order to show this, let us first derive the SG counterpart on the perturbed light-cone background.
We have that the non-perturbative definition of the SG is given by $\phi^{(n)}=0$, $B^{(n)}=0$ and $B^{(n)}_i=0$ for any perturbative order $n$. As a consequence, using Eqs.~\eqref{eq:dictionary}, the non-perturbative version of the SG in the perturbative light-cone framework is given by
\begin{align}
& a^2 L^{(n)} + N^{(n)}  + 2a M^{(n)}  = 0 \, , \nex 
& N^{(n)} =-a M^{(n)} \, , \nex 
&v^{(n)}  = -\frac{1}{a} u^{(n)} \, , \nex 
&\hat{v}^{(n)}  = -\frac{1}{a} \hat{u}^{(n)} \, . 
\label{eq:SG-lightcone}
\end{align}
Let us now make a brief remark and first look at the gauge transformations needed to fix the GLC gauge.
Starting from Eqs.~\eqref{eq:dictionary}, and using the expression of $\phi^{(1)}$ in Eqs.~\eqref{eq:xi-1st-order}, we obtain the first-order gauge mode to go to GLC gauge written in standard perturbation theory:
\begin{equation}
\xi^\tau_{(1)} = \int_{\tau_{\text{in}}}^\tau \text{d}\tau^\prime \, \phi^{(1)} \big ( \tau^\prime, w-\eta(\tau)+ \eta (\tau^\prime)\big ) = \int_{\eta_{\text{in}}}^\eta \text{d}\eta^\prime a \phi^{(1)}(\eta^\prime, r) \, .
\label{eq:xi0}
\end{equation}
This shows how $\xi^\tau_{(1)}$ corresponds to same time-shift of the SG, obtained by requiring $\tilde{\phi}^{(1)}=0$ and  $\tilde{B}^{(1)}=0$ in Eqs.~\eqref{eq:gauge-transf-scalars-FLRW}.
This first result already shows at first order how the GLC proper time and the SG one are identical. 

In general, taking the limit $\tau\rightarrow \tau \o$ in Eqs.~(\ref{eq:xi-1st-order}) at first order and in Eqs.~\eqref{eq:xi-w-2nd-order}, \eqref{eq:chi-2nd-order} and \eqref{eq:xi-0-2nd-order} at second order,  we see that the only non-zero gauge modes are the time ones, modulo a non-zero integration constant in $\xi^{w}_{\t{o}}$ and $\xi^{a}_\t{o}$. 
Thus, \textit{a priori} the only difference between a general gauge and the GLC one would be in the definition of the proper time at the observer.
Then, as a matter of fact, with a little of algebra one can prove that the first- and second-order time gauge modes needed to move from the SG to the GLC one (or the OSG) are zero. Indeed, using the first of Eqs.~\eqref{eq:SG-lightcone}, we immediately see that $\xi^\tau_{(1)}=0$. 

At second order, we can just use Eqs.~\eqref{eq:gauge-inv-1st-order} and \eqref{eq:SG-lightcone} into \eqref{eq:F2-def} and \eqref{eq:xi-0-2nd-order}, to show that, after some algebraic manipulations, we get
\begin{align}
\dot{\xi}^\tau_{(2)}+\frac{\partial_w \xi^\tau_{(2)}}{a} &=0 \, .
\end{align}
This proves that the solution for $\xi^\tau_{(2)}$ is a constant, which can be safely taken to be zero.
In this way, the conditions  $\xi^\tau_{(1, \t{SG} \to \t{OSG})}=\xi^\tau_{(2, \t{SG} \to \t{OSG})}=0$ are valid in general, not only in the limit $\tau\rightarrow \tau \o$. This is a consequence of the fact  that both the observers in OSG and  SG foliate the space-time into the same $\tau = \t{const.}$ hyper-surfaces. 

On the other hand, the light-cone and angular gauge modes to go from the SG to the GLC gauge are non-zero and given at first order by
\begin{align}
\xi^w_{(1)}&= -\frac{1}{2}\int_\tau^{\tau_\t{o}}\t{d}\tau^\pr\, M + w^{(1)}_0 \, , \nex
\chi_{(1)}&= \int_\tau^{\tau_\t{o}}\frac{\t{d}\tau^\pr}{a}\, \bigg ( u^{(1)}+\frac{1}{2r^2}\int_{\tau^\pr}^{\tau_\t{o}}\t{d}\tau^{\pr\pr} 
\, M - \frac{w^{(1)}_0}{r^2}\bigg ) + \chi^{(1)}_0 \,  , \nex
\hat{\chi}_{(1)}&= \int_\tau^{\tau_\t{o}}\frac{\t{d}\tau^\pr}{a}\,  \hat{u}^{(1)}+ \hat{\chi}^{(1)}_0 \, , 
\end{align}
which is consequence of the fact that the two gauges share the same proper time and are physically equivalent at the observer position, but differ at a generic time different from the one of the observer. At  second order, we also have non-zero expressions for $\xi^w_{(2)}$, $\chi_{(2)}$ and $\hat{\chi}_{(2)}$, that we do not show here for the sake of brevity.


\section{Gauge-Invariant Variables and Observables}
\label{sec:obs-glc}
In this section, we will first recall that cosmological observables admit non-perturbative expressions once the GLC gauge is fixed. As mentioned, this is ensured by the fact that, in the GLC gauge, observer and source are connected through $\tilde{\theta}^a =\t{constant}$ geodesics on a $w=\t{constant}$ past light-cone of a free-falling observer. We will then compute two sets of gauge-invariant variables which correspond to the first- and second-order versions of the GLC gauge. Finally, we will provide a  unified treatment of all the GLC approaches to cosmological perturbations currently available in the literature.

\subsection{Cosmological observables in the Geodesic Light-Cone gauge}
\label{eq:cosmobs}
\paragraph{Redshift.} Considering that in the GLC gauge the photon 4-momentum and the 4-velocity of a free-falling observer are given by
\begin{equation}
    k_\mu = \pa_\mu w \qquad  , \qquad  u_\mu = -\pa_\mu \tau \, , 
\end{equation}
in \cite{Gasperini:2011us} it was shown that the  redshift  in GLC coordinates is exactly given by
\begin{equation}
    1+z = \frac{(k^\mu u_\mu)_\text{s}}{(k^\mu u_\mu)_\text{o}} = \frac{\Upsilon (\tau_\text{o}, w, \tilde{\theta}^a_\text{o})}{\Upsilon (\tau_\text{s}, w, \tilde{\theta}^a_\text{s})} \, , 
    \label{eq:z-fully-non-linear-glc}
\end{equation}
where we have accounted for the fact that the source and the observer are lying on the same light-cone, namely $w_\text{s}=w_\text{o}\equiv w$. The subscripts \quotes{s} and \quotes{o} stand for evaluation at the source and observer position.

\paragraph{Angular distance.} Another late-time cosmological observable admitting a simple, non-perturbative expression within the GLC gauge fixing is the angular distance, which we call $d_\text{A}$. In a generic framework, the latter can be computed as the square root of the determinant of the Jacobi map $J^A_B$, connecting a source to a given observer via the relation (see \cite{Schneider:1992bmb}) 
\begin{equation}
    \zeta^A (\la_\text{s}) = J^A_B (\la_\text{s}, \la_\text{o}) \bigg ( \frac{k^\mu \pa_\mu \xi^B}{k^\nu u_\nu}\bigg )_\text{o} \qquad ,  \qquad A,B=1,2\, , 
    \label{eq:jacobi-map}
\end{equation}
where $\zeta^\mu \equiv \zeta^A s^\mu_A$ is the displacement between 2 geodesics due to the geodesic deviation effect,  $\lambda$ is the affine parameter along the geodesics, $\zeta^A = g_{\mu \nu} \zeta^\mu s^\nu_A$ and the vectors $s^\mu_A$ are defined by the conditions (see \cite{Fleury:2013sna, Pitrou:2015iya})
\begin{equation}
    g_{\mu \nu} s^\mu_A s^\nu_B = \delta_{AB} \qquad  ,  \qquad s^\mu_A u_\mu =0 = s^\mu_A k_\mu = \Pi^\mu_{\nu}\nabla_\la s^\nu_A  \, , 
\end{equation}
with 
\begin{equation}
    \nabla_\la \equiv k^\mu \nabla_\mu \qquad  ,  \qquad \Pi^\mu_\nu \equiv \delta^\mu_\nu - \frac{k^\mu k_\nu}{(u^\rho k_\rho)^2} - \frac{k^\mu u_\nu + u^\mu k_\nu}{u^\rho k_\rho} \, .
\end{equation}
If we then fix the GLC gauge, in \cite{Fanizza:2013doa} it was first shown that, in this case, we have
\begin{equation}
    d^2_\text{A} = \text{det}[J^A_B] = \frac{\sqrt{g}}{\left (\frac{\text{det}[\dot{g}_{ab}]}{4\sqrt{g}} \right )_\text{o}}  \, .
    \label{eq:dA-fully-non-linear-glc}
\end{equation}
The determinant at the observer has a simple limit in the so-called observational gauge. This was proved in \cite{Fanizza:2018tzp}, and we will also demonstrate this in the next section using perturbation theory.

\subsection{Gauge-invariant variables on the light-cone}
\label{subsec:gaugeinvariance-LC}
\paragraph{GLC gauge-invariant variables.} Here we report a set of first- and second-order gauge-invariant variables  which will be exploited while evaluating the angular distance–redshift relation in a gauge-invariant way within the perturbative scheme defined in the previous sections. 

In particular, starting from  Eqs.~\eqref{eq:xi-1st-order}, \eqref{eq:xi-w-2nd-order}, \eqref{eq:chi-2nd-order} and \eqref{eq:xi-0-2nd-order}, which give  the dependence of the gauge fields $\xi^\mu_{(1)}$ and $\xi^\mu_{(2)}$ on the metric perturbations in order to move 
from a general gauge to the GLC one, we can  perform a gauge transformation on the original perturbative light-cone variables using  these fields. In this way, we obtain a set of variables which are automatically gauge invariant. 
 
Proceeding as above, the set of gauge-invariant variables corresponding to the linearized version of the fully non-linear GLC gauge of Eq.~\eqref{eq:GLCmetric-start} was already obtained in \cite{Fanizza:2020xtv} with the following results: 
\begin{align}
\mathscr{V}^{(1)} &\equiv \nu^{(1)} - \frac{1}{2}D^2 \chi_{(1)} - \xi^\tau_{(1)} \bigg (H-\frac{1}{ar} \bigg )- \frac{\xi^w_{(1)}}{r} \, , \notag \\[1ex]
\mathscr{N}^{(1)} & \equiv N^{(1)}-2H\xi^\tau_{(1)} +\frac{2}{a}\partial_w \xi^\tau_{(1)} - 2\partial_w \xi^w_{(1)} \, , \notag \\[1ex]
\mathscr{M}^{(1)} &\equiv  \mu^{(1)} - \chi_{(1)} \, , \notag \\[1ex]
\mathscr{\hat{M}}^{(1)} &\equiv \hat{\mu}^{(1)} -\hat{\chi}_{(1)} \, , \notag \\[1ex]
\mathscr{U}^{(1)} &\equiv u^{(1)} + \frac{\xi^\tau_{(1)}}{ar^2} - \frac{\xi^w_{(1)}}{r^2}-\partial_w \chi_{(1)} \, , \notag \\[1ex]
\mathscr{\hat{U}}^{(1)} &\equiv \hat{u}^{(1)} - \partial_w \hat{\chi}_{(1)} \, .
\label{eq:gauge-inv-quantities-1st-order}
\end{align}
The above quantities are gauge invariant since they are the values of  the perturbations of the GLC gauge written in a general gauge.  Here, the gauge modes  are the ones obtained in Eqs.~\eqref{eq:xi-1st-order}, unlike the formulae written in Eqs.~\eqref{eq:gauge-transf-1st-order} and \eqref{eq:gauge-transf-1st-decom}. The same reasoning holds for the second-order gauge-invariant variables that we report in the following,
\begin{align}
\mathscr{V}^{(2)} & \equiv \nu^{(2)} -\frac{1}{4}D^2 \chi_{(2)}- \frac{1}{2}\xi^\tau_{(2)} \bigg (H-\frac{1}{ra} \bigg )-\frac{1}{2}\frac{\xi^w_{(2)}}{r}+\mathbb{V}^{(2)} \, , \notag \\[1ex]
\mathscr{N}^{(2)} & \equiv N^{(2)}-H\xi^\tau_{(2)}+\frac{1}{a}\partial_w \xi^\tau_{(2)}-\partial_w \xi^w_{(2)}+\mc{N}^{(2)} \, , \notag \\[1ex]
\mathscr{M}^{(2)} & \equiv \mu^{(2)}-\frac{1}{2}\chi_{(2)} +\frac{1}{r^2} \mathcal{D}^{-1} D^{ab} \mc{M}^{(2)}_{ab} \notag \, , \\[1ex]
\hat{\mathscr{M}}^{(2)} & \equiv \hat{\mu}^{(2)}-\frac{1}{2}\hat{\chi}_{(2)} +\frac{1}{r^2} \mathcal{D}^{-1} \tilde{D}^{ab} \mc{M}^{(2)}_{ab} \, , 
\notag \\[1ex]
\mathscr{U}^{(2)} & \equiv u^{(2)}+ \frac{1}{2ar^2}\xi^\tau_{(2)}-\frac{1}{2r^2} \xi^w_{(2)}-\frac{1}{2} \partial_w \chi_{(2)}-\frac{1}{2ar^2}\xi^\mu_{(1)} \partial_\mu \xi^\tau_{(1)}+ \frac{1}{r^2D^2}\mc{U}^{(2)} \, , 
\notag \\[1ex]
\hat{\mathscr{U}}^{(2)} & \equiv \hat{u}^{(2)}- \frac{1}{2} \partial_w \hat{\chi}_{(2)}+ \frac{1}{r^2D^2}\hat{\mc{U}}^{(2)} \, ,
\label{eq:glc-gauge-inv-2nd-order}
\end{align}
where, for convenience of notation,  we have defined
\begin{align}
\mathbb{V}^{(2)} &\equiv -\frac{1}{a^2r^2}\xi^\tau_{(1)} \partial_\tau (a^2r^2\nu^{(1)})  -\frac{1}{r^2}\xi^w_{(1)} \partial_w (r^2 \nu^{(1)})+ \frac{1}{2}(\dot{H}+2H^2)\big (\xi^\tau_{(1)} \big )^2 \notag \\[1ex]
& \quad + \frac{H}{2}\xi^\mu_{(1)}\partial_\mu \xi^\tau_{(1)}+ 2 H\xi^\tau_{(1)} \bigg (-\frac{\xi^\tau_{(1)}}{ar}+\frac{\xi^w_{(1)}}{r} \bigg ) + \frac{1}{4r^2} q^{ab} \mc{V}^{(2)}_{ab} \, , \notag \\[1ex]
\mc{N}^{(2)} & \equiv  r^2 \big [-2 (D_a u^{(1)}+\tilde{D}_a \hat{u}^{(1)})+ \partial_w (D_a \chi_{(1)}+\tilde{D}_a \hat{\chi}_{(1)}) \big ] \big [q^{ab}\partial_w (D_b \chi_{(1)}+\tilde{D}_b \hat{\chi}_{(1)})] \notag \\[1ex]
& \quad + \frac{\dot{a}}{a^2}\xi^\tau_{(1)} \big (\dot{a} \xi^\tau_{(1)}-2\partial_w  \xi^\tau_{(1)} \big ) -2 \big (M^{(1)}\partial_w \xi^\tau_{(1)}+  N^{(1)} \partial_w \xi^w_{(1)} \big ) + \big (\partial_w \xi^w_{(1)} \big )^2  \notag \\[1ex]
& \quad + \partial_w \big (\xi^\mu_{(1)}\partial_\mu \xi^w_{(1)} \big ) - \xi^\mu_{(1)}\partial_\mu N^{(1)}-\frac{1}{a} \bigg \{-\ddot{a} (\xi^\tau_{(1)})^2-\dot{a} \xi^\mu_{(1)}\partial_\mu \xi^\tau_{(1)}+2 \partial_w \xi^\tau_{(1)} \partial_w \xi^w_{(1)}\notag \\[1ex]
& \quad + \partial_w \big (\xi^\mu_{(1)} \partial_\mu \xi^\tau_{(1)} \big )+ \dot{a}\xi^\tau_{(1)} \big (2N^{(1)}-4 \partial_w \xi^w_{(1)} \big ) \bigg \} \, .
\label{eq:definitions-glc-gauge-inv-2nd}
\end{align}
 
\paragraph{Evaluation in terms of the Bardeen potentials.} To conclude, we now want to rewrite the first- and second-order gauge modes needed to fix the GLC gauge in terms of the gravitational potentials $\Phi^{(n)}$ and $\Psi^{(n)}$ at first and second order (see Eqs.~\eqref{eq:bardeen-1st} and \eqref{eq:bardeen-2nd}). These are also the values that the standard perturbations $\phi^{(n)}$ and $\psi^{(n)}$ acquire in the  Poisson gauge, defined by the conditions $E^{(n)}=0=B^{(n)}$. Therefore, to do so we   fix the gauge corresponding to PG within our new perturbation theory.  By neglecting vector and tensor fluctuations and by imposing the two aforementioned relations in Eqs.~\eqref{eq:SPS-in-terms-SVT}, valid at any order, we obtain the following non-perturbative definition of the PG:
\begin{align}
    L &= -\frac{2}{a^2} (\Phi + \Psi) \quad  , \quad  M = \frac{2}{a}\Psi \quad  , \quad N = -2\Psi \quad  , \quad \nu = -\Psi \, ,  \notag \\[1ex]
    v &=0 \quad  , \quad \hat{v} =0 \quad 
    ,\quad u = 0
    \quad ,\quad \hat{u} = 0
    \quad ,\quad \mu = 0
    \quad ,\quad \hat{\mu} =0 
    \, .
\label{eq:SPS-NG}
\end{align}
From now on, following the notation of \cite{Fanizza:2015swa,Marozzi:2014kua,BenDayan:2012wi}, we will use the definitions
\begin{equation}
  P \equiv \frac{1}{a(\tau)}\int_{\tau_{\t{in}}}^\tau \t{d}\tau^\pr \, \Phi^{(1)}\big ( w-\eta(\tau)+\eta(\tau^\pr)\big ) \qquad 
   , \qquad Q \equiv \int_{\tau}^{\tau_\t{o}} \frac{\t{d}\tau^\pr}{a} (\Psi^{(1)}+\Phi^{(1)} ) \, ,
  \label{eq:P-Q-defined}
\end{equation}
and we will omit the superscript \quotes{1} for the first-order perturbations $\Phi$ and $\Psi$. Moreover, we define the isotropic and anisotropic parts of the Bardeen potentials as
\begin{align}
    \Psi^\t{I} \equiv \frac{\Psi +\Phi}{2} \qquad  ,  \qquad  \Psi^\t{A} \equiv \frac{\Psi - \Phi}{2} \, .
    \label{eq:IA-potentials-defined}
\end{align}
Keeping $\Phi \neq \Psi$ already at first order allows us to consider also the case of non-vanishing anisotropic stress in the Einstein equations, which would be relevant to dark energy or modified gravity models (see, for example, \cite{Sawicki:2012re, Amendola:2012ky, Amendola:2013qna, Sobral-Blanco:2021cks, Castello:2022uuu}). 

Then, the forms that the first-order gauge modes   \eqref{eq:xi-1st-order} acquire using Eqs.~\eqref{eq:SPS-NG} are
\begin{align}
    \xi^\tau_{(1)} &= aP \, , \notag \\[1ex]
    \xi^w_{(1)} &= -Q + w^{(1)}_0 (w) \, , \notag \\[1ex]
    \chi_{(1)} &= \int_\tau^{\tau_\t{o}} \frac{\text{d}\tau^\prime}{ar^2}\,\big [Q-w^{(1)}_0 (w)\big ] + \chi^{(1)}_0 (w) \, , \notag \\[1ex]
     \hat{\chi}_{(1)} &= \hat{\chi}^{(1)}_0 (w) \, .
     \label{eq:gauge-modes-NG-1st}
\end{align}
Since we are adopting  a gauge-invariant procedure, we can already fix the PG at the level of the gauge modes. For the sake of completeness, we also report the first-order GLC gauge-invariant variables \eqref{eq:gauge-inv-quantities-1st-order} written in this notation:
\begin{align}
    \mathscr{V}^{(1)} &= -(\Psi^\t{I} +\Psi^\t{A})- \frac{1}{2}\int_\tau^{\tau_\t{o}} \frac{\t{d}\tau^\pr}{ar^2}\,D^2 Q 
 -  a \bigg (H - \frac{1}{ar} \bigg )P  + \frac{1}{r}Q - \frac{1}{r}w^{(1)}_0 (w)\, , \notag \\[1ex]
    \mathscr{N}^{(1)} &= -2(\Psi^\t{I}+\Psi^\t{A})-2aH P + 2 \pa_w P+2 \pa_w Q -2 \pa_w w^{(1)}_0(w)\, , \notag \\[1ex]
    \mathscr{M}^{(1)} &= -  \int_\tau^{\tau_\t{o}} \frac{\text{d}\tau^\prime}{ar^2} \, \big [Q-w^{(1)}_0 (w)\big ]- \chi^{(1)}_0 (w)\, , \notag \\[1ex]
    \hat{\mathscr{M}}^{(1)} &=- \hat{\chi}^{(1)}_0 (w)\, , \notag \\[1ex]
    \mathscr{U}^{(1)} &= \int_\tau^{\tau_\t{o}} \frac{\text{d}\tau^\prime}{ar} \, \bigg [\frac{2(Q-w^{(1)}_0(w))}{r}-\pa_w (Q-w^{(1)}_0(w)) \bigg ] \notag \\[1ex]
    &\quad  + \frac{1}{r^2} P+ \frac{1}{r^2} Q - \frac{1}{r^2}w^{(1)}_0 (w)- \pa_w \chi^{(1)}_0 (w)  \, , \notag \\[1ex]
    \hat{\mathscr{U}}^{(1)} &= - \pa_w \hat{\chi}^{(1)}_0 (w)\, . 
\end{align}
At second order, it is convenient to start from the differential equations for $\xi^w_{(2)}$ and $\xi^a_{(2)}$ written in terms of the Bardeen potentials:
\begin{align}
    \partial_{\tau}\big [\xi_{(2)}^{w}-(\xi^{\mu}\partial_{\mu}\xi_{(1)}^{w})\big ]&= \frac{4}{a}\Psi_{(2)}^{\t{I}}-\frac{1}{a}\bar{\gamma}^{ab}\partial_{a}\xi_{(1)}^{w}\partial_{b}\xi_{(1)}^{w}+\frac{4}{a^{2}}\Psi^{\t{I}}\partial_{w}\xi_{(1)}^{\tau}+\frac{4H}{a}\Psi^{\t{I}}\xi_{(1)}^{\tau}\nex
    & \quad -\frac{4}{a}\xi^\mu_{(1)}\partial_{\mu}\Psi^\t{I}+\frac{12}{a}\Psi^{\t{I}}\Psi^{\t{A}} \, ,\nex
    \partial_{\tau}\big [\xi_{(2)}^{a}-(\xi^{\mu}\partial_{\mu}\xi^{a})\big ]&= \frac{1}{a} \bar{\gamma}^{ab}\bigg \{ \partial_{b}\big [\xi_{(2)}^{w}-(\xi^{\mu}_{(1)}\partial_{\mu}\xi^{w}_{(1)})\big ]+ \frac{4}{a}\Psi^{\t{I}}\partial_{b}\xi^{\tau}_{(1)}+2H\xi^{\tau}_{(1)}\partial_{b}\xi^{w}_{(1)}\nex
    & \quad + \frac{4}{r}\bigg (\xi^{w}_{(1)}-\frac{\xi^{\tau}_{(1)}}{a}\bigg )\partial_{b}\xi^{w}_{(1)}+2\partial_{c}\xi^{w}_{(1)}(\partial_{b}\xi^{c}_{(1)}+\bar{\gamma}^{dc}\xi^{e}_{(1)}\partial_{e}\bar{\gamma}_{bd})\nex
    & \quad + 2(\Psi^{\t{I}}+\Psi^{\t{A}})\partial_{b}\xi^{w}_{(1)}+\frac{2}{a}\partial_{w}\xi^{\tau}_{(1)}\partial_{b}\xi^{w}_{(1)}\bigg \}   \, . 
    \label{eq:dif-eqns-NG-wa}
\end{align}
Next, using the differential equations \eqref{eq:gauge-inv-1st-order} for the first-order gauge modes we have
\begin{align}
    \frac{\t{d}P}{\t{d}\eta} &= - a H P+\Psi^\t{I}-\Psi^\t{A} \, ,\nex
    \pa_w Q &= \frac{\t{d}Q}{\t{d}\eta} - \pa_\eta Q = \frac{\t{d}Q}{\t{d}\eta} + 2\Psi^{\t{I}} \, , \nex
    \pa_w \xi^a_{(1)} &=  \frac{\t{d} \xi^a_{(1)}}{\t{d}\eta} - \pa_\eta \xi^a_{(1)}  = \frac{\t{d} \xi^a_{(1)}}{\t{d}\eta} - \bar{\ga}^{ab}\pa_b Q  \, . 
\end{align}
Then, using  $\xi^\tau_{(1)}=a P$ (see Eqs.~\eqref{eq:gauge-modes-NG-1st}), we obtain 
\begin{equation}
    \dot{\xi}^\tau_{(1)} = \Psi^\t{I}-\Psi^\t{A}-\pa_w P 
   \quad  , \quad 
    \ddot{\xi}^\tau_{(1)} = \dot{\Psi}^{\t{I}}-\dot{\Psi}^{\t{A}} + H \pa_w P-\frac{1}{a}\pa_w (\Psi^\t{I}-\Psi^\t{A}) +\frac{1}{a}\pa^2_w P \, 
\end{equation}
and, from Eq.~\eqref{eq:xi-0-2nd-order} we can rewrite $\xi^\tau_{(2)}$ in a more compact form after repeated integrations by parts. Putting everything together, our second-order gauge modes in terms of the gravitational potentials can be expressed as
\begin{align}
    \xi^\tau_{(2)}  &= \int_{\tau_{\t{in}}}^\tau \t{d}\tau^\pr \,  \bigg [2\Phi^{(2)}-\Phi^2 + (\pa_w P)^2 + \bar{\ga}^{ab}\pa_a P \pa_b P \bigg ]   - \pa_w P \o  \int_{\tau_{\t{in}}}^\tau \t{d}\tau^\pr \, \pa_w P \nex
    & \quad -a(\Psi^\t{I}-\Psi^\t{A}) P + aP\pa_w P - a (P\o - Q) \pa_w P - a\xi^a_{(1)}\pa_a P \, , \nex
    \xi^w_{(2)}&= \int_{\tau}^{\tau_\t{o}}\, \frac{\t{d}\tau^\pr}{a} \bigg [-4\Psi^\t{I}_{(2)}-\bar{\ga}^{ab} \pa_a Q \pa_b Q +4 \Psi^\t{I}\pa_w Q  - 8 \Psi^\t{I}\Psi^\t{A}-4 (\Psi^\t{I})^2 \bigg ]+2\Psi^\t{I}_\t{o} P\o -2  \Psi^\t{I}P\nex
    & \quad  +(P\o -2Q)\pa_w Q - Q (\pa_w P\o -\pa_w Q) + \xi^a_{(1)}\pa_a Q+w^{(2)}_0(w, \tilde{\theta}^a)\,  ,\nex
    \xi^a_{(2)} &= q^{ab}\int_{\tau}^{\tau_\t{o}}\frac{\t{d}\tau^\pr}{ar^2} \, \bigg \{\int_{\tau^\pr}^{\tau_\t{o}}\frac{\t{d}\tau^{\pr \pr}}{a} \, \pa_b \Big [4\Psi^\t{I}_{(2)} + \bar{\ga}^{cd}\pa_c Q \pa_d Q - 4\Psi^\t{I}\pa_w Q +8\Psi^\t{I}\Psi^\t{A}+4(\Psi^\t{I})^2 \Big ]\nex
    & \quad -2 \pa_b (\xi^c\pa_c Q)- \pa_b \big [(P\o -2Q)\pa_w Q \big ]+\frac{2}{r}(P\o -2Q)\pa_b Q + 2 \pa_c Q (\pa_b \xi^c + \bar{\ga}^{dc}\xi^e\pa_e \bar{\ga}_{bd})\nex
    & \quad +4\Psi^\t{I}\pa_b Q \bigg \} + \bar{\ga}^{ab}P \pa_b Q +\xi^w_{(1)}\pa_w \xi^a_{(1)}+\xi^b_{(1)}\pa_b \xi^a_{(1)}+ \xi^a_{(2)0}(w, \tilde{\theta}^a)  \, , 
    \label{eq:gauge-modes-NG-2nd}
\end{align}
where $w^{(2)}_0 (w, \tilde{\theta}^a)$ and $\xi^a_{(2)0} (w, \tilde{\theta}^a)$ are the second-order residual gauge freedoms associated with the observer position.

Let us remark that Eqs.~\eqref{eq:gauge-modes-NG-1st} and \eqref{eq:gauge-modes-NG-2nd} are relevant outcomes of the general second-order perturbation theory developed so far. Indeed, any light-like cosmological observable can be first computed at a fully non-linear level using  GLC coordinates. Then, it can be expanded up to the desired perturbative order within the framework of light-cone perturbation theory. After that, in order to have  an expression valid in any gauge, we can use Eqs.~\eqref{eq:gauge-inv-quantities-1st-order} and ~\eqref{eq:glc-gauge-inv-2nd-order}. Finally,  thanks to the gauge modes of Eqs.~\eqref{eq:gauge-modes-NG-1st} and \eqref{eq:gauge-modes-NG-2nd}, we can  write the general relativistic corrections in terms of the gauge-invariant gravitational potentials.

\subsection{A unification of all the Geodesic Light-Cone approaches}\label{sec:unifying}
Before proceeding with the computation of the angular distance–redshift relation, we first clarify the connection between our approach and the coordinate transformation methods employed in \cite{BenDayan:2012wi,BenDayan:2012pp,Fanizza:2013doa, Fanizza:2015swa,Marozzi:2014kua}. 

In the gauge transformation given by Eq.~\eqref{eq:gauge-trans-coord}, we have derived the conditions required to transform from a generic gauge fixing (later identified with the PG) to the GLC gauge. Hence, the two systems of coordinates are related by
\begin{equation}
\tilde{y}^{\mu}=y^{\mu}+\xi^{\mu}_{(1)}+\frac{1}{2}\left ( \xi_{(2)}^{\mu}+ \xi^{\nu}_{(1)}\partial_{\nu}\xi^{\mu}_{(1)}\right ) \,,\label{eq:coordinate-gauge-relations}
\end{equation}
where we recall that $y^{\mu}$ denote the coordinates in the PG, while $\tilde{y}^{\mu}$ correspond to the GLC coordinates. In other words, Eq.~\eqref{eq:coordinate-gauge-relations} explicitly provides the relations between our gauge modes and the fluctuations $\tau^{(1,2)},\,w^{(1,2)}$ and $\tilde{\theta}^{a(1,2)}$ first derived in  \cite{Ben-Dayan:2012uam} through a full coordinate transformation stemming from 
\begin{equation}
g_{\t{GLC}}^{\rho\sigma}\left(\tilde{y}\right)=\frac{\partial\tilde{y}^{\rho}}{\partial y^{\mu}}\frac{\partial\tilde{y}^{\sigma}}{\partial y^{\nu}}g^{\mu\nu}_{\t{PG}}\left(y\right)\,.\label{eq:coordinate-transformation}
\end{equation}
Indeed, using Eq.~\eqref{eq:coordinate-gauge-relations}, and neglecting terms at the observer position,  at first order these relations are
\begin{equation}
    \tau^{(1)} = \xi^\tau_{(1)} \qquad , \qquad w^{(1)} = \xi^w_{(1)} \qquad , \qquad \tilde{\theta}^{a(1)} = \xi^a_{(1)}\label{eq:firstorder-primoapproach}
\end{equation}
whereas at second order they are
\begin{align}
\tau^{(2)}&= \frac{1}{2}\left(\xi_{(2)}^{\tau}+\xi^{\mu}_{(1)}\partial_{\mu}\xi^{\tau}_{(1)}\right)\,,\nonumber \\[1ex]
w^{(2)}&=  \frac{1}{2}\left(\xi_{(2)}^{w}+\xi^{\mu}_{(1)}\partial_{\mu}\xi^{w}_{(1)}\right)\,,\nonumber \\[1ex]
\tilde{\theta}^{a(2)}&=  \frac{1}{2}\left(\xi_{(2)}^{a}+\xi^{\mu}_{(1)}\partial_{\mu}\xi^{a}_{(1)}\right)\,.\label{eq:relations-first-approach}
\end{align}
On the other hand, the second approach described for the first time in \cite{Fanizza:2015swa} is based on  the inverse coordinate transformation
\begin{align}
    g^{\mu \nu}_{\t{PG}}(y) =\frac{\pa y^\mu }{\pa \tilde{y}^\rho}\frac{\pa y^\nu}{\pa \tilde{y}^\sigma} g^{\rho \sigma}_{\t{GLC}}(\tilde{y}) \, , 
    \label{eq:NG-GLCmetricrelation}
\end{align}
where now $\tilde{y}^\mu$  are the GLC coordinates and $y^\mu = (\eta, \eta^+, \theta^a)$, with $\eta^+ \equiv \eta +r$, are the PG background coordinates. At this point, we can expand $y^\mu$ in terms of $\tilde{y}^\mu$ up to the desired order as
\begin{align}
    y^\mu = (y^\mu)^{(0)}(\tilde{y})+(y^\mu)^{(1)}(\tilde{y}) + (y^\mu)^{(2)}(\tilde{y}) + \cdots  \, , \label{eq:coordinate-transf}
\end{align}
and then  expand the PG components of \eqref{eq:NG-GLCmetricrelation} around $(y^\mu)^{(0)}(\tilde{y})$. In this way, we  get differential equations for the fluctuations $(\eta^{(1)}, \eta^{+(1)}, \theta^{a(1)})$, whose solutions, from \cite{Fanizza:2015swa},  are
\begin{align}
    \eta^{(1)} &= -\frac{1}{a}\int_{\tau_\t{in}}^{\tau} \t{d}\tau^\pr \,  \Psi^\t{I}\big (\tau^\pr, w-\eta(\tau)+\eta(\tau^\pr), \tilde{\theta}^a\big )\, , \nex
    \eta^{+(1)} &=2\int_{\tau}^{\tau_\t{o}} \frac{\t{d}\tau^\pr}{a(\tau^\pr)} \, \Psi^{\t{I}}(\tau^\pr, w, \tilde{\theta}^a)\, , \nex
    \theta^{a(1)}&=-2 \int_{\tau}^{\tau_\t{o}} \frac{\t{d}\tau^\pr}{a(\tau^\pr)} \bar{\ga}^{ab}(\tau^\pr, w,\tilde{\theta}^a ) \int_{\tau^\pr}^{\tau_\t{o}} \frac{\t{d}\tau^{\pr\pr}}{a(\tau^{\pr \pr})} \pa_b \Psi^\t{I}(\tau^{\pr\pr}, w,\tilde{\theta}^a) \, . 
\end{align}
Hence, comparing the above formulae with Eqs.~\eqref{eq:gauge-modes-NG-1st}, we  notice that \begin{align}
    \eta^{(1)} = -\frac{\xi^\tau_{(1)}}{a} = -P\, ,  \qquad\qquad \eta^{+(1)} = -\xi^w_{(1)} \, , \qquad\qquad \theta^{a(1)} = -\xi^a_{(1)}\, , 
    \label{eq:GLC-NG-gaugemodes-1storder}
\end{align}
provided that we neglect the terms at the observer position $w^{(1)}_0(w)$,  $\chi^{(1)}_0(w)$ and $\hat{\chi}^{(1)}_0(w)$.  

At second order, the comparison of Eqs.~\eqref{eq:dif-eqns-NG-wa} and \eqref{eq:gauge-modes-NG-2nd} to Eqs.~(4.19), (4.21) and (A.5) of \cite{Fanizza:2015swa} entails the following dictionary between our  gauge modes $(\xi^\tau_{(2)}, \xi^w_{(2)}, \xi^a_{(2)})$ and the fluctuations $(\eta^{(2)}, \eta^{+(2)},\theta^{a(2)})$:
\begin{align}
    \eta^{(2)} &= -\frac{1}{2a}\xi^\tau_{(2)} + \frac{1}{2}\xi^\mu_{(1)} \pa_\mu \bigg ( \frac{\xi^\tau_{(1)}}{a}\bigg ) \, ,\nex
    \eta^{+(2)} &= -\frac{1}{2}\xi^w_{(2)} + \frac{1}{2}\xi^\mu_{(1)}\pa_\mu \xi^w _{(1)}\, , \nex 
    \theta^{a(2)} &= -\frac{1}{2}\xi^a_{(2)} + \frac{1}{2}\xi^\mu_{(1)} \pa_\mu \xi^a_{(1)} \, . 
    \label{eq:GLC-NG-gaugemodes-2ndorder}
\end{align}
These are indeed the expected relations in our formalism because the gauge modes to go from the PG  coordinates to the GLC ones can be computed considering that, in general, we have
\begin{equation}
y^{\mu}=\tilde{y}^{\mu}+\epsilon^{\mu}_{(1)}+\frac{1}{2}\epsilon_{(2)}^{\mu}+\frac{1}{2}\epsilon^{\nu}_{(1)}\partial_{\nu}\epsilon^{\mu}_{(1)}\,, \label{eq:inverse-gauge-transform-1}
\end{equation}
where now $\ka^\mu_{(1,2)}$ are the gauge modes in the opposite transformation wrt \eqref{eq:coordinate-gauge-relations}.
At  first order, the relation between the perturbations of the two approaches  -- namely, those of Eqs.~\eqref{eq:coordinate-transformation} which are equivalent to \cite{BenDayan:2012wi,Marozzi:2014kua} and \eqref{eq:NG-GLCmetricrelation} which are equivalent to the inverse approach developed in \cite{Fanizza:2015swa} -- is simply $\xi^{\mu}_{(1)}=-\epsilon^{\mu}_{(1)}$.
At second order, Lie derivatives also come into play, which is why the relations between the fluctuations of the two methods described in \cite{Fanizza:2015swa} are  
\begin{align}
a\eta^{(2)}& =  -\tau^{(2)}+\xi^{\mu}_{(1)}\partial_{\mu}\xi^{\tau}_{(1)}\,,\nonumber \\[1ex]
\eta^{+(2)}&= -w^{(2)}+\xi^{\mu}_{(1)}\partial_{\mu}\xi^{w}_{(1)}\,,\nonumber \\[1ex]
\theta^{a(2)}&=  -\tilde{\theta}_{(2)}^{a}+\xi^{\mu}_{(1)}\partial_{\mu}\xi^{a}_{(1)}\, . \label{eq:relations-two-approaches}
\end{align}
The  above analysis clarifies the connections between our approach, which is manifestly written as a gauge transformation, and the previous approaches of
the GLC literature. In our case, we perform a background coordinate transformation,
followed by a gauge transformation of the metric perturbations
directly on the past light-cone. Instead, the previous approaches directly performed 
a perturbative coordinate transformation.

\section{The Observational Angular Distance–Redshift Relation}
\label{sec:dA}
Hereafter, we will apply the new perturbative scheme developed so far to compute the angular distance–redshift relation $d_\text{A}(z)$,
as measured by a free-falling observer and in a perturbative and gauge-invariant approach, starting from the fully non-linear formulae \eqref{eq:z-fully-non-linear-glc} and \eqref{eq:dA-fully-non-linear-glc} valid within the GLC gauge. In the evaluation of such an angular distance, as anticipated at the end of Subsect.~\ref{subsec:gaugeinvariance-LC}, we will go  through the following steps:
\begin{enumerate}
\item First, we take the fully non-linear formula \eqref{eq:dA-fully-non-linear-glc} valid in the GLC gauge and we expand it up to second order in the light-cone perturbations;
\item Second, we replace each perturbation with its gauge-invariant counterpart provided in Eqs.~\eqref{eq:gauge-inv-quantities-1st-order} and \eqref{eq:glc-gauge-inv-2nd-order};
\item Then, using the observed redshift to define a new time coordinate, we add to $d_\t{A}$  the corrections coming from expanding each quantity around the time at the source corresponding to the observed redshift, thus obtaining the fully gauge-invariant formula for the  angular distance–redshift relation;
\item We completely fix the observational gauge around the observer position by  fixing  the residual gauge freedom \eqref{eq:glc-residual-freedom} of the GLC gauge. This allows for the cancellation of all the would-be divergent terms scaling as $\sim r^{-n}$ (for $n=1,2,3$) at the observer. This gauge fixing also results in  further new observer terms, not present in the literature, which  are needed to make the angular distance–redshift relation to be the one  truly measured by a free-falling observer;
\item Finally, we express each light-cone perturbation in terms of the gauge-invariant Bardeen potentials $\Phi^{(n)}$ and $\Psi^{(n)}$, and compare the results with the ones already present in the literature.
\end{enumerate}
Not only are the techniques described above useful to evaluate the angular distance–redshift relation, but they also constitute a general procedure to compute  linear and non-linear relativistic effects of any kind of light-like cosmological observable.

\subsection{Perturbative expansion}
\label{subsec:perturb-calc-dA}

\paragraph{Comparison between perturbed and unperturbed metrics.} 
As a starting point, we  consider the fully non-linear GLC metric introduced in Eq.~\eqref{eq:GLCmetric-start} 
and we  express the functions $\Upsilon$ and $\mathcal{U}^a$ in terms of the GLC perturbations  appearing in the metric \eqref{eq:metricGLC}, working up to second order.  In general, we can write
\begin{equation}
    \Upsilon = \bar{\Upsilon} (1+ \Upsilon^{(1)} + \Upsilon^{(2)})    \qquad  , \qquad 
    \mathcal{U}^a = \mathcal{U}^a_{(1)}+ \mathcal{U}^a_{(2)} \, ,
\label{eq:decomposition-LC-functions}
\end{equation}
where we have accounted for the fact that  $\mathcal{U}^a$ has zero background value, as we have specified after Eqs.~\eqref{eq:GLC-background-FLRW}. Then, we can conformally re-scale the 2-dimensional metric of Eq.~\eqref{eq:GLCmetric-start} as $g_{ab} \equiv a^2 \ga_{ab}$. In this way, comparing the off-diagonal terms $\text{d}w \text{d}\tilde{\theta}^a$ of Eqs.~\eqref{eq:metricGLC} and \eqref{eq:GLCmetric-start}, we can straightforwardly obtain that
\begin{equation}
    2a^2 \gamma_{ab} (\mathcal{U}^a_{(1)}+\mathcal{U}^a_{(2)}) \text{d} w  \text{d}\tilde{\theta}^b = 2a^2  (U^{(1)}_a+U^{(2)}_a) \text{d} w  \text{d}\tilde{\theta}^a \, .
\end{equation}
By explicitly writing the perturbations of $\ga_{ab}$, we have
\begin{align}
\big ( \bar{\gamma}_{ab}+ \gamma^{(1)}_{ab}+ \gamma^{(2)}_{ab}\big )  (\mathcal{U}^b_{(1)}+ \mathcal{U}^b_{(2)}) &= \bar{\ga}_{ab}\mathcal{U}^b_{(1)}+\ga^{(1)}_{ab}\mathcal{U}^b_{(1)} + \bar{\ga}_{ab} \mathcal{U}^b_{(2)} \notag \\
& = U^{(1)}_a + U^{(2)}_a \, .
\end{align}
Then, by equaling first- and second-order perturbations among themselves, we are simply left with 
\begin{equation}
\begin{cases}
U^{(1)}_a = \bar{\ga}_{ab} \mathcal{U}^b_{(1)} \\[1ex]
U^{(2)}_a = \ga^{(1)}_{ab} \mathcal{U}^b_{(1)}+ \bar{\ga}_{ab} \mathcal{U}^b_{(2)}
\end{cases} \qquad \Rightarrow \qquad
\begin{cases}
\mathcal{U}^a_{(1)} =  \bar{\gamma}^{ab} U^{(1)}_b \\[1ex]
\mathcal{U}^a_{(2)} =\bar{\gamma}^{ab}  (U^{(2)}_b - \gamma^{(1)}_{bc} \bar{\gamma}^{cd} U^{(1)}_d )
\end{cases} \,\,\,\,\,\,\,\, .
 \label{eq:U-perturbations}
\end{equation}
Next,  taking the terms proportional to $\text{d}w^2$, we have
\begin{equation}
    \Upsilon^2 + \mathcal{U}^a \mathcal{U}_a = a^2  (1+N^{(1)}+N^{(2)}  ) \, .
    \label{eq:dictionary-perturb-step}
\end{equation}
Considering that the $2\times 2$ spatial metric of Eq.~\eqref{eq:metricGLC} is $g_{ab}=a^2\ga_{ab}$, from a perturbative point of view we have 
\begin{equation}
    \mathcal{U}^a\mathcal{U}_a = g_{ab}\mathcal{U}^a \mathcal{U}^b = a^2 \bar{\ga}_{ab} \mathcal{U}^a_{(1)}\mathcal{U}^b_{(1)} = a^2 U^a_{(1)}U^{(1)}_a \equiv a^2 U^2_{(1)} \, .
\end{equation}
Therefore, Eq.~\eqref{eq:dictionary-perturb-step} becomes
\begin{equation}
    \begin{split}
       \bar{\Upsilon}^2 \Big [1+ 2 \Upsilon^{(1)} + \big (\Upsilon^{(1)} \big )^2 + 2\Upsilon^{(2)}  \Big ] +  a^2U^2_{(1)} = a^2 \big (1+N^{(1)}+N^{(2)} \big ) \, . 
    \end{split}
\end{equation}
By respectively equaling terms of first and second order together, we get the relations\footnote{We choose $\bar{\Upsilon}^2 = a^2 \, \Rightarrow \, \bar{\Upsilon} = + a$ consistently with the requirement that $u^w > 0$ (see  discussion after Eq.~\eqref{eq:4-velocity-GLC}).} 
\begin{equation}
\bar{\Upsilon} = a\quad , \quad 
\Upsilon^{(1)} = \frac{1}{2}N^{(1)} \quad , \quad 
\Upsilon^{(2)} = \frac{1}{2}N^{(2)}- \frac{1}{8} \big (N^{(1)} \big )^2 - \frac{1}{2}U^2_{(1)} \, . 
\label{eq:Upsilon_pertubations}
\end{equation}

\paragraph{Perturbative formula for the redshift and redshift parameterization.} As first remarked in \cite{Biern:2016kys}, the angular distance is a bi-scalar, being affected by the gauge transformations both at the source and at the observer positions. So, to completely fix the gauge, we need to account for both these effects (see also \cite{Fanizza:2020xtv} for a more detailed discussion). Regarding the latter, thanks to what we have discussed at the end of Subsect.~\ref{subsec:map-FLRW-GLC}, the usage of the sets of gauge-invariant variables \eqref{eq:gauge-inv-quantities-1st-order} and \eqref{eq:glc-gauge-inv-2nd-order} entirely fixes the observer's gauge modes. Regarding the former, since $d_\text{A}$ does not depend on the angles at the background level, we are left with 2 conditions to be fixed in order to eliminate the gauge freedom: (i) as for the $w$ gauge mode, since formula \eqref{eq:dA-fully-non-linear-glc} is valid within the GLC gauge, we are already imposing that the photons emitting the luminous signal  propagate on the past light-cone; (ii) moreover, we use the observed redshift of the source to fix the time gauge mode, as we  are about to show hereafter.

Starting from the exact formula \eqref{eq:z-fully-non-linear-glc} for the redshift  and using Eqs.~\eqref{eq:Upsilon_pertubations}, we obtain
\begin{align}
    1+z & = (1+\bar{z})(1+\delta^{(1)}z+\delta^{(2)}z)\nex
    &=\frac{a_\text{o}}{a_\text{s}} \bigg [ 1+\Upsilon^{(1)}|^\t{o}_\t{s} +\Upsilon^{(2)}|^\t{o}_\t{s} + (\Upsilon^{(1)}_\t{s})^2 - \Upsilon^{(1)}_\t{o}\Upsilon^{(1)}_\t{s} \bigg ] \nex
    & = \frac{a_\text{o}}{a_\text{s}} \bigg [ 1+\frac{N^{(1)}|^\text{o}_\text{s}}{2}+\frac{N^{(2)}|^\text{o}_\text{s}}{2}-\frac{(N^{(1)})^2|^\text{o}_\text{s}}{8}+ \bigg (\frac{N^{(1)}_\text{s}}{2}\bigg )^2- \frac{N^{(1)}_\text{o}N^{(1)}_\text{s}}{4}-\frac{1}{2}\left[U_{(1)}^2\right]^\text{o}_\text{s} \bigg ] \, .
    \label{eq:perturbed-z}
\end{align}
Cosmological surveys typically provide us with data sets of the form $\big (z, d_\t{A}(z)\big )$, $z$ being the redshift of the source. Thus, we  perturbatively redefine the proper time at the source as
\begin{equation}
    \tau = \tau_z + \tau^{(1)}_z + \tau^{(2)}_z \, , 
    \label{eq:time-source-expanded}
\end{equation}
where $\tau_z$ is the proper time of the source evaluated at the observed redshift $z$ and $\tau^{(1,2)}_z$ are the distorsions at first and second order caused by the presence of inhomogeneities and anisotropies  along the line of sight (see, for example,  \cite{Scaccabarozzi:2017ncm, Fanizza:2018qux}). As a side remark, note that we do not have to do the same also for the coordinates $w$ and $\tilde{\theta}^a$ because, since we are working within the GLC gauge, observations are made on the past light-cone, namely $w = \text{const.}$, and the angles are precisely those of the background. Following this strategy, we first Taylor expand 
\begin{equation}
    a(\tau) = a_z \bigg [1+H_z \tau^{(1)}_z + H_z \tau^{(2)}_z -\frac{1}{2}q_z (\tau^{(1)}_z)^2\bigg ] \, , 
    \label{eq:a-s-expanded}
\end{equation}
where
\begin{equation}
    a_z \equiv a(\tau_z) \qquad , \qquad H_z \equiv H(\tau_z) \qquad , \qquad q_z \equiv -\frac{\ddot{a}_z}{a_z} \, .
\end{equation}
More in general, any scalar quantity $X$ can be expressed in terms of the observed redshift working up to second order in perturbation theory, in the  simplest case in which its background counterpart only depends on the time $\tau$, namely $\bar{X}(x^\mu) \equiv \bar{X} (\tau)$. Indeed, starting from
\begin{equation}
    X(x^\mu) = \bar{X} (\tau) + X^{(1)}(x^\mu) + X^{(2)}(x^\mu) 
\end{equation}
and considering Eq.~\eqref{eq:time-source-expanded}, we can Taylor expand as follows,
\begin{equation}
\begin{split}
\bar{X}(\tau) &= \bar{X}(\tau_z) + \dot{\bar{X}}(\tau_z) \tau^{(1)}_z + \dot{\bar{X}}(\tau_z) \tau^{(2)}_z + \frac{1}{2}\ddot{\bar{X}}(\tau_z) (\tau^{(1)}_z)^2 \, ,  \\[1ex]
X^{(1)} (x^\mu) &= X^{(1)}(\tau_z) + \dot{X}^{(1)} (\tau_z) \tau^{(1)}_z \, ,  \\[1ex]
X^{(2)} (x^\mu) &= X^{(2)} (\tau_z) \, ,
\end{split}
\end{equation}
where, for the sake of brevity, we omit to also report the dependence of the perturbations on the coordinates $(w, \tilde{\theta}^a)$.
In this way, we can write
\begin{equation}
    X(x^\mu) = \bar{X}_z + X^{(1)}_z + X^{(2)}_z 
    \label{eq:expansion-obs-redshift}
\end{equation}
with
\begin{align}
    \bar{X}_z &\equiv \bar{X}(\tau_z) \, ,  \nex
    X^{(1)}_z &\equiv  \dot{\bar{X}}(\tau_z) \tau^{(1)}_z + X^{(1)}(\tau_z) \, , \nex
     X^{(2)}_z &\equiv \dot{\bar{X}}(\tau_z) \tau^{(2)}_z + \frac{1}{2}\ddot{\bar{X}}(\tau_z) (\tau^{(1)}_z)^2 + \dot{X}^{(1)}(\tau_z) \tau^{(1)}_z \, .
     \label{eq:perturbations-with-z}
\end{align}
Then, Eqs.~\eqref{eq:a-s-expanded} and \eqref{eq:perturbations-with-z} can be applied to Eq.~\eqref{eq:perturbed-z} to obtain
\begin{align}
    1+z &=\frac{a_\text{o}}{a_z} \bigg \{1+\frac{N^{(1)}|^\text{o}_z}{2} -\frac{\dot{N}^{(1)}_z}{2}\tau^{(1)}_z + \frac{N^{(2)}|^\text{o}_z}{2} + \frac{3}{8}(N^{(1)}_z)^2 - \frac{(N^{(1)}_\text{o})^2}{8}-\frac{1}{2}[U^2_{(1)}]^\text{o}_z-\frac{N^{(1)}_\text{o}N^{(1)}_z}{4}\notag \\[1ex]
    & \quad -H_z \tau^{(1)}_z -H_z \tau^{(2)}_z +  \bigg (\frac{1}{2}q_z + H^2_z \bigg ) (\tau^{(1)}_z)^2 -\frac{1}{2}H_z \tau^{(1)}_z N^{(1)}|^\text{o}_z 
    \bigg \}   \, . 
    \label{eq:z-z-expansion}
\end{align}
Thus, following \cite{Fanizza:2020xtv}, by requiring that the redshift is the \textit{observed} one, \textit{i.e.} $1+z = a_\text{o}/a_z$, we obtain at first and second order that
\begin{align}
    \tau^{(1)}_z &= \frac{1}{2H_z} N^{(1)}|^\text{o}_z \, , \notag \\[1ex]
    \tau^{(2)}_z &= \frac{1}{2H_z} \Bigg \{ N^{(2)}|^\text{o}_z-\frac{1}{4}\left(N^{(1)}_\text{o}\right)^2+ \frac{3}{4}\left(N^{(1)}_z\right)^2 -\frac{1}{2}N^{(1)}_\text{o}N^{(1)}_z  \nex
    &  \quad -\left[U^2_{(1)} \right]^\text{o}_z + \frac{1}{4}\frac{q_z}{H^2_z} \left(N^{(1)}|^\text{o}_z\right)^2 - \frac{N^{(1)}\oz \dot{N}^{(1)}_z}{2H_z}\Bigg \} \, .
    \label{eq:time-z-perturbations}
\end{align}

\paragraph{Perturbative formula for the angular distance–redshift relation.} We are now ready to obtain an up to the second-order perturbative expression for $d_\t{A}$ and then expand it according to the redshift parameterization outlined in the previous paragraph, thus getting the  relation for $d_\t{A}(z)$.

Looking back at Eq.~\eqref{eq:dA-fully-non-linear-glc}, we first have to compute  the (square root of the) determinant of the metric, which, using the GLC coordinates, is exactly given by
\begin{equation}
\sqrt{-g_{(4)}} =  \Upsilon \sqrt{g} \qquad ,  \qquad g \equiv \text{det}[g_{ab}] = a^4 \t{det}[\ga_{ab}] \equiv a^4 \ga \, .
\label{eq:detformula}
\end{equation}
In order to calculate $\gamma$, we simply use that $\ga_{ab}$ is a $2 \times 2$ matrix: exploiting the SPS decomposition \eqref{eq:SPS-vec-tensor} and the fact that, from the definitions given in \eqref{eq:D-ab-def}, $q^{ab}D_{ab} =0$, $q^{ab}\tilde{D}_{ab}=0$, we can obtain
\begin{equation}
    \ga = \bar{\ga} (1+ \ga^{(1)}+\ga^{(2)})
    \label{eq:det-perturbed}
\end{equation}
with
\begin{equation}
\begin{split}
    \bar{\ga} &= r^4 \sin^2 \tilde{\theta}^1 \, , \\
    \ga^{(1)} &= 4\nu^{(1)} \, ,  \\
    \ga^{(2)} &= 4\nu^{(2)} + 4 (\nu^{(1)})^2 +  \mathfrak{g}^{(2)}\left(\mu^{(1)}, \hat{\mu}^{(1)}\right) \, , 
\end{split}
\end{equation}
having defined
\begin{align}
    \mathfrak{g}^{(2)}\left(\mu^{(1)}, \hat{\mu}^{(1)}\right) & \equiv
\frac{2}{\sin^2 \tilde{\theta}^1}\varepsilon^{ab}\varepsilon^{cd} \bigg [D_{ac}\mu^{(1)} D_{bd}\mu^{(1)}+2 D_{ac}\mu^{(1)} \tilde{D}_{bd}\hat{\mu}^{(1)} +\tilde{D}_{ac}\hat{\mu}^{(1)}\tilde{D}_{bd}\hat{\mu}^{(1)}\bigg ]\, .
    \label{eq:tilde-gamma-def}
\end{align}
It follows that
\begin{equation}
    \sqrt{\ga} = r^2 \sin \tilde{\theta}^1 \bigg (1+2\nu^{(1)}+2\nu^{(2)}+ \frac{1}{2} \mathfrak{g}^{(2)}(\mu^{(1)}, \hat{\mu}^{(1)}) \bigg ) \, .
    \label{eq:sqrt-gamma}
\end{equation}
Note that the only genuine second-order contribution to $\sqrt{\ga}$ stems from $2\nu^{(2)}$, while the term $\mathfrak{g}^{(2)}(\mu^{(1)}, \hat{\mu}^{(1)})$ contains quadratic combinations of the perturbations $\mu^{(1)}$ and $\hat{\mu}^{(1)}$.

Finally, just for the sake of completeness, by combining Eq.~\eqref{eq:detformula} with the above formula, we find that $\sqrt{-g_{(4)}}$ perturbed up to second order is
\begin{align}
    \sqrt{-g_{(4)}}
& = a^4 \sqrt{\bar{\gamma}} \, \bigg [1+2\nu^{(1)} + \Upsilon^{(1)} +2\nu^{(2)}+ \Upsilon^{(2)} + 2\Upsilon^{(1)} \nu^{(1)} + \frac{1}{2} \mathfrak{g}^{(2)}\left(\mu^{(1)}, \hat{\mu}^{(1)}\right)\bigg ]\,,
 \label{eq:perturbed-determinant}
\end{align}
and, by replacing $\Upsilon^{(n)}$ with their forms given in Eqs.~\eqref{eq:Upsilon_pertubations}, we end up with the perturbed $\sqrt{-g_{(4)}}$ written in terms of SPS variables:
\begin{align}
    \sqrt{-g_{(4)}} &= a^4 \sqrt{\bar{\ga}} \bigg[ 1+2\nu^{(1)}+\frac{1}{2}N^{(1)}+2\nu^{(2)}+\frac{1}{2}N^{(2)}-\frac{1}{8}(N^{(1)})^2+\nu^{(1)}N^{(1)}-\frac{1}{2}U^2_{(1)}\notag \\[1ex]
    & \quad + \frac{1}{2} \mathfrak{g}^{(2)}\left(\mu^{(1)}, \hat{\mu}^{(1)}\right) \bigg] \, .
\end{align}
We now expand the angular distance in Eq.~\eqref{eq:dA-fully-non-linear-glc} up to second order in the light-cone perturbations. As a first step, remembering that
\begin{equation}
    g_{ab} = a^2 r^2 \big [(1+2\nu )q_{ab} +2D_{ab} \mu + \tilde{D}_{ab} \hat{\mu} \big ] \ , 
\end{equation}
we find
\begin{equation}
    \dot{g}_{ab} = 2ar \Big \{ \big [(aHr-1)(1+2\nu) + ar\dot{\nu} \big ]q_{ab}  + 2(aHr-1)(D_{ab}\mu + \tilde{D}_{ab}\hat{\mu}) + D_{ab}\dot{\mu}+\tilde{D}_{ab}\dot{\hat{\mu}}\Big \} \, ,
\end{equation}
and, using that $\nu=\nu^{(1)}+\nu^{(2)}$, $\mu=\mu^{(1)}+\mu^{(2)}$ and $\hat{\mu}=\hat{\mu}^{(1)}+\hat{\mu}^{(2)}$, we obtain
\begin{align}
    \text{det}[\dot{g}_{ab}] &= 4a^2 r^2 \sin^2 \tilde{\theta}^1 (aHr-1)^2 \bigg [ 1+4\nu^{(1)}+\frac{2ar\dot{\nu}^{(1)}}{aHr-1} + 4\nu^{(2)}+\frac{2ar\dot{\nu}^{(2)}}{aHr-1}\notag \\[1ex]
    & \quad + \bigg (2\nu^{(1)}+\frac{ar\dot{\nu}^{(1)}}{aHr-1}\bigg )^2 + \tilde{\mathfrak{g}}^{(2)}\left(\mu^{(1)}, \hat{\mu}^{(1)}\right) \bigg ] \, ,
\end{align}
having defined
\begin{align}    \tilde{\mathfrak{g}}^{(2)}\left(\mu^{(1)}, \hat{\mu}^{(1)}\right) & \equiv 
\frac{\varepsilon^{ab}\varepsilon^{cd}}{\sin^2 \tilde{\theta}^1} \bigg \{2D_{ac}\mu^{(1)} \left (D_{bd}\mu^{(1)}+2\tilde{D}_{bd}\hat{\mu}^{(1)}\right ) + 2\tilde{D}_{ac}\hat{\mu}^{(1)}\tilde{D}_{bd}\hat{\mu}^{(1)} \notag \\[1ex]
& \quad +\frac{2}{aHr-1} \left  [D_{ac}\mu^{(1)}\left(D_{bd}\dot{\mu}^{(1)}+\tilde{D}_{bd}\dot{\hat{\mu}}^{(1)}\right) + D_{ac}\dot{\mu}^{(1)}\tilde{D}_{bd}\hat{\mu}^{(1)}\right. \notag \\[1ex]
& \left. \quad + \tilde{D}_{ac}\hat{\mu}^{(1)}\tilde{D}_{bd}\dot{\hat{\mu}}^{(1)} \right ]+\frac{1}{(aHr-1)^2} \bigg [D_{ac}\dot{\mu}^{(1)}\bigg (\frac{1}{2}D_{bd}\dot{\mu}^{(1)} + \tilde{D}_{bd}\dot{\hat{\mu}}^{(1)}\bigg )\notag \\[1ex]
& \quad + \frac{1}{2}\tilde{D}_{ac}\dot{\hat{\mu}}^{(1)}\tilde{D}_{bd}\dot{\hat{\mu}}^{(1)} \bigg ] \bigg \}\, .
\label{eq:new-tilde-g-2}
\end{align}
Thus, combining the above results with Eqs.~\eqref{eq:tilde-gamma-def} and \eqref{eq:sqrt-gamma} for $\sqrt{\ga}$, we get that
\begin{align}
    \bigg (\frac{\text{det}[\dot{g}_{ab}]}{4\sqrt{g}}\bigg )_\text{o} &= \sin \tilde{\theta}^1 \cdot \lim_{\tau \to \tau_\text{o}} \bigg \{(aHr-1)^2 \bigg [1+2\nu^{(1)}+\frac{2ar\dot{\nu}^{(1)}}{aHr-1}+ 2\nu^{(2)}-4(\nu^{(1)})^2 \notag \\[1ex]
    & \quad + \frac{2ar}{aHr-1} (\dot{\nu}^{(2)}-2\nu^{(1)}\dot{\nu}^{(1)}) + \bigg (2\nu^{(1)}+\frac{ar\dot{\nu}^{(1)}}{aHr-1} \bigg )^2 - \frac{1}{2}\mathfrak{g}^{(2)} +  \tilde{\mathfrak{g}}^{(2)}\bigg ] \bigg \} \, ,
\end{align}
where we recall that $\tau_\text{o}$ indicates the proper time of the observer.
Thus, the angular distance written in \eqref{eq:dA-fully-non-linear-glc} is given by
\begin{equation}
    d_\text{A} =  ar \frac{1+\nu^{(1)}+\nu^{(2)}-\frac{1}{2}(\nu^{(1)})^2+\frac{1}{4}\mathfrak{g}^{(2)}}{\Big |\lim_{\tau \to \tau_{\text{o}}} \big [ (aHr-1) \mathfrak{D} \big ] \Big |}
    \label{eq:dA-perturbed}
\end{equation}
where 
\begin{align}
    \mathfrak{D} & \equiv  1+\nu^{(1)} + \frac{ar\dot{\nu}^{(1)}}{aHr-1}+\nu^{(2)}-\frac{1}{2}(\nu^{(1)})^2 +\frac{ar\dot{\nu}^{(2)}}{aHr-1}-\frac{ar\nu^{(1)}\dot{\nu}^{(1)}}{aHr-1}-\frac{1}{4}\mathfrak{g}^{(2)}+\frac{1}{2}\tilde{\mathfrak{g}}^{(2)} \, .
\end{align}
As a next step, using Eqs.~\eqref{eq:perturbations-with-z} and  \eqref{eq:time-z-perturbations} we expand  the background and linear expression for the angular distance around the observed redshift as:
\begin{align}
    a(\tau) r(\tau) &= a_z r_z \bigg \{1+ \bigg (H_z - \frac{1}{a_z r_z} \bigg )\tau^{(1)}_z + \bigg (H_z - \frac{1}{a_z r_z} \bigg )\tau^{(2)}_z  -\frac{1}{2}\bigg ( q_z + \frac{H_z}{a_zr_z}\bigg ) (\tau^{(1)}_z)^2 \bigg \} \notag \\[1ex]
    & =a_z r_z \bigg \{1+\frac{1}{2} \bigg (1-\frac{1}{a_zH_zr_z}\bigg ) N^{(1)}|^\text{o}_z +\frac{1}{2} \bigg (1-\frac{1}{a_zH_zr_z}\bigg ) \bigg [ N^{(2)}|^\text{o}_z- \frac{(N^{(1)}_\text{o})^2}{4} \notag \\[1ex]
    & \quad + \frac{3}{4}(N^{(1)}_z)^2 - \frac{1}{2}N^{(1)}_\text{o}N^{(1)}_z -  [U^2_{(1)}]^\text{o}_z  - \frac{N^{(1)}\oz \dot{N}^{(1)}_z}{2H_z}
    \bigg ]  +\frac{\dot{H}_z}{8a_z H^3_z r_z} (N^{(1)}\oz)^2\bigg \} \, , \nex
    \nu^{(1)}(\tau) &= \nu^{(1)}_z + \tau^{(1)}_z \dot{\nu}^{(1)}_z\, .
    \label{eq:ar-z-expanded}
\end{align} 
We can then obtain a formula for $d_\t{A}(z)$ which is gauge invariant both at the source and at the observer positions. Indeed, as explained in the previous paragraph, the gauge at the source is fixed once we adopt the redshift parameterization, while the one at the observer is fixed once, in the  expression \eqref{eq:dA-perturbed} 
of $d_\text{A}$,  we substitute each variable with its gauge-invariant counterpart provided in Eqs.~\eqref{eq:gauge-inv-quantities-1st-order} and \eqref{eq:glc-gauge-inv-2nd-order} at first and second order respectively (this amounts to requiring that the observer is geodesic).
By doing this, the result, as function of the redshift, can be written as
\begin{equation}
    d_\text{A}(z) =  a_zr_z \frac{d^{(\text{num})}_\text{A} (z)}{\big |\lim_{\tau \to \tau_\text{o}}d^{(\text{den})}_\text{A} (z) \big |} \, , 
    \label{eq:dA-z-limit}
\end{equation}
where 
\begin{align}
    d^{(\text{num})}_\text{A} (z) &= 1+\mathscr{V}^{(1)}_z+\frac{1}{2}\bigg (1-\frac{1}{a_z H_z r_z} \bigg )\mathscr{N}^{(1)}|^\text{o}_z + \mathscr{V}^{(2)}_z + \frac{1}{2}\bigg ( 1-\frac{1}{a_z H_zr_z} \bigg ) \ms{N}^{(2)}|^\t{o}_z\nex
    & \quad -\frac{1}{2}(\mathscr{V}^{(1)}_z)^2 + \frac{1}{2H_z}\dot{\mathscr{V}}^{(1)}_z \mathscr{N}^{(1)}|^\text{o}_z +\frac{1}{4}\mathfrak{g}^{(2)}_z(\mathscr{M}^{(1)}, \hat{\mathscr{M}}^{(1)})\nex
    & \quad +\frac{1}{2}\bigg (1-\frac{1}{a_z H_z r_z} \bigg ) \bigg \{ \mathscr{N}^{(1)}|^\text{o}_z \mathscr{V}^{(1)}_z  - \frac{1}{4}(\mathscr{N}^{(1)}_\text{o})^2  + \frac{3}{4}(\mathscr{N}^{(1)}_z)^2 - \frac{1}{2}\mathscr{N}^{(1)}_\text{o}\mathscr{N}^{(1)}_z \nex
    & \quad - \big [\bar{\ga}^{ab} (D_a \ms{U}^{(1)}+ \tilde{D}_a \hat{\ms{U}}^{(1)})(D_b \ms{U}^{(1)}+ \tilde{D}_b \hat{\ms{U}}^{(1)}) \big ]^{\t{o}}_z- \frac{\ms{N}^{(1)}\oz \dot{\ms{N}}^{(1)}_z}{2H_z}
    \bigg \}  \nex
    & \quad +\frac{\dot{H}_z}{8a_zH^3_zr_z}(\ms{N}^{(1)}\oz)^2 \, ,  \nex
    d^{(\text{den})}_\text{A} (z) &=  (aHr-1) \bigg [ 1+\mathscr{V}^{(1)} + \frac{ar\dot{\mathscr{V}}^{(1)}}{aHr-1}+\mathscr{V}^{(2)}-\frac{1}{2}(\mathscr{V}^{(1)})^2 +\frac{ar\dot{\mathscr{V}}^{(2)}}{aHr-1}\notag \\[1ex]
    & \quad -\frac{ar\ms{V}^{(1)}\dot{\ms{V}}^{(1)}}{aHr-1}-\frac{1}{4}\mathfrak{g}^{(2)}(\mathscr{M}^{(1)}, \hat{\mathscr{M}}^{(1)})+\frac{1}{2}\tilde{\mathfrak{g}}^{(2)}(\mathscr{M}^{(1)}, \hat{\mathscr{M}}^{(1)}) \bigg ] \, .
    \label{eq:d_A-num-den}
\end{align}
Now, let us compute the limit of $d^{(\t{den})}_\t{A}(z)$ for $\tau \rightarrow \tau_\text{o}$. Since in this limit  we have $r \rightarrow 0$, it follows that
\begin{align}
    \lim_{\tau \to \tau_\text{o}} d^{(\text{den})}_\text{A}(z) &=\bigg [ -1-\mathscr{V}^{(1)} + ar\dot{\mathscr{V}}^{(1)}-\mathscr{V}^{(2)}+\frac{1}{2}(\mathscr{V}^{(1)})^2 +ar\dot{\mathscr{V}}^{(2)}\notag \\[1ex]
    & \quad -ar\ms{V}^{(1)}\dot{\ms{V}}^{(1)}+ \frac{1}{4}\mathfrak{g}^{(2)}(\mathscr{M}^{(1)}, \hat{\mathscr{M}}^{(1)})-\frac{1}{2}\tilde{\mathfrak{g}}^{(2)}(\mathscr{M}^{(1)}, \hat{\mathscr{M}}^{(1)}) \bigg ]_{\text{o}} \, .
    \label{eq:den-dA}
\end{align}
The terms $(ar\dot{\ms{V}}^{(1)})_\t{o}$,  $(ar\dot{\ms{V}}^{(2)})_\t{o}$ and $(ar\ms{V}^{(1)}\dot{\ms{V}}^{(1)})_\t{o}$ can not be straightforwardly set to zero despite the fact that $r_\t{o} =0$. Indeed, from Eqs.~\eqref{eq:gauge-inv-quantities-1st-order} and \eqref{eq:glc-gauge-inv-2nd-order}, we see that the gauge-invariant variables $\ms{V}^{(n)}$ contain terms diverging with negative powers of $r$, and we will analyze this issue in the next subsection.

Then, by Taylor expanding up to second order formula \eqref{eq:dA-z-limit} using the previous result, we have
\begin{align}
    d_{\text{A}}(z) &= a_z r_z \big [ 1+d^{(1)}_\text{A}(z)+ d^{(2)}_\text{A}(z) \big ]
    \label{eq:dA(z)-split}
\end{align}
where
\begin{align}
d^{(1)}_\text{A}(z) &= -\mathscr{V}^{(1)}|^\text{o}_z+\frac{1}{2}\bigg (1-\frac{1}{a_zH_zr_z} \bigg ) \mathscr{N}^{(1)}|^\text{o}_z +(a r\dot{\mathscr{V}}^{(1)})_\t{o} \, , \notag \\[1ex]
d^{(2)}_\text{A}(z) &= -\ms{V}^{(2)}|^\t{o}_z+\frac{1}{2}\bigg (1-\frac{1}{a_zH_zr_z} \bigg ) \ms{N}^{(2)}|^\t{o}_z +(ar\dot{\ms{V}}^{(2)})_\t{o}+\frac{1}{2}(\ms{V}^{(1)})^2|^\t{o}_z + (\ms{V}^{(1)}_\t{o})^2  +\frac{1}{4}\mathfrak{g}^{(2)}_\t{o}\notag \\[1ex]
& \quad + \frac{1}{4}\mathfrak{g}^{(2)}_z-\frac{1}{2}\tilde{\mathfrak{g}}^{(2)}_\t{o} +\frac{1}{2H_z} \dot{\ms{V}}^{(1)}_z \ms{N}^{(1)}|^\t{o}_z -\ms{V}^{(1)}_\t{o}\ms{V}^{(1)}_z + (ar\dot{\ms{V}}^{(1)})_\t{o} \big [\ms{V}^{(1)}_z+ (ar\dot{\ms{V}}^{(1)})_\t{o}\nex
& \quad -3\ms{V}^{(1)}\o \big ] + \frac{1}{2}\bigg ( 1- \frac{1}{a_zH_zr_z} \bigg ) \bigg \{-\ms{N}^{(1)}|^\t{o}_z \ms{V}^{(1)}|^\t{o}_z-\frac{1}{4}(\ms{N}^{(1)}_\t{o})^2 +\frac{3}{4}(\ms{N}^{(1)}_z)^2 \nex
& \quad -\frac{1}{2}\ms{N}^{(1)}_\t{o}\ms{N}^{(1)}_z  -\big [\bar{\ga}^{ab} (D_a \ms{U}^{(1)}+ \tilde{D}_a \hat{\ms{U}}^{(1)})(D_b \ms{U}^{(1)}+ \tilde{D}_b \hat{\ms{U}}^{(1)}) \big ]^\t{o}_z + \nex
& \quad + \ms{N}^{(1)}|^\t{o}_z (ar\dot{\ms{V}}^{(1)})_\t{o}-\frac{1}{2H_z} \ms{N}^{(1)}\oz \dot{\ms{N}}^{(1)}_z \bigg \}+\frac{\dot{H}_z}{8a_zH^3_zr_z}(\ms{N}^{(1)}\oz)^2 \, .
\label{eq:perturbations-dA-z}
\end{align}
Finally, once we replace in the above equations the explicit forms of the gauge-invariant perturbations \eqref{eq:gauge-inv-quantities-1st-order} and \eqref{eq:glc-gauge-inv-2nd-order}, and we plug the expressions of the gauge modes \eqref{eq:xi-1st-order}, \eqref{eq:xi-w-2nd-order}, \eqref{eq:chi-2nd-order} and \eqref{eq:xi-0-2nd-order}, we  obtain the fully gauge-invariant form of the angular distance–redshift relation on the past light-cone within the perturbative scheme developed in this paper.

\subsection{Gauge fixing  at the observer position}
\label{subsec:eliminationIRdiver}
As we have already stressed, the GLC gauge at the observer position is not totally specified: this residual gauge freedom is encoded in the functions $w^{(n)}_0(w, \tilde{\theta}^a)$, $\chi^{(n)}_0(w, \tilde{\theta}^a)$ and $\hat{\chi}^{(n)}_0(w, \tilde{\theta}^a)$ arising  in the first- and second-order gauge modes needed to fix the GLC gauge in perturbation theory. 

In this subsection, we will demonstrate how a well-motivated choice of these free functions is capable at once of completely fixing the \quotes{observational} gauge and  eliminating potentially diverging terms at the observer position.

\paragraph{First-order analysis.} Let us first discuss about the choice of $w^{(1)}_0 (w, \tilde{\theta}^a)$. Looking back at Eqs.~\eqref{eq:perturbations-dA-z}, one can see that
\begin{equation}
    \dot{\ms{V}}^{(1)} = \dot{\nu}^{(1)}-\frac{D^2 v^{(1)}}{2}-H\dot{\xi}^\tau_{(1)}-\dot{H}\xi^\tau_{(1)}+\frac{1}{r}\pa_\tau \bigg ( \frac{\xi^\tau_{(1)}}{a}-\xi^w_{(1)} \bigg ) +\frac{1}{ar^2} \bigg ( \frac{\xi^\tau_{(1)}}{a}-\xi^w_{(1)}-\frac{D^2\xi^w_{(1)}}{2}\bigg ) \, .
\end{equation}
Thus, 
\begin{align}
    \left (ar\dot{\ms{V}}^{(1)}\right )\o = \bigg [ a \pa_\tau \bigg ( \frac{\xi^\tau_{(1)}}{a}-\xi^w_{(1)} \bigg )+\frac{1}{r} \bigg ( \frac{\xi^\tau_{(1)}}{a}-\xi^w_{(1)}-\frac{D^2\xi^w_{(1)}}{2}\bigg )\bigg ] \o  \, , 
\end{align}
which diverges as $r^{-1}$ when $r\rightarrow 0$ at the observer position. In order to eliminate such a divergence, following \cite{Fanizza:2020xtv}, we impose the condition
\begin{equation}
    \bigg ( 1+ \frac{D^2}{2}\bigg ) w^{(1)}_0 = \frac{\left (\xi^\tau_{(1)}\right )_\t{o}}{a_\t{o}} \, .
    \label{eq:cancel-diverg-w0}
\end{equation}
Since $(\xi^\tau_{(1)})_\t{o}$ is a monopole (see Eq.~\eqref{eq:xi0}), the  general solution is 
\begin{equation}
    w^{(1)}_0(w,\tilde{\theta}^a) = \frac{\left (\xi^\tau_{(1)}\right )_\t{o}}{a_\t{o}} + d^{(1)}_\t{o}(w)P_1(\cos \tilde{\theta}^1) \,  , 
\end{equation}
with $d^{(1)}_\t{o}(w)$ a free function of $w$ and $P_1$ the Legendre polynomial of order 1. However, at a fully non-linear order, the GLC gauge admits the symmetry $w\rightarrow w^\pr (w)$, as we have recalled in Eqs.~\eqref{eq:glc-residual-freedom}. Thus, in order to be consistent with this symmetry, we require that $d^{(1)}_\t{o} =0$, namely $w^{(1)}_0$ is a pure monopole:
\begin{equation}
    w^{(1)}_0 = \frac{\left (\xi^\tau_{(1)}\right )_\t{o}}{a_\t{o}} \, .
    \label{eq:w-0-fixing}
\end{equation}
The relevant consequence of this choice is that
\begin{equation}
   \ka^r_{(1)} = \frac{\pa r}{\pa x^\mu} \xi^\mu_{(1)} = -\frac{\xi^\tau_{(1)}}{a}+\xi^w_{(1)} \, \qquad \Rightarrow \qquad \left (\ka^r_{(1)}\right )_\t{o} =0 \, .
\end{equation}
Therefore, from a physical point of view, we are requiring that the observer is set at the center of the polar coordinates.

Moreover, we can also reason in the opposite way wrt what done so far, that is, we start by requiring that the observer is set at the center of the polar frame, \textit{i.e.} $r\o = w\o-\eta(\tau \o)=0$. Then, in order to preserve the condition $w\o=\eta(\tau \o)$ in any other gauge, by taking the first-order gauge transformations
\begin{equation}
    \tau \rightarrow \tilde{\tau} = \tau + \xi^\tau_{(1)} \quad , \quad w \rightarrow \tilde{w} = w + \xi^w_{(1)} \quad , \quad  \eta \rightarrow \tilde{\eta} = \eta + \ka^\eta_{(1)} \,  , 
\end{equation}
we have to demand that, locally around the observer, it holds
\begin{equation}
    \left(\xi^w_{(1)}\right)\o \equiv w^{(1)}_0 = \left(\ka^\eta_{(1)}\right)\o = \frac{\left(\xi^\tau_{(1)}\right)\o}{a\o} \, ,
\end{equation}
which is exactly condition \eqref{eq:w-0-fixing}. Therefore, at the observer we are left with 
\begin{align}
    \ms{V}^{(1)}\o = \nu^{(1)}\o-H\o\left(\xi^\tau_{(1)}\right)\o \qquad ,  \qquad \ms{N}^{(1)}\o = N^{(1)}\o - 2H\o\left(\xi^\tau_{(1)}\right)\o \, .
\end{align}
Also the residual angular gauge modes $\chi^{(n)}_0(w, \tilde{\theta}^a)$ and $\hat{\chi}^{(n)}_0(w, \tilde{\theta}^a)$ can be fixed to require the observational gauge, where the angles $\tilde{\theta}^a$ coincide with the \textit{observed} angular directions of incoming photons in the sky, as in the so-called Fermi Normal Coordinates (see \cite{Fanizza:2018tzp}). Let us consider the first-order gauge transformation mapping the angles as $\tilde{\theta}^a_\t{o} \rightarrow \tilde{\theta}^a_\t{o} + (\xi^a_{(1)})_\t{o}$, with $(\xi^a_{(1)})_\t{o} = q^{ab} (D_b \chi^{(1)}_0+\tilde{D}_b \hat{\chi}^{(1)}_0)$ at the observer position. In the GLC gauge, the angles are precisely the background ones of the polar frame, namely $\tilde{\theta}^a = \theta^a_\t{o}$ (see also \cite{Fanizza:2015swa}), hence $\left(\xi^a_{(1)}\right)_\t{o}=0$, which is satisfied upon assuming that
\begin{equation}
    \chi^{(1)}_0 = \chi^{(1)}_0(w) \qquad ,  \qquad \hat{\chi}^{(1)}_0 = \hat{\chi}^{(1)}_0(w) \, .
    \label{eq:angular-modes-obs-fixing}
\end{equation}
More in general, since no preferable angular  direction at the center of the polar frame is allowed, any quantity $X$ evaluated at the observer position must satisfy $D_a X \o \equiv 0$.

\paragraph{Second-order analysis.} The techniques described above can be conveniently extended also to second order. In particular, upon requiring the monopole conditions
\begin{align}
    w^{(n)}_0= \frac{\left (\xi^\tau_{(n)}\right )\o}{a\o} \qquad , \qquad \chi^{(n)}_0 = \chi^{(n)}_0(w)\qquad ,  \qquad \hat{\chi}^{(n)}_0=\hat{\chi}^{(n)}_0(w) \, , 
    \label{eq:obs-gauge-fixing}
\end{align}
we are able to eliminate all the divergences around the observer position. This is one of the key results of our work, showing for the first time in the literature that, both at first and second order, these cancellations can be achieved only by requiring the \quotes{geometrical} or \quotes{kinematic} conditions \eqref{eq:obs-gauge-fixing} (namely, no assumption on the underlying cosmological model is needed).

To prove that this is indeed the case also at second order, we take
\begin{align}
     -\ms{V}^{(2)} \o + \left(ar\dot{\ms{V}}^{(2)}\right)\o &= -\nu^{(2)}\o +\frac{1}{2}\left(\dot{\xi}^\tau_{(2)}\right)\o -\frac{a\o}{2}\dot{\xi}^w_{(2)} -\mathbb{V}^{(2)}\o+\left(ar\dot{\mathbb{V}}^{(2)}\right)\o .
     \label{eq:noIRdiv-obs}
\end{align}
Using the first of Eqs.~\eqref{eq:definitions-glc-gauge-inv-2nd}, one can compute
\begin{align}
    -\mathbb{V}^{(2)}\o+\left(ar\dot{\mathbb{V}}^{(2)}\right)\o &= \bigg [ \dot{\nu}^{(1)}\xi^\tau_{(1)}+ \xi^w_{(1)}\pa_w \nu^{(1)}+\frac{H^2}{2}\left(\xi^\tau_{(1)}\right)^2 -H \xi^\tau_{(1)} \dot{\xi}^\tau_{(1)}+2\nu^{(1)}\dot{\xi}^\tau_{(1)}-2a\nu \dot{\xi}^w_{(1)}\nex
    & \quad +2aH\xi^\tau \dot{\xi}^w_{(1)} -\frac{1}{2}\left(\dot{\xi}^\tau_{(1)}\right)^2-\frac{1}{2}\xi^\tau_{(1)}\ddot{\xi}^\tau_{(1)}-\frac{1}{2}\xi^w_{(1)}\pa_w \dot{\xi}^\tau_{(1)} +\frac{a}{2}\dot{\xi}^\tau_{(1)}\dot{\xi}^w_{(1)}\nex
    & \quad +\frac{a}{2}\xi^\tau_{(1)}\ddot{\xi}^w_{(1)}+\frac{a}{2}\xi^w \pa_w \dot{\xi}^w_{(1)}\bigg ]\o \, ,
    \label{eq:elimination-div}
\end{align}
devoid of any  divergence.

\subsection{Detailed expressions for the angular distance–redshift relation}
We are now ready to fully open formulae \eqref{eq:perturbations-dA-z} for $d_\t{A}(z)$ and compare our results to the ones obtained directly in standard perturbation theory. This is also a consistency check of the perturbative scheme that we have developed in this paper.

In order to do so, we express each light-cone perturbation in terms of PG quantities through the relations \eqref{eq:SPS-NG}, which can be directly plugged into Eqs.~\eqref{eq:perturbations-dA-z} because the latter are valid in any gauge.

\subsubsection{First-order angular distance–redshift relation}
Following the strategy outlined above, for $d_\t{A}(z)$ at first order we obtain
\begin{align}
    d^{(1)}_\t{A}(z) &=-\Psi_z 
    -\bigg (1-\frac{1}{a_zr_zH_z} \bigg ) \Phi|^\t{o}_z + \frac{1}{r_z} Q_z -\int_{\tau_z}^{\tau_\t{o}}\frac{\t{d}\tau}{2ar^2}  D^2 Q \nex
    & \quad -\bigg ( 1- \frac{1}{a_zH_zr_z} \bigg ) \big (\pa_w Q_z +\pa_w P_z\big )\notag \\[1ex]
    & \quad - \frac{1}{a_z H_zr_z}   \pa_w P\o-\bigg ( a_\t{o}H_\t{o} -\frac{a_\t{o}H_\t{o}}{a_zH_zr_z}+ \frac{1}{r_z}\bigg )  P\o \, .
    \label{eq:dA-glc-1st-ng}
\end{align}
Before moving to standard coordinates, we recall that the partial derivatives transform as
\begin{equation}
    \pa_\tau = \frac{1}{a}(\pa_\eta - \pa_r) \quad , \quad \pa_w = \pa_r \quad , \quad \pa_{\tilde{\theta}^a} = \pa_{\theta^a}  \, , 
\end{equation}
and that, for a generic quantity $X$, we have (see \cite{Fanizza:2020xtv,Scaccabarozzi:2017ncm})
\begin{align}
    \pa_w X \big (\tau^\prime, w-\eta(\tau)+\eta(\tau^\prime) \big ) &= \pa_r X(\eta^\prime, \eta_\t{o}-\eta) \,   \,    \label{eq:map-deriv} 
    \end{align}
 and 
 \begin{align}
    \pa_w X(\tau^\prime, w) &= \pa_r X(\eta^\prime, \eta_\t{o}-\eta^\prime)  = \pa_{\eta^\prime} X(\eta^\prime, \eta_\t{o}-\eta^\prime) -\frac{\t{d}}{\t{d}\eta^\prime}X(\eta^\prime, \eta_\t{o}-\eta^\prime)  \, .
    \label{eq:map-deriv-2}
\end{align}
Moreover, let us define the radial peculiar velocity
\begin{align}
    v_{|| \alpha} \equiv \pa_r P_\alpha = \frac{1}{a_\alpha} \int_{\eta_{\t{in}}}^{\eta_\alpha} \t{d}\eta \, a \pa_r \Phi \qquad  , \qquad \alpha = \t{o}, z \, .
    \label{eq:pecvel-1st-defined}
\end{align}
Then, using Eqs. \eqref{eq:map-deriv} and \eqref{eq:map-deriv-2} and the definition \eqref{eq:pecvel-1st-defined} into the first-order gauge modes \eqref{eq:gauge-modes-NG-1st}, we see that each gauge mode corresponds to a general relativistic effect. In particular, $\xi^\tau_{(1)}$ is the velocity potential, the term $\pa_w \xi^w_{(1)z}$, namely
\begin{align}
    \pa_w \xi^w_{(1)z} = -2\int_{\eta_z}^{\eta_\t{o}} \t{d}\eta \, \pa_\eta \Psi^\t{I} + 2 \Psi^\t{I}\o - 2\Psi^\t{I}_z \, , 
\end{align}
gives the local and integrated Sachs-Wolfe effects, and finally 
\begin{align}
D_a \xi^a_{(1)z}=D^2 \chi_{(1)z} = \frac{2}{r_z} \int_{\eta_z}^{\eta_\t{o}} \t{d}\eta \, \frac{\eta - \eta_z}{\eta_\t{o}-\eta} D^2 \Psi^\t{I}
\end{align}
amounts to lensing.

Exploiting the above results, the first-order perturbation to the angular distance–redshift relation  in standard coordinates  is
\begin{align}
    d^{(1)}_\t{A}(z) &=-(\Psi^\t{I}_z+\Psi^\t{A}_z)
    +\bigg (1-\frac{1}{r_z\Hcal_z} \bigg ) \big [ \Psi^\t{I}\o - \Psi^\t{A}\o - (\Psi^\t{I}_z - \Psi^\t{A}_z)\big ] \notag \\[1ex]
    & \quad - \frac{1}{r_z}\int_{\eta_z}^{\eta_\t{o}}\t{d}\eta \, \frac{\eta - \eta_z}{\eta \o - \eta} D^2 \Psi^\t{I}+\frac{2}{r_z} \int_{\eta_z}^{\eta_\t{o}}\t{d}\eta \, \Psi^\t{I}\notag \\[1ex]
    & \quad -\bigg ( 1- \frac{1}{r_z\Hcal_z} \bigg ) \bigg (2\int_{\eta_z}^{\eta_\t{o}} \t{d}\eta \, \pa_{\eta}\Psi^\t{I}+v_{|| z} \bigg )- \frac{1}{r_z \Hcal_z}  v_{|| \t{o}}\notag \\[1ex]
    & \quad 
-\bigg (\Hcal_\t{o}-\frac{\Hcal_\t{o}}{r_z\Hcal_z} + \frac{1}{r_z}\bigg ) \frac{1}{a_\t{o}}\int_{\eta_{\t{in}}}^{\eta_\t{o}} \t{d}\eta \, a \Phi \, ,
    \label{eq:dA-1st-standard-NG}
\end{align}
where  we have  rewritten
\begin{align}
\int_{\eta_z}^{\eta_\t{o}} \frac{\t{d}\eta}{r^2} \int_\eta^{\eta_\t{o}} \t{d}\eta^\pr D^2 \Psi^\t{I}(\eta^\pr) &=\int_0^{\eta_\t{o}-\eta_z} \frac{\t{d}x}{x^2} \, \int_0^x \t{d}x^\pr D^2 \Psi^\t{I}(\eta_\t{o}-x^\pr) \notag \\[1ex]
& = \frac{1}{\eta_\t{o}-\eta_z} \int_0^{\eta_\t{o}-\eta_z} \t{d}x \, \frac{\eta_\t{o}-\eta_z-x}{x} D^2 \Psi^\t{I}(\eta \o - x)\notag \\[1ex]
& = \frac{1}{\eta_\t{o}-\eta_z} \int_{\eta_z}^{\eta_\t{o}} \t{d}\eta \, \frac{\eta-\eta_z}{\eta_\t{o}-\eta}D^2 \Psi^\t{I}(\eta) \, .
\end{align}
In the first line we have introduced the auxiliary variables $x \equiv \eta_\t{o}-\eta$ and $x^\pr \equiv \eta_\t{o}-\eta^\pr$ and in the second one we have exploited the fact that, for a generic function $f(x)$, it holds
\begin{equation}
   \int_0^{a}\frac{\t{d}x}{x^2} \int_0^x \t{d}x^\pr f(x^\pr) = \int_0^a \t{d}x^\pr f(x^\pr) \int_{x^\pr}^a \frac{\t{d}x}{x^2}  = \frac{1}{a}\int_{0}^a \t{d}x \, \frac{a-x}{x}f(x) \, .
\end{equation}
As a first consistency check of Eq.~\eqref{eq:dA-1st-standard-NG} with \cite{Fanizza:2020xtv}, one can easily verify that the same result is recovered by first moving to standard coordinates and then imposing the PG fixing conditions $E=0=B$. However, from a computational point of view, it is more convenient to fix the PG conditions still within the GLC formalism, so as to write $d^{(1)}_{\t{A}}(z)$ only in terms of $\Phi$ and $\Psi$, and finally go back to standard coordinates, as we have done so far.

As a final step regarding the first-order evaluation of $d_\t{A}(z)$ in terms of the Bardeen potentials $\Phi$ and $\Psi$, we can now compare our result \eqref{eq:dA-1st-standard-NG} with the ones present in the literature. In particular, all but the last line of our Eq.~\eqref{eq:dA-1st-standard-NG} match with the works \cite{BenDayan:2012wi,Bonvin:2005ps} for vanishing anisotropic stress and with  \cite{Marozzi:2014kua} for the case of case of non-vanishing anisotropic stress.

Moreover, the last term of \eqref{eq:dA-1st-standard-NG} at the observer position matches with the quantity that the authors of \cite{Biern:2016kys,Scaccabarozzi:2017ncm} called the \textit{time-shift}. This is needed to re-express the angular distance–redshift relation of \cite{Marozzi:2014kua,BenDayan:2012wi} in a gauge where the observer is free-falling.

\subsubsection{Second-order angular distance–redshift relation} 
In this final subsection, we will present our complete result for the  angular distance–redshift relation as measured by a free-falling observer. Before reporting the full formulae, we will separately focus on the numerator and denominator of Eq.~\eqref{eq:dA-z-limit}. In this way, we will be able to show that the purely source terms of $d_\t{A}(z)$ agree with the ones of the literature as they are gauge invariant/observer independent. On the other hand, we will provide the explicit form of all the new terms at the observer position due to our complete gauge fixing procedure associated with the geodesic observer.

\paragraph{Numerator of the angular distance–redshift relation.} Let us start from the numerator of $d^{(2)}_\t{A}(z)$ as written in \eqref{eq:dA-z-limit}, namely the first of Eqs.~\eqref{eq:d_A-num-den}. We replace the GLC gauge-invariant variables with their explicit forms \eqref{eq:glc-gauge-inv-2nd-order} and write everything in terms of $\Psi^\t{I}$ and $\Psi^\t{A}$. By defining $\Xi \equiv 1-\frac{1}{\Hcal_zr_z}$, and omitting the superscript \quotes{1} for first-order gauge modes, the result is\footnote{Here, our formula is obtained in terms of a set of fiducial coordinates, namely $\big (z_{\t{obs}},\,w_{\t{obs}},\tilde{\theta}^a_{\t{obs}}\big)$. In the following we will omit it and just write  $d_{\t{A}}(z)$.}
\begin{align}
    d^{(2, \t{num})}_{\t{A}}(z)&= \frac{1}{H_{z}}\left[-\Psi^\t{I}-\Psi^\t{A}-H\xi^{\tau}+\partial_{w}\left(\frac{\xi^{\tau}}{a}-\xi^{w}\right)\right]_{z}^\t{o} \times\nonumber \\[1ex]
 &\quad \times\left[-\dot{\Psi}^{\t{I}}- \dot{\Psi}^\t{A}-\partial_{\tau}\left(\Xi H\right)\xi^{\tau}-\Xi H\left(\Psi^\t{I}-\Psi^\t{A}-\frac{1}{a}\partial_{w}\xi^{\tau}\right)-\frac{2\Psi^{\t{I}}}{ar}\right.\nonumber \\[1ex]
 & \quad \left.-\frac{1}{ar^{2}}\xi^{w}+\frac{1}{ar^{2}}\int_{\tau}^{\tau_{\t{o}}}\frac{\t{d}\tau^\pr}{a}\, D^{2}\Psi^{\t{I}}\right]_{z}\nonumber \\[1ex]
 & \quad -\Xi\left[\Psi^\t{I}+\Psi^\t{A}+H\xi^{\tau}-\partial_{w}\left(\frac{\xi^{\tau}}{a}-\xi^{w}\right)\right]_{z}^{\t{o}}\left[-\Psi^\t{I}-\Psi^\t{A}-\Xi H\xi^{\tau}-\frac{\xi^{w}}{r}-\frac{1}{2}D_{a}\xi^{a}\right]_{z}\nonumber \\[1ex]
 & \quad +\frac{1}{2}\Xi\bigg [ -2\Psi_{(2)}-H\left(\xi_{(2)}^{\tau}-\xi^{\mu}\partial_{\mu}\xi^{\tau}\right)+\frac{1}{a}\partial_{w}\left(\xi_{(2)}^{\tau}-\xi^{\mu}\partial_{\mu}\xi^{\tau}\right)\nonumber \\[1ex]
 & \quad -\partial_{w}\left(\xi_{(2)}^{w}-\xi^{\mu}\partial_{\mu}\xi^{w}\right) +\frac{\ddot{a}}{a}\left(\xi^{\tau}\right)^{2}+H^{2}\left(\xi^{\tau}\right)^{2}-2H\frac{\xi^{\tau}}{a}\partial_{w}\xi^{\tau} \nonumber \\[1ex]
 & \quad -4(\Psi^\t{I}+\Psi^\t{A})\left(\partial_{w}\frac{\xi^{\tau}}{a}-\partial_{w}\xi^{w}\right)+\left(\partial_{w}\xi^{w}\right)^{2}+\bar{\gamma}_{ab}\partial_{w}\xi^{a}\partial_{w}\xi^{b}+2\xi^{\mu}\partial_{\mu}(\Psi^\t{I}+\Psi^\t{A}) \nonumber\\[1ex]
 &  \quad -2\partial_{w}\frac{\xi^{\tau}}{a}\partial_{w}\xi^{w} +4H\xi^{\tau}(\Psi^\t{I}+\Psi^\t{A})+4H\xi^{\tau}\partial_{w}\xi^{w}\bigg ]_{z}^{\t{o}}  \nonumber\\[1ex]
 & \quad  +\frac{1}{2}\Xi\Bigg\{-\left[-\Psi^\t{I}-\Psi^\t{A}-H\xi^{\tau}+\partial_{w}\left(\frac{\xi^{\tau}}{a}-\xi^{w}\right)\right]_{\t{o}}^{2} \nonumber \\[1ex]
 & \quad +3\left[-\Psi^\t{I}-\Psi^\t{A}-H\xi^{\tau}+\partial_{w}\left(\frac{\xi^{\tau}}{a}-\xi^{w}\right)\right]_{z}^{2}\nonumber \\[1ex]
 & \quad -\left[\bar{\gamma}^{ab}\left(\partial_{a}\frac{\xi^{\tau}}{a}-\partial_{a}\xi^{w}-\bar{\gamma}_{ac}\partial_{w}\xi^{c}\right)\left(\partial_{b}\frac{\xi^{\tau}}{a}-\partial_{b}\xi^{w}-\bar{\gamma}_{bd}\partial_{w}\xi^{d}\right)\right]_{z}^{\t{o}}\nonumber \\[1ex]
 & \quad -2\left[-\Psi^\t{I}-\Psi^\t{A}-H\xi^{\tau}+\partial_{w}\left(\frac{\xi^{\tau}}{a}-\xi^{w}\right)\right]_{\t{o}}\times \nonumber \\[1ex]
 & \quad \times \left[-\Psi^\t{I}-\Psi^\t{A}-H\xi^{\tau}+\partial_{w}\left(\frac{\xi^{\tau}}{a}-\xi^{w}\right)\right]_{z}\Bigg\}\nonumber \\[1ex]
 & \quad -\frac{\Xi}{H_{z}}\left[-\Psi^\t{I}-\Psi^\t{A}-H\xi^{\tau}+\partial_{w}\left(\frac{\xi^{\tau}}{a}-\xi^{w}\right)\right]_{z}^{\t{o}}\times \nonumber \\[1ex]
 & \quad \times \left[-\dot{\Psi}^\t{I}-\dot{\Psi}^\t{A}-\dot{H}\xi^{\tau}-H(\Psi^\t{I}-\Psi^\t{A})+\frac{1}{a}\partial_{w}\left(\Psi^\t{I}-\Psi^\t{A}-\frac{1}{a}\partial_{w}\xi^{\tau}\right)-\frac{2}{a}\partial_{w}\Psi^{\t{I}}\right]_{z}\nonumber \\[1ex]
 & \quad +\frac{\dot{H}_{z}}{2a_{z}r_{z}H_{z}^{3}}\left\{-[\Psi^\t{I}+\Psi^\t{A}]_{z}^{\t{o}}-\left[H\xi^{\tau}\right]_{z}^{\t{o}}+\partial_{w}\left[\frac{\xi^{\tau}}{a}-\xi^{w}\right]_{z}^{\t{o}}\right\}^{2}-\Psi^{(2)}_z\nonumber \\[1ex]
 & \quad -\frac{1}{2}\Xi H_z\left(\xi_{(2)}^{\tau}-\xi^{\mu}\partial_{\mu}\xi^{\tau}\right)_z-\frac{1}{2r_z}\left(\xi_{(2)}^{w}-\xi^{\mu}\partial_{\mu}\xi^{w}\right)_z-\frac{1}{4}D_{a}\left(\xi_{(2)}^{a}-\xi^{\mu}\partial_{\mu}\xi^{a}\right)_z\nonumber \\[1ex]
 & \quad +\frac{1}{2}(\Psi^\t{I}_z+\Psi^\t{A}_z) D_{a}\xi^{a}_z+\xi^{\mu}_z\partial_{\mu}(\Psi^\t{I}_z+\Psi^\t{A}_z)+2\Xi H_z\xi^{\tau}_z(\Psi^\t{I}_z+\Psi^\t{A}_z)+\frac{1}{r_z}\xi^{w}_z(\Psi^\t{I}_z+\Psi^\t{A}_z)\nonumber \\[1ex]
 & \quad +\frac{1}{2}\left(H_z\xi^{\tau}_z\right)^{2}\left(2+\frac{\dot{H}_z}{H^{2}_z}\right)+\frac{1}{4}\left ( \bar{\gamma}^{ab}\partial_{a}\xi^{w}\partial_{b}\xi^{w}\right)_z-\frac{1}{2}\left(\bar{\gamma}^{ab}\partial_{a}\frac{\xi^{\tau}}{a}\partial_{b}\xi^{w}\right)_z \nonumber \\[1ex]
 & \quad + H_z\xi^{\tau}_zD_{a}\xi^{a}_z-\frac{2}{a_zr_zH_z}\left(H_z\xi^{\tau}_z\right)^{2}+\frac{2H_z}{r_z}(1-\Xi)\xi^{\tau}_z\xi^{w}_z-\frac{1}{r_z}\left(\frac{\xi^{\tau}_z}{a_z}-\xi^{w}_z\right)D_{a}\xi^{a}_z\nonumber \\[1ex]
 & \quad +\frac{1}{2}\left ( \bar{\gamma}^{ab}\xi^{c}\partial_{c}\bar{\gamma}_{ad}\partial_{b}\xi^{d}\right)_z +\frac{1}{4}\left ( \bar{\gamma}^{ab}\bar{\gamma}_{cd}\partial_{a}\xi^{c}\partial_{b}\xi^{d}\right)_z+\frac{1}{2a_zr_z}\left(\xi^{\tau}_z\right)^{2}\left(H_z+\frac{1}{a_zr_z}\right)\nonumber \\[1ex]
 & \quad -\frac{1}{a_zr^{2}_z}\xi^{\tau}_z\xi^{w}_z +\frac{1}{8}\left (\bar{\gamma}^{ab}\xi^{c}\xi^{d}\partial_{c}\partial_{d}\bar{\gamma}_{ab}\right)_z-\frac{1}{2}(\Psi^\t{I}_z+\Psi^\t{A}_z)^{2}-\frac{1}{2}\left(\Xi H_z \xi^{\tau}_z\right)^{2}-\frac{1}{8}\left(D_{a}\xi^{a}_z\right)^{2}\nonumber \\[1ex]
 & \quad -\frac{\Xi}{2}H_z\xi^{\tau}_zD_{a}\xi^{a}_z-\frac{1}{2r_z}\xi^{w}_zD_{a}\xi^{a}_z\nonumber \\[1ex]
 & \quad + \frac{1}{2}\left[\frac{1}{\sin^2 \tilde{\theta}^{1}}\varepsilon^{ac}\varepsilon^{bd}\left(D_{ab}\chi+\tilde{D}_{ab}\hat{\chi}\right)\left(D_{cd}\chi+\tilde{D}_{cd}\hat{\chi}\right) \right]_z\, , 
\label{eq:dA-source-comparison}
\end{align}
where, using $\hat{\chi}= \hat{\chi}_0(w)$ from  \eqref{eq:gauge-modes-NG-1st}, the last term on the last line can be recast as
\begin{align}
 \frac{1}{2\sin^2 \tilde{\theta}^{1}}\varepsilon^{ac}\varepsilon^{bd}\left(D_{ab}\chi+\tilde{D}_{ab}\chi\right)\left(D_{cd}\chi+\tilde{D}_{cd}\chi\right) =  \frac{1}{4}(D_a \xi^a)^2 -\frac{1}{2}D_a \xi^b D_b \xi^a\, .  
\end{align}
Eq.~\eqref{eq:dA-source-comparison} also represents a validation of the new perturbation theory developed in this paper.
Indeed, by removing all the terms at the observer position from \eqref{eq:dA-source-comparison} and taking the limit of vanishing anisotropic stress, this equation agrees with Eq.~(A.17)  of \cite{Fanizza:2015swa}.

In order to prove this, we just need to  recall the relations connecting the PG quantities to the GLC gauge ones,   obtained in \cite{Fanizza:2015swa} starting from the diffeomorphism law \eqref{eq:diff} for the inverse metrics,  as we have done in Subsect.~\ref{sec:unifying}.
Indeed, by plugging into Eq.~\eqref{eq:dA-source-comparison} the expressions of each gauge mode in terms of $(\eta^{(1,2)}, \eta^{+(1,2)}, \theta^{a(1,2)})$ at first and second order using Eqs.~\eqref{eq:GLC-NG-gaugemodes-1storder} and \eqref{eq:GLC-NG-gaugemodes-2ndorder}, we precisely recover Eq.~(A.17) of \cite{Fanizza:2015swa},  which is the angular distance–redshift relation expressed in terms of observed quantities using the approach of \cite{Fanizza:2015swa} applied at the next-to-leading order.

As a  technical remark, in doing the comparison with \cite{Fanizza:2015swa}, we noticed that the terms
\begin{align}
    -\frac{\dot{\Psi}^\t{I}_z}{a_zH^2_zr_z}\left (\Psi^\t{I}-\pa_w \eta^{+(1)}+\pa_w \eta^{(1)}-aH\eta^{(1)}\right )_z
\end{align}
were accidentally missed in (A.17) of \cite{Fanizza:2015swa}, but consistently included in the comparison of that work with \cite{Bonvin:2005ps,Pyne:2003bn} for first-order results and \cite{BenDayan:2012wi,Marozzi:2014kua} for second-order ones. 

Therefore, the rigorous comparison of the source terms obtained with our approach and in \cite{Fanizza:2015swa} provides a first confirmation of the correctness of our light-cone perturbation theory to second order. Moreover, Eq.~\eqref{eq:dA-source-comparison} also represents an improvement of \cite{Fanizza:2015swa}, as we have not neglected the effects of the anisotropic potential, as also done in \cite{Marozzi:2014kua}. 

We emphasize that the terms at the observer position in \eqref{eq:dA-source-comparison} can not be directly matched with the literature. In particular, all the terms with the velocity potential $\xi^\tau_\t{o}$ represents genuine new contributions to the already existing formulae.

\paragraph{Denominator of the angular distance–redshift relation.} Thanks to the matching with the literature described in the previous paragraph, we have now validated our methodology to compute $d^{(2)}_\t{A}(z)$. Thus, we can systematically proceed with the evaluation of the observer terms which do not have to perfectly match with the literature, as they are due to the observational gauge fixing, correspondent to the fact that we are computing $d_\t{A}(z)$ as measured by a free-falling observer. 

The first observer contribution is contained in Eq.~\eqref{eq:dA-source-comparison}, as explained at the end of the previous paragraph. The second stems from the denominator of $d^{(2)}_\t{A}(z)$. By Taylor expanding  Eq.~\eqref{eq:den-dA} and making use of 
\begin{align}
   (ar\dot{\ms{V}}^{(1)})_\t{o}& = \bigg [ -\frac{1}{a}\pa_w \xi^\tau -H \xi^\tau - \Psi^\t{I}-\Psi^\t{A}\bigg ] \o 
\end{align}
and of Eqs.~\eqref{eq:noIRdiv-obs} and \eqref{eq:elimination-div}, we get 
\begin{align}
    \frac{1}{d^{(\t{den})}_\t{A}(z)} 
    & = \left [ 1- \ms{V}^{(1)}+ar\dot{\ms{V}}^{(1)}-\ms{V}^{(2)}+ar\dot{\ms{V}}^{(2)}+\frac{3}{2}(\ms{V}^{(1)})^2-3ar\ms{V}^{(1)}\dot{\ms{V}}^{(1)}\right. \nex
    & \left. \quad +\frac{\mathfrak{g}^{(2)}}{4}-\frac{\tilde{\mathfrak{g}}^{(2)}}{2}+(ar\dot{\ms{V}}^{(1)})^2\right ]\o \nex
    & = 1-\frac{1}{a\o}\pa_w \xi^\tau\o -\frac{1}{a\o}(\Psi^\t{I}\o+\Psi^\t{A}\o)\pa_w \xi^\tau \o-\frac{H\o}{2a\o}\xi^\tau\o \pa_w \xi^\tau \o +\frac{\xi^\tau\o}{a}\pa_w (\Psi^\t{I}\o-\Psi^\t{A}\o)
    \nex 
    & \quad -\frac{1}{2a\o}\int_{\tau_{\t{in}}}^{\tau_\t{o}} \t{d}\tau^\pr \,  \pa_w \bigg [2\Phi^{(2)}-\Phi^2 + (\pa_w P)^2 + \bar{\ga}^{ab}\pa_a P \pa_b P \bigg ] \nex
    & \quad +\frac{1}{2a^2\o}\pa_w\xi^\tau \o\int_{\tau_{\t{in}}}^{\tau_\t{o}} \t{d}\tau^\pr \, \pa^2_w P  \nex
    & \equiv 1+d^{(1,\t{den})}_{\t{A}}(z)+d^{(2,\t{den})}_{\t{A}}(z) \, .
\end{align}
By multiplying the above equation with $d^{(\t{num})}_\t{A}(z)$ of Eqs.~\eqref{eq:d_A-num-den} up to second order, we obtain 
\begin{align}
    d^{(2,\t{new})}_\t{A}(z)&= d^{(1,\t{num})}_{\t{A}}(z)d^{(1,\t{den})}_{\t{A}}(z) + d^{(2,\t{den})}_{\t{A}}(z)\nex
    & =\frac{1}{a\o} \pa_w \xi^\tau_{\t{o}}\bigg [ \Psi^\t{I}_z +\Psi^\t{A}_z +\frac{1}{2}D_a \xi^a_{z} +\frac{1}{r_z}\xi^w_{z}\nex
    & \quad +\bigg ( 1-\frac{1}{a_z H_z r_z}\bigg ) \big (\Psi^\t{I}\o + \Psi^\t{A}\o +H\o \xi^\tau_{(1)\t{o}} -\Psi^\t{I}_z - \Psi^\t{A}_z\nex
    & \quad + \frac{1}{a_z}\pa_w \xi^\tau_{z}-\pa_w \xi^w_{z}\big ) \bigg ]  -\frac{1}{a\o}(\Psi^\t{I}\o+\Psi^\t{A}\o)\pa_w \xi^\tau_{\t{o}}\nex
    & \quad -\frac{H\o}{2a\o}\xi^\tau_{\t{o}}\pa_w \xi^\tau_{\t{o}} +\frac{\xi^\tau_{\t{o}}}{a\o}\pa_w (\Psi^\t{I}\o-\Psi^\t{A}\o)
    \nex 
    & \quad -\frac{1}{2a\o}\int_{\tau_{\t{in}}}^{\tau_\t{o}} \t{d}\tau^\pr \,  \pa_w \bigg [2\Phi^{(2)}-\Phi^2 + (\pa_w P)^2 + \bar{\ga}^{ab}\pa_a P \pa_b P \bigg ] \nex
    & \quad +\frac{1}{2a^2\o}\pa_w \xi^\tau_{\t{o}}\int_{\tau_{\t{in}}}^{\tau_\t{o}} \t{d}\tau^\pr \, \pa^2_w P \, . 
    \label{eq:new-dA-1}
\end{align}
Therefore, the complete gauge-invariant formula for the angular distance–redshift relation is obtained once we sum Eqs.~\eqref{eq:dA-source-comparison} and \eqref{eq:new-dA-1}, by replacing the explicit forms of the second-order gauge modes. This is what we are going to do in the next final paragraph.

\paragraph{Full expressions for the angular distance–redshift relation.} To conclude, we are now ready to report the final, fully opened formulae for $d^{(2)}_\t{A}(z)$ as measured by a free-falling observer, by collecting all the terms from Eqs.~\eqref{eq:d_A-num-den}, \eqref{eq:dA-source-comparison} and \eqref{eq:new-dA-1}. The purely source terms will be matched with \cite{Marozzi:2014kua}, where analogous formulae were provided also with anisotropic corrections. 

Following the notation of \cite{Marozzi:2014kua}, we define the first-order orthogonal velocity and the second-order peculiar velocity
\begin{align}
    v_{a \perp \alpha} &\equiv \pa_a P_\alpha \, , \nex
  v^{(2)}_{||\alpha} &\equiv \Psi_\alpha v_{|| \alpha} +\frac{1}{2a_\alpha}\int_{\eta_{\t{in}}}^{\eta_\alpha} \t{d}\eta \,  a \pa_r \big [2\Phi^{(2)}-\Phi^2 + (\pa_r P)^2 + \bar{\ga}^{ab}\pa_a P \pa_b P \big ]
  \label{eq:vperp-v2par}
\end{align}
(for $\alpha = \t{o}, z$) and we decompose 
\begin{align}
d^{(2)}_{\t{A}} (z) \equiv  d^{(2)}_{\t{pos}} (z) + d^{(2)}_{\t{mixed}} (z) +d^{(2)}_{\t{path}} (z) + d^{(2)}_{P, \t{pos}} (z)+d^{(2)}_{P, \t{mixed}} (z)+d^{(2)}_{v,\t{int}} (z)\,  , 
\end{align}
where, in terms of GR effects, 
\begin{itemize}
    \item \quotes{pos} indicates terms generated by peculiar velocities;
    \item \quotes{mixed} refers to the cross effects \quotes{peculiar velocities $\times$ (SW/ISW/lensing)};
    \item \quotes{path} contains all the effects due to SW, ISW and lensing;
    \item The subscript \quotes{$P$} indicates all the corrections due to the monopole $P\o$;
    \item Finally, \quotes{$v,\t{int}$} refers to terms with peculiar velocities  integrated between $\eta_{\t{in}}$ and $\eta_{\alpha}$.
\end{itemize}
We further split the contributions \quotes{mixed} and \quotes{path} into two subgroups of terms: the subscript \quotes{I} refers to couplings to the only isotropic potential $\Psi^\t{I}$, while \quotes{A} to couplings involving the anisotropic one $\Psi^\t{A}$. 

First we report the terms originating from the Taylor expansion of the denominator, Eq.~\eqref{eq:new-dA-1}. Once fully opened and written in terms of GR effects, they amount to
\begin{align}
    d^{(2,\t{new})}_\t{A}(z)&= d^{(1,\t{num})}_{\t{A}}(z)d^{(1,\t{den})}_{\t{A}}(z) + d^{(2,\t{den})}_{\t{A}}(z)\nex
    & =v_{||\t{o}}\bigg [ \Psi^\t{I}_z +\Psi^\t{A}_z +\frac{1}{r_z}\int_{\eta_z}^{\eta_\t{o}}\text{d}\eta \, \frac{\eta - \eta_z}{\eta_\t{o}-\eta}D^2 \Psi^\t{I} -\frac{2}{r_z} \int_{\eta_z}^{\eta_\t{o}}\text{d}\eta \, \Psi^\t{I}    \nex
    & \quad +\bigg ( 1-\frac{1}{\Hcal_z r_z}\bigg ) \bigg (-\Psi^\t{I}\o + \Psi^\t{A}\o +\Hcal \o P_\t{o} +\Psi^\t{I}_z - \Psi^\t{A}_z\nex
    & \quad + v_{||z}-v_{||\t{o}}+2\int_{\eta_z}^{\eta_\t{o}}\text{d}\eta \, \pa_\eta \Psi^\t{I}    \bigg ) \bigg ]  -
   v^{(2)}_{|| \t{o}}\nex
    & \quad -\frac{\Hcal \o}{2}P\o v_{||\t{o}} +P\o \pa_r (\Psi^\t{I}\o-\Psi^\t{A}\o)
   +\frac{v_{||\t{o}}}{2a\o}\int_{\eta_{\t{in}}}^{\eta_\t{o}} \t{d}\eta \, a \pa_r v_{||} \, . 
    \label{eq:new-dA-1-GReffects}
\end{align}
Finally, our results for the full angular distance–redshift relation are the following ones:
\begin{align}
   d^{(2)}_{\t{pos}}(z) &= \bigg [ 1-\frac{1}{\Hcal_z r_z}\bigg ] \bigg [ \frac{1}{2}v^a_{\perp z}v_{a\perp z}-\frac{1}{\Hcal_z}(v_{||z}-v_{||\t{o}})\pa_r v_{||z}  -v^{(2)}_{|| z}\bigg ]\nex
   & \quad -\frac{1}{\Hcal_z r_z}v^{(2)}_{||\t{o}}-\frac{1}{2}v^2_{|| z} +\frac{\Hcal^\pr_z}{2\Hcal^3_zr_z }(v_{||z}-v_{||\t{o}})^2+\bigg ( 1-\frac{2}{\Hcal_z r_z}\bigg )v_{||\t{o}}v_{||z}+\frac{1}{2}v^2_{||\t{o}}\, ,
   \label{eq:dpos}\\[1ex]   
   d^{(2)}_{\t{mixed,I}}(z) &= \left[1-\frac{1}{ \Hcal_z r_z}\right]\left[+ \Psi^\t{I}_z  v_{||z}+
\frac{1}{r_z} v_{||z} \int_{\eta_z}^{\eta_{\text{o}}} \t{d} \eta \, \frac {\eta - \eta_z}{\eta_{\text{o}} - \eta} D^2\Psi^\t{I}
+2  v^a_{\perp z} \int_{\eta_z}^{\eta_{\text{o}}}\t{d} \eta\, \partial_a \Psi^\t{I}
\right. \nex
 & \quad \left.
+\frac{1}{ \Hcal_z} \left(\Psi^\t{I}_{\text{o}}-\Psi^\t{I}_z-2\int_{\eta_z}^{\eta_{\text{o}}}\t{d} \eta\, \partial_{\eta}\Psi^\t{I}
-2\Hcal_z \int_{\eta_z}^{\eta_{\text{o}}}\t{d} \eta \Psi^\t{I}\right)\partial_r v_{||z} 
\right.\nex
 & \quad \left.
+2 \partial_a v_{||z} \int_{\eta_z}^{\eta_{\text{o}}}\t{d} \eta\, \bar{\gamma}^{ab} \partial_b \int_{\eta}^{\eta_{\text{o}}}\t{d}\eta^\pr \, \Psi^\t{I}\right]
- \frac{2}{r_z} v_{||z}\int_{\eta_z}^{\eta_{\text{o}}} \t{d} \eta\, \Psi^\t{I}\nex
&\quad +\left(v_{||z}-v_{||\t{o}}\right)\left[\frac{1}{ \Hcal_z r_z}\left(1-\frac{\Hcal_z'}{\Hcal^2_z}\right)\left(\Psi^\t{I}_{\text{o}}-\Psi^\t{I}_z-2\int_{\eta_z}^{\eta_{\text{o}}}\t{d} \eta\, \partial_{\eta}\Psi^\t{I}\right)
\right. \nex
& \quad \left.
+ \frac{2}{\Hcal_z r_z} \Psi^\t{I}_z+\frac{1}{\Hcal^2_z r_z} \partial_\eta \Psi^\t{I}_z-\frac{1}{\Hcal_z} \partial_r \Psi^\t{I}_z
-\frac{1}{\Hcal_z r^2_z}\int_{\eta_z}^{\eta_{\text{o}}}\t{d} \eta\, D^2\Psi^\t{I}
\right]
\nex
 &\quad  +v_{||\t{o}}\left[\frac{1}{2}\bigg (-7+\frac{9}{\Hcal_zr_z} \bigg )\Psi^\t{I}_z+5\bigg (1-\frac{1}{\Hcal_zr_z} \bigg )\Psi^\t{I}\o-\frac{4}{r_z}\int_{\eta_z}^{\eta_\t{o}}\t{d}\eta \, \Psi^\t{I}\right. \nex
 & \left. \quad +\frac{1}{r_z}\bigg (\frac{1}{\Hcal_zr_z}-\frac{1}{2}\bigg ) \int_{\eta_z}^{\eta_{\text{o}}} \t{d} \eta\,  \frac {\eta - \eta_z}{\eta_{\text{o}} - \eta} D^2 \Psi^\t{I} -3\bigg (1-\frac{1}{\Hcal_zr_z} \bigg )\int_{\eta_z}^{\eta_\t{o}}\t{d}\eta \, \pa_\eta \Psi^\t{I} \right.\nex
 & \left.\quad  -\frac{1}{2}\int_{\eta_z}^{\eta_\t{o}}\t{d}\eta \,  \int_{\eta}^{\eta_\t{o}}\t{d}\eta^\pr \, D^2 \Psi^\t{I} \right]
\label{eq:dmixedI} \, , \\[1ex]
   d^{(2)}_{\t{mixed,A}}(z) &= \left(1-\frac{1}{ \Hcal_z r_z}\right)\left[
\Psi^\t{A}_z  v_{||z}+
\frac{1}{ \Hcal_z} \partial_r v_{||z} (\Psi^\t{A}_z-\Psi^\t{A}\o)\right]
-\frac{\Hcal^\pr_z}{  \Hcal^3_z r_z }(\Psi^\t{A}_z-\Psi^\t{A}\o)
 \left(v_{||z}-v_{||\t{o}}\right)\nex
& \quad  
+\frac{1}{ \Hcal_z r_z}(2 \Psi^\t{A}_z-\Psi^\t{A}\o)v_{||z}
{+\left(\frac{1}{ \Hcal_z r_z}  \partial_\eta \Psi^\t{A}_z-\partial_r \Psi^\t{A}_z\right) \frac{1}{ \Hcal_z} \left(v_{||z}-v_{||\t{o}}\right)} +v_{||\t{o}}\Psi^\t{A}\o\, , 
\label{eq:dmixedA}\\[1ex]
d^{(2)}_{\t{path,I}}(z)&=\left(1-\frac{1}{ \Hcal_z r_z}\right) \left\{
\Phi_{\text{o}}^{(2)}-\Phi_z^{(2)}-\int_{\eta_z}^{\eta_{\text{o}}}\t{d} \eta\, \partial_{\eta}\left(
\Psi^{(2)}+\Phi^{(2)}\right)\right\}-\Psi_z^{(2)}
\nex
 &\quad 
-\frac{1}{2r_z} \int_{\eta_z}^{\eta_{\text{o}}}\t{d} \eta \,  \frac {\eta - \eta_z}{\eta_{\text{o}} - \eta} D^2 \left(
\Psi^{(2)}+\Phi^{(2)}\right)
+ \frac{1}{r_z}\int_{\eta_z}^{\eta_{\text{o}}}\t{d} \eta \, \left(
\Psi^{(2)}+\Phi^{(2)}\right)
\nex
&\quad 
+ \left(1-\frac{1}{ \Hcal_z r_z}\right) \left\{
\bigg[-2\partial_r\Psi^\t{I}_{\text{o}}-2\partial_\eta\Psi^\t{I}_{\text{o}}+\partial_r\Psi^\t{I}_z +2 \partial_\eta\Psi^\t{I}_z \right. \nonumber \\[1ex]
& \quad  \left. +2 \int_{\eta_z}^{\eta_{\text{o}}}\t{d} \eta\,  \partial^2_{\eta}\Psi^\t{I}\bigg ]
\left(-2 \int_{\eta_z}^{\eta_{\text{o}}}\t{d} \eta\, \Psi^\t{I}\right)+\frac{1}{2}(\Psi^\t{I}_z)^2+\frac{7}{2}(\Psi^\t{I}_{\text{o}})^2
\right. \nonumber \\[1ex]
 & \quad \left.
-\left(\Psi^\t{I}_{\text{o}}-\Psi^\t{I}_z-2 \int_{\eta_z}^{\eta_{\text{o}}}\t{d} \eta\, \partial_{\eta}\Psi^\t{I}\right)
\frac{1}{r_z}\int_{\eta_z}^{\eta_{\text{o}}}\t{d} \eta \, \frac {\eta - \eta_z}{\eta_{\text{o}} - \eta} D^2
\Psi^\t{I} +2 \Psi^\t{I}_{\text{o}} \int_{\eta_z}^{\eta_\t{o}} \t{d}\eta\,  \partial_{\eta} \Psi^\t{I}
\right.\nonumber \\[1ex] 
 & \quad \left.
-4\int_{\eta_z}^{\eta_\t{o}} \t{d}\eta\, \left[
- \Psi^\t{I} \partial_{\eta}\Psi^\t{I}
-\partial_{\eta}\Psi^\t{I} \int_{\eta}^{\eta_{\text{o}}} \t{d} \eta^\pr\, \partial_{\eta^\pr}\Psi^\t{I}- \Psi^\t{I} \int_{\eta}^{\eta_{\text{o}}} \t{d} \eta^\pr \, \partial^2_{\eta^\pr}\Psi^\t{I}
\right.\right.
\nonumber \\[1ex]
 &\quad  \left.\left.
+ \bar{\gamma}^{ab}\partial_a \left( \int_{\eta}^{\eta_{\text{o}}} \t{d} \eta^\pr \,  \Psi^\t{I} \right) 
\partial_b \left( \int_{\eta}^{\eta_{\text{o}}} \t{d} \eta^\pr \, \partial_{\eta^\pr}\Psi^\t{I} \right) \right]
\right. \nonumber \\[1ex] 
 & \quad \left.
+2\partial_a \Psi^\t{I}_z
\!\int_{\eta_z}^{\eta_{\text{o}}}\t{d}\eta\, \bar{\gamma}^{ab} \partial_b \int_{\eta}^{\eta_{\text{o}}}\t{d}\eta^\pr \, \Psi^\t{I}
+4 \int_{\eta_z}^{\eta_{\text{o}}}\t{d}\eta\, \partial_a \left(\partial_{\eta}\Psi^\t{I}\right)
\int_{\eta_z}^{\eta_{\text{o}}}\t{d}\eta^\pr \, \bar{\gamma}^{ab} \partial_b \int_{\eta^\pr}^{\eta_{\text{o}}}\t{d}\eta^{\pr \pr \, }\Psi^\t{I}
\right\}
\nonumber \\[1ex] 
 & \quad +4 \Psi^\t{I}_z \int_{\eta_z}^{\eta_{\text{o}}}\t{d} \eta\,  \partial_{\eta}\Psi^\t{I}
+\frac{3}{2}(\Psi^\t{I}_z)^2+\left ( \frac{9}{2}-\frac{13}{2\Hcal_zr_z}\right ) \Psi^\t{I}_z \Psi^\t{I}_{\text{o}}\nonumber\\[1ex]
& \quad + 
\frac{1}{\Hcal_z}\left(\partial_r \Psi^\t{I}_z-\frac{1}{ \Hcal_z r_z}\partial_\eta \Psi^\t{I}_z\right)\left(\Psi^\t{I}_{\text{o}}-\Psi^\t{I}_z-2 \int_{\eta_z}^{\eta_{\text{o}}}\t{d} \eta\,  \partial_{\eta}\Psi^\t{I}\right)
-2\partial_r \Psi^\t{I}_z\int_{\eta_z}^{\eta_{\text{o}}}\t{d}  \eta\, \Psi^\t{I}
\nonumber \\ 
 &\quad 
-\frac{1}{ 2\Hcal_z r_z}\left(1-\frac{\Hcal^\pr_z}{\Hcal^2_z}\right)
\left[\left(\Psi^\t{I}_z-\Psi^\t{I}_{\text{o}}\right)^2+2\left(\Psi^\t{I}_z-\Psi^\t{I}_{\text{o}}\right)\left(2 \int_{\eta_z}^{\eta_{\text{o}}}\t{d} \eta\,  \partial_{\eta}\Psi^\t{I}\right)\right.
\nonumber \\[1ex]
 &\quad \left. +4 \left(\int_{\eta_z}^{\eta_{\text{o}}}\t{d} \eta\, \partial_{\eta} \Psi^\t{I}\right)^2\right]
+\frac{2}{r_z}\int_{\eta_z}^{\eta_{\text{o}}}\t{d} \eta\, \left[\Psi^\t{I}\left(\Psi^\t{I}_{\text{o}}-\Psi^\t{I}
-2 \int_{\eta}^{\eta_{\text{o}}}\t{d}\eta^\pr \,  \partial_{\eta^\pr}\Psi^\t{I}\right)\right. \nonumber \\[1ex]
& \quad \left. 
+\bar{\gamma}^{ab}\partial_a \left(\int_{\eta}^{\eta_{\text{o}}}\t{d}\eta^\pr \, \Psi^\t{I}\right)\!\partial_b \left(\int_{\eta}^{\eta_{\text{o}}}\t{d}\eta^\pr\, \Psi^\t{I}\right)\right] 
\nonumber 
\\[1ex] 
 & \quad 
+\left(\Psi^\t{I}_z-\frac{2}{r_z}\int_{\eta_z}^{\eta_{\text{o}}}\t{d} \eta\, \Psi^\t{I}\right)
\frac{1}{r_z}\int_{\eta_z}^{\eta_{\text{o}}}\t{d} \eta \,  \frac {\eta - \eta_z}{\eta_{\text{o}} - \eta} D^2\Psi^\t{I}
+\frac{1}{2}\left(\frac{1}{r_z}\int_{\eta_z}^{\eta_{\text{o}}}\t{d} \eta \,  \frac {\eta - \eta_z}{\eta_{\text{o}} - \eta} D^2\Psi^\t{I}\right)^2\nonumber \\[1ex]
&\quad 
+\left[\frac{1}{ \Hcal_z r_z}\left(\Psi^\t{I}_{\text{o}}-\Psi^\t{I}_z-2 \int_{\eta_z}^{\eta_{\text{o}}}\t{d} \eta \, \partial_{\eta}\Psi^\t{I}\right)
-\frac{1}{r_z}\int_{\eta_z}^{\eta_{\text{o}}}\t{d} \eta\, \Psi^\t{I}\right]\frac{1}{r_z}\int_{\eta_z}^{\eta_{\text{o}}}\t{d} \eta\,  D^2\Psi^\t{I}
\nonumber \\[1ex] 
& \quad 
+2\partial_a \Psi^\t{I}_z
\int_{\eta_z}^{\eta_{\text{o}}}\t{d}\eta\, \bar{\gamma}^{ab}\partial_b \int_{\eta}^{\eta_{\text{o}}}\t{d}\eta^\pr \Psi^\t{I}
-\frac{4}{r_z}\left[\int_{\eta_z}^{\eta_{\text{o}}}\t{d} \eta\,  \partial_a \Psi^\t{I}
\int_{\eta_z}^{\eta_{\text{o}}}\t{d}\eta^\pr\, \bar{\gamma}^{ab}\partial_b\int_{\eta^\pr}^{\eta_{\text{o}}}\t{d}\eta^{\pr \pr}\, \Psi^\t{I}\right]
\nonumber
\\[1ex]
& \quad 
+\left(\partial_a \int_{\eta_z}^{\eta_{\text{o}}}\t{d}\eta \, \Psi^\t{I}\right)
\left[4 \int_{\eta_z}^{\eta_{\text{o}}}\t{d}\eta\,  \frac{1}{\eta_{\text{o}}-\eta}\bar{\gamma}^{ab}
\int_{\eta}^{\eta_{\text{o}}}\t{d}\eta^\pr \, \partial_b \Psi^\t{I} - 2 \int_{\eta_z}^{\eta_{\text{o}}}\t{d}\eta\, \bar{\gamma}^{ab}\partial_b\Psi^\t{I}
\right.
\nonumber \\[1ex]
& \quad 
\left.
-4\int_{\eta_z}^{\eta_{\text{o}}}\t{d}\eta\, \bar{\gamma}^{ab}
\int_{\eta}^{\eta_{\text{o}}}\t{d}\eta^\pr\,\partial_b\partial_{\eta^\pr}\Psi^\t{I}
\right] \nonumber \\[1ex]
& \quad 
+\partial_a\left(\int_{\eta_z}^{\eta_{\text{o}}}\t{d}\eta\, \bar{\gamma}^{bc}\partial_c \int_{\eta}^{\eta_{\text{o}}}\t{d}\eta^\pr\, \Psi^\t{I}\right)
\partial_b\left(\int_{\eta_z}^{\eta_{\text{o}}}\t{d}\eta \, \bar{\gamma}^{ad}\partial_d \int_{\eta}^{\eta_{\text{o}}}\t{d}\eta^\pr\Psi^\t{I}\right)
\nonumber \\[1ex]
& \quad 
-2\left(\int_{\eta_z}^{\eta_{\text{o}}}\t{d} \eta\, \Psi^\t{I}\right) \int_{\eta_z}^{\eta_{\text{o}}}\t{d} \eta\, \left[-\frac{1}{(\eta_{\text{o}}-\eta)^3}
\!\int_{\eta}^{\eta_{\text{o}}}\t{d} \eta^\pr\, D^2\Psi^\t{I}\right. \nonumber \\[1ex]
& \quad \left. 
+\frac{1}{(\eta_{\text{o}}-\eta)^2}\left(\frac{1}{2} D^2\Psi^\t{I}+\int_{\eta}^{\eta_{\text{o}}}\t{d}\eta^\pr \, \partial_{\eta^\pr}\left(D^2\Psi^\t{I}
\right)\right )
\right]
\nonumber \\[1ex] 
& \quad 
+\frac{2}{r_z} \int_{\eta_z}^{\eta_{\text{o}}}\t{d}\eta \,   \frac{\eta-\eta_z}{\eta_{\text{o}}-\eta} \partial_b 
\left( D^2 \Psi^\t{I}\right)
\int_{\eta_z}^{\eta_{\text{o}}} \t{d} \eta\, 
\bar{\gamma}^{ab} \partial_a \int_{\eta}^{\eta_{\text{o}}}\t{d} \eta^\pr   \Psi^\t{I}
\nonumber
\\[1ex]
& \quad 
+\frac{1}{\left(\sin\tilde{\theta}\right)^2}\left[\frac{1}{r_z}\int_{\eta_z}^{\eta_{\text{o}}}\t{d} \eta\,  \frac{\eta-\eta_z}{\eta_{\text{o}}-\eta}
\partial_{\tilde{\theta}}\Psi^\t{I}\right]^2 \nonumber \\[1ex]
& \quad -\frac{1}{r_z} \int_{\eta_z}^{\eta_{\text{o}}}\t{d}\eta\,  \frac{\eta-\eta_z}{\eta_{\text{o}} - \eta} D^2\left[
\Psi^\t{I}\left(\Psi^\t{I}_{\text{o}}-\Psi^\t{I}-2\int_{\eta}^{\eta_{\text{o}}}\t{d}\eta\,  \partial_{\eta}\Psi^\t{I}\right)
\right.
\nonumber \\ 
& \quad \left.
+\bar{\gamma}^{ab}\partial_a\left( \int_{\eta}^{\eta_{\text{o}}}\t{d} \eta^\pr \,  \Psi^\t{I} \right) 
\partial_b \left( \int_{\eta}^{\eta_{\text{o}}} \t{d} \eta^\pr \,  \Psi^\t{I} \right) \right]\nonumber \\[1ex]
& \quad - \int_{\eta_z}^{\eta_{\text{o}}} \t{d} \eta\, \left\{ 
\Psi^\t{I}\, \lim_{\bar{\eta}\rightarrow \eta_{\text{o}}}\left(\frac{1}{(\eta_{\text{o}}-\bar{\eta})^2}  \int_{\bar{\eta}}^{\eta_{\text{o}}} \t{d} \eta^\pr \,  D^2
\Psi^\t{I}\right) 
\right. \nonumber \\[1ex] 
 & \quad \left. 
-\frac{2}{\eta_{\text{o}}-\eta}\Psi^\t{I}\int_{\eta}^{\eta_{\text{o}}} \t{d} \eta^\pr\,   \frac {\eta^\pr - \eta}{\eta_{\text{o}} - \eta^\pr} \partial_{\eta^\pr} \left ( D^2 \Psi^\t{I}\right )\right. \nonumber \\[1ex]
& \quad \left. +2 \bar{\gamma}^{ab}\partial_b \left(\int_{\eta}^{\eta_{\text{o}}} \t{d} \eta^\pr \, \Psi^\t{I}\right) \frac{1}{\eta_{\text{o}}-\eta}\!\int_{\eta}^{\eta_{\text{o}}}\t{d} \eta^\pr \,   \frac {\eta^\pr-\eta}{\eta_{\text{o}}-\eta^\pr}\partial_a \left (D^2 \Psi^\t{I}\right )
\right. \nonumber \\[1ex] 
 & \quad \left.
-\left(\Psi^\t{I}_{\text{o}}-2\Psi^\t{I}-2\int_{\eta}^{\eta_{\text{o}}}\t{d}\eta^\pr \, \partial_{\eta^\pr}\Psi^\t{I}\right)
\frac{1}{(\eta_{\text{o}}-\eta)^2}\int_{\eta}^{\eta_{\text{o}}}\t{d}\eta^\pr\, D^2 \Psi^\t{I}\right.\nonumber \\[1ex]
& \quad \left. +\partial_a \Psi^\t{I}\left[
 \lim_{\bar{\eta}\rightarrow \eta_{\text{o}}}\left(\bar{\gamma}^{ab} \partial_b\!\!\int_{\bar{\eta}}^{\eta_{\text{o}}} \t{d} \eta^\pr\, 
\Psi^\t{I}\right) 
-2 \int_{\eta}^{\eta_{\text{o}}}\t{d} \eta^\pr\,  \bar{\gamma}^{ab} \partial_b \int_{\eta^\pr}^{\eta_{\text{o}}}\t{d}  \eta^{\pr \pr}\, \partial_{\eta^{\pr \pr }}\Psi^\t{I}
\right] \right. \nonumber \\[1ex]
& \quad \left. +2 \partial_a\left(\bar{\gamma}^{bc}\partial_c\int_{\eta}^{\eta_{\text{o}}}\t{d}\eta^\pr\, \Psi^\t{I}\right)\int_{\eta}^{\eta_{\text{o}}}\t{d}\eta^\pr \,  \partial_b
\left(\bar{\gamma}^{ad}\partial_d\int_{\eta^\pr}^{\eta_{\text{o}}}\t{d}\eta^{\pr \pr }\Psi^\t{I}\right)
\right. \nonumber \\[1ex]
 & \quad \left.
+2\bar{\gamma}^{ab}\partial_a\left(\Psi^\t{I}+\int_{\eta}^{\eta_{\text{o}}}\t{
d}\eta^\pr\, \partial_{\eta^\pr }\Psi^\t{I}\right)
\partial_b\int_{\eta}^{\eta_{\text{o}}}\t{d}\eta^\pr\, \Psi^\t{I}
\right\}\, , 
\label{eq:pathASa}\\[1ex]
d^{(2)}_{\t{path,A}}(z)&=\left(1-\frac{1}{ \Hcal_z r_z}\right) \left\{
-\frac{1}{r_z}(\Psi^\t{A}_z- \Psi^\t{A}\o)\int_{\eta_z}^{\eta_\t{o}} \t{d} \eta\,  \frac {\eta - \eta_z}{\eta_\t{o} - \eta}D^2  \Psi^\t{I}
-\Psi^\t{A}_z \Psi^\t{A}\o+(\Psi^\t{A}_z)^2\right. \nex
& \quad \left. + \Psi^\t{A}_z \Psi^\t{I}\o
+\Psi^\t{A}\o \Psi^\t{I}_z-2 \Psi^\t{A}_z \Psi^\t{I}_z
-4 \int_{\eta_z}^{\eta_\t{o}} \t{d} \eta \, \partial_{\eta} (\Psi^\t{I}\Psi^\t{A})
\right\}\nex
& \quad - \frac{1}{\Hcal_z}
{\left(\frac{1}{ \Hcal_z r_z}\partial_\eta  \Psi^\t{I}_z-\partial_r  \Psi^\t{I}_z\right)(\Psi^\t{A}_z- \Psi^\t{A}_\t{o})
-\frac{2}{\Hcal_z r_z} \partial_r \Psi^\t{A}_z \int_{\eta_z}^{\eta_\t{o}}\t{d} \eta \, \Psi^\t{I}}
\nex
& \quad 
{ -\frac{1}{\Hcal_z}
\left(\frac{1}{ \Hcal_zr_z}\partial_\eta  \Psi^\t{A}_z-\partial_r  \Psi^\t{A}_z\right)\left[\Psi^\t{I}\o-\Psi \o^\t{A}-(\Psi^\t{I}_z-
\Psi_z^\t{A})-2\int_{\eta_z}^{\eta_\t{o}}\t{d}\eta \, \partial_{\eta}\Psi^\t{I}\right]
}
\nex
& \quad 
-\frac{2}{r_z} \Psi^\t{A}\o \int_{\eta_z}^{\eta_\t{o}} \t{d} \eta \, \Psi^\t{I}
+\frac{1}{r_z}\Psi^\t{A}_z \int_{\eta_z}^{\eta_\t{o}} \t{d} \eta\,  \frac {\eta - \eta_z}{\eta_\t{o} - \eta}D^2  \Psi^\t{I}\nex
& \quad 
 +\frac{1}{ \Hcal_z r^2_z} (\Psi^\t{A}_z- \Psi^\t{A}_\t{o})\int_{\eta_z}^{\eta_\t{o}}\t{d} \eta\, D^2 \Psi^\t{I}
 \nex
& \quad 
+\left[\frac{1}{ \Hcal_z r_z}(-\Psi_z^\t{A}+2 \Psi_\t{o}^\t{A})+
\frac{\Hcal^\pr_z}{\Hcal_z^3r_z}(\Psi^\t{A}_z- \Psi^\t{A}_\t{o})-\Psi \o^\t{A}\right]\left(-2\int_{\eta_z}^{\eta_\t{o}}\t{d}\eta\,\partial_{\eta}\Psi^\t{I}\right)\nex
& \quad -\frac{1}{2} (\Psi^\t{A}_\t{o})^2+\Psi^\t{A}_z \Psi^\t{A}_\t{o} 
-(\Psi^\t{A}_z)^2+\Psi^\t{A}_\t{o} \Psi^\t{I}_\t{o}-\Psi^\t{A}_z \Psi^\t{I}_\t{o}+\Psi^\t{A}_\t{o} \Psi^\t{I}_z-2 \Psi^\t{A}_z \Psi^\t{I}_z\nex
& \quad 
+
\frac{\Hcal^\pr_z}{\Hcal^3_zr_z}\left[\frac{1}{2}(\Psi^\t{A}_z- \Psi^\t{A}_\t{o})^2-\Psi^\t{A}_\t{o} \Psi^\t{I}_\t{o}+\Psi^\t{A}_z\Psi^\t{I}_\t{o}+
\Psi^\t{A}_\t{o} \Psi^\t{I}_z-\Psi^\t{A}_z \Psi^\t{I}_z\right]
\nex
& \quad 
+\frac{2}{\Hcal_z  r_z}\partial_a  \Psi^\t{A}_z
\int_{\eta_z}^{\eta_\t{o}}\t{d}\eta\, \bar{\gamma}^{ab} \partial_b \int_{\eta}^{\eta_\t{o}}\t{d}\eta^\pr \, \Psi^\t{I}
\nex
& \quad 
+\frac{4}{r_z} \int_{\eta_s}^{\eta_o} \t{d} \eta\,  (\Psi^\t{I}\Psi^\t{A})
-\frac{2}{r_z} \int_{\eta_z}^{\eta_\t{o}}\t{d} \eta \, \frac {\eta - \eta_z}{\eta_\t{o} - \eta}
D^2  (\Psi^\t{I}\Psi^\t{A}) \, . 
\label{eq:dpathA}
\end{align}
We thoroughly compared Eqs.~\eqref{eq:dpos}-\eqref{eq:dpathA} to Eqs.~(3.29)-(3.32) of \cite{Marozzi:2014kua} and found an almost perfect agreement for the purely source terms, with only few differences. One concerns the last term on the first line of \eqref{eq:dmixedI} (the correspondent term of \cite{Marozzi:2014kua} has coefficient $-a$ which is most likely a typo, as our result agree with \cite{BenDayan:2012wi, Fanizza:2013doa}), whereas  others are probably related to the way to manipulate nested integrals and deal with integration by parts. We would also like to stress that, in \cite{Fanizza:2015swa}, the authors use their Eq. (A.24) to facilitate the comparison of their result (A.17) with the previous ones in the literature \cite{BenDayan:2012wi, Fanizza:2013doa}. Instead, in our case, we have shown that the corresponding relations between the coordinate perturbations in the two approaches are explicitly given in Eqs.~\eqref{eq:relations-two-approaches}. Indeed, at first order, our gauge modes coincide with the coordinate perturbations obtained in the first approach of \cite{BenDayan:2012wi, Fanizza:2013doa}, see our Eqs.~\eqref{eq:firstorder-primoapproach}.

The most relevant outcome of the above results is that the systematic cross-check of our second-order source terms  with the literature \cite{BenDayan:2012wi, Fanizza:2013doa, Marozzi:2014kua} provides a definite validation of the light-cone perturbation theory developed in this paper. 

To conclude,  the  terms with the monopole $P\o$ due to the OSG fixing are:
\begin{align}
    d^{(2)}_{P, \t{pos}}(z)&=-\frac{1}{2}\bigg (\Hcal_\t{o}-\frac{\Hcal_\t{o}}{r_z\Hcal_z} + \frac{1}{r_z}\bigg ) \frac{1}{a_\t{o}}\bigg [ \int_{\eta_{\t{in}}}^{\eta_\t{o}} \t{d}\eta \, a \big (\Phi^{(2)}-\Phi^2 + v^2_{||}+v^a_\perp v_{a\perp}\big ) \bigg ]\nex
    & \quad + P\o \bigg \{ v_{||\t{o}}\bigg (\frac{3}{2r_z}+\frac{\Hcal \o}{2}-\frac{\Hcal \o \Hcal^\pr_z}{ \Hcal^3_z r_z} \bigg )  + v_{||z}\bigg [\frac{1}{r_z}+\frac{\Hcal \o }{\Hcal_z r_z}\bigg (\frac{\Hcal^\pr_z}{\Hcal^2_z} -1\bigg ) \bigg ]\nex
    & \quad + 2\bigg ( 1-\frac{1}{\Hcal_zr_z}\bigg ) \pa_r v_{||\t{o}} +\bigg ( 1-\frac{1}{\Hcal_z r_z}\bigg ) \bigg (1-\frac{\Hcal \o}{\Hcal_z} \bigg ) \pa_r  v_{||z}\bigg \} \nex
    & \quad +P^2\o \bigg [\frac{\Hcal \o}{r_z}+\frac{\Hcal^\pr\o}{2}\bigg ( 1-\frac{1}{\Hcal_z r_z}\bigg ) +\frac{\Hcal^2\o}{2\Hcal_z r_z}\bigg (\frac{\Hcal^\pr_z}{\Hcal^2_z}-1 \bigg ) \bigg ] \, ,  \label{eq:dPpos} \\[1ex] 
d^{(2)}_{P, \t{mixed,I}}(z)&= P\o \bigg \{\bigg (-1+\frac{1}{\Hcal_z r_z}+\frac{\Hcal \o}{\Hcal^2_z r_z} \bigg ) \pa_\eta \Psi^\t{I}_z  +\frac{1}{\Hcal_z}\bigg ( \frac{1}{r_z}-\Hcal \o\bigg ) \pa_r \Psi^\t{I}_z\nex
    & \quad +2 \bigg (1-\frac{1}{\Hcal_z r_z} \bigg )\pa_\eta \Psi^\t{I}\o + \pa_r \Psi^\t{I}\o +2 \bigg (\frac{1}{r_z}+\Hcal \o \bigg ) \Psi^\t{I}_z\nex
    & \quad + \bigg (-\frac{3}{2r_z}-\frac{1}{\Hcal_zr^2_z}-\Hcal \o + \frac{\Hcal \o}{\Hcal_z r_z} \bigg )\Psi^\t{I}\o -\frac{2\Hcal \o}{r_z}\int_{\eta_z}^{\eta_\t{o}}\t{d}\eta \, \Psi^\t{I}\nex
    & \quad + \bigg (\Hcal \o -\frac{\Hcal \o}{\Hcal_z r_z}+\frac{1}{r_z} \bigg )\bigg (\frac{1}{r_z}\int_{\eta_z}^{\eta_\t{o}}\t{d}\eta \,\frac{\eta - \eta_z}{\eta_\t{o}-\eta} \, D^2 \Psi^\t{I}+2\int_{\eta_z}^{\eta_\t{o}}\t{d}\eta \, \pa_\eta \Psi^\t{I}\bigg )\nex
    & \quad +\frac{1}{r^2_z}\bigg (\frac{1}{2}-\frac{\Hcal \o}{\Hcal_z} \bigg ) \int_{\eta_z}^{\eta_\t{o}}\t{d}\eta \, D^2 \Psi^\t{I}+\frac{\Hcal \o}{\Hcal_zr_z}\bigg ( \frac{\Hcal^\pr_z}{\Hcal^2_z}-1\bigg ) \bigg (\Psi^\t{I}_z -\Psi^\t{I}\o +2 \int_{\eta_z}^{\eta_\t{o}}\t{d}\eta \, \pa_\eta \Psi^\t{I} \bigg )\nex
    & \quad +\int_{\eta_z}^{\eta_\t{o}}\t{d}\eta \, \bigg [-\frac{1}{2r^3_z}\int_{\eta}^{\eta_\t{o}}\t{d}\eta^\pr \, D^2 \Psi^\t{I} +\frac{1}{2r^2_z}\bigg (\frac{1}{2}D^2 \Psi^\t{I}+ \int_{\eta}^{\eta_\t{o}}\t{d}\eta^\pr \, \pa_{\eta^\pr} D^2 \Psi^\t{I}\bigg )\bigg ] \bigg \}\, , \\[1ex]
d^{(2)}_{P, \t{mixed,A}}(z)&=P\o \bigg \{ \frac{\Hcal \o}{\Hcal^2_z r_z}\pa_\eta \Psi^\t{A}_z +\bigg (1-\frac{\Hcal \o}{\Hcal_z} \bigg )\pa_r \Psi^\t{A}_z-\bigg (4-\frac{3}{\Hcal_z r_z} \bigg )\pa_r \Psi^\t{A}_\t{o}\nex
    & \quad + \bigg (1-\frac{1}{\Hcal_z r_z} \bigg )\pa_\eta \Psi^\t{A}_\t{o} +\frac{\Hcal \o}{\Hcal_z r_z} \bigg ( 1-\frac{\Hcal^\pr_z}{\Hcal^2_zr_z}\bigg ) \Psi^\t{A}_z\nex
    & \quad + \bigg (\frac{1}{2r_z}-\frac{1}{\Hcal^2_zr_z}+\Hcal \o -\frac{2\Hcal \o}{\Hcal_z r_z}+\frac{\Hcal \o \Hcal^\pr_z}{\Hcal^3_z r_z} \bigg )\Psi^\t{A}_\t{o} \bigg \} \, , \label{eq:dPmixedA}\\[1ex]
d^{(2)}_{v,\t{int}}(z)&= -\frac{1}{2}\bigg (\Hcal_\t{o}-\frac{\Hcal_\t{o}}{r_z\Hcal_z} + \frac{1}{r_z}\bigg ) \frac{1}{a_\t{o}}v_{||\t{o}}\int_{\eta_{\t{in}}}^{\eta_\t{o}} \t{d}\eta \, a v_{||} \nex
    & \quad  +\frac{1}{2} \bigg( 1-\frac{1}{\Hcal_z r_z}\bigg ) \pa_r v_{||\t{o}}\bigg (\frac{1}{a_z}\int_{\eta_{\t{in}}}^{\eta_z} \t{d}\eta \, a v_{||}-\frac{1}{a_\t{o}}\int_{\eta_{\t{in}}}^{\eta_\t{o}} \t{d}\eta \, a v_{||} \bigg )\nex
    & \quad  +\frac{1}{2} \bigg( 1-\frac{1}{\Hcal_z r_z}\bigg ) v_{||\t{o}}\bigg (\frac{1}{a_z}\int_{\eta_{\t{in}}}^{\eta_z} \t{d}\eta \, a \pa_r v_{||}-\frac{1}{a_\t{o}}\int_{\eta_{\t{in}}}^{\eta_\t{o}} \t{d}\eta \, a \pa_r v_{||} \bigg )\nex
    & \quad + \frac{1}{2a\o}v_{||\t{o}} \int_{\eta_{\t{in}}}^{\eta_\t{o}} \t{d}\eta \, a \pa_r v_{||} +\frac{1}{r_z a\o}v_{||\t{o}} \int_{\eta_{\t{in}}}^{\eta_\t{o}} \t{d}\eta \, a  v_{||}\, .
    \label{eq:dvint}
\end{align}
These last four equations represent one of the most remarkable  results of this paper,  because they confirm that the GLC gauge is a very convenient gauge to compute cosmological observables to higher orders in perturbation theory, having  a divergences-free result.


\section{Summary and Conclusions}
\label{sec:conclusions}
 In this manuscript, we have developed a second-order cosmological perturbation theory adapted to the observer past light-cone, extending previous first-order results \cite{Fanizza:2020xtv}.
Starting from a homogeneous and isotropic background in the Geodesic Light-Cone (GLC) coordinates, we have constructed general perturbations and analyzed their gauge-transformation properties. The corresponding GLC gauge-fixing conditions have been derived at second order, both in terms of light-cone perturbations and within standard perturbation theory. In particular, following \cite{Fanizza:2020xtv}, the standard counterpart of the GLC gauge has been dubbed Observational Synchronous Gauge (OSG).

Within the GLC gauge, the past light-cone coincides with the background one, and observables are factorized in terms of products of perturbations evaluated at the source and observer positions. Consequently, light-like cosmological observables expressed in the GLC gauge, or in the OSG one, are manifestly bi-scalar (as already stated in \cite{Biern:2016kys}) and depend solely on local metric perturbations. When rewritten in terms of gauge-invariant Bardeen variables, the usual integrated effects such as gravitational lensing, the ISW effect and the Shapiro time-delay naturally emerge. This provides a systematic and gauge-invariant framework to compute cosmological observables to non-linear orders using the exact relations of \cite{Gasperini:2011us,Fanizza:2013doa} and the gauge transformations connecting the GLC gauge to the Poisson Gauge (PG). 

By applying this framework, we have derived the second-order expression for the angular distance–redshift relation,  as an informative application and with the aim of validating our perturbative framework and method through the comparison of our results obtained with the ones of the present literature. In particular, our procedure has been based on the following steps:
\begin{itemize}
    \item First we identify the fully non-linear formulae  for the redshift  and for the angular distance valid in the GLC gauge and we expand them up to second order within our perturbative framework;
    \item Next, we  express the angular distance in terms of the observed redshift, the observer past light-cone and the observed angles;
    \item Thereafter, we promote each perturbations to its gauge-invariant counterpart;
    \item Finally, we fix the PG to express the general relativistic effects in terms of the gravitational potentials.
\end{itemize}
The above techniques represent a new approach to compute higher-order relativistic effects to cosmological observables, validated by the fact that
our results exactly reproduce\footnote{ A part from few terms out of hundreds, probably due to typos or differences in integration by parts and/or the way to expand terms.} all known source terms from the literature based on the GLC gauge (see \cite{BenDayan:2012wi,Marozzi:2014kua,Fanizza:2013doa}). They also provide additional observer contributions which, to our knowledge, have not been previously identified. Indeed, our observer terms come from the fact that we have considered the angular distance–redshift relation as measured by a free-falling observer, where, working in the GLC gauge, the residual gauge freedom is exploited together with the additional requirement that the center of the coordinates coincides with the observer position. Thanks to this careful construction,  non-physical divergences at the observer position have been eliminated, thus confirming the  predictions and results of \cite{Biern:2016kys,Fanizza:2020xtv}. 

Another important point is that, while the GLC gauge provides a simple geometrical interpretation of cosmological observables, the associated perturbation dynamics is more intricate than in standard perturbation theory. On the other hand, the OSG is fully equivalent to the GLC gauge and it is entirely formulated  within the standard perturbative framework. Consequently, the dynamics and all related physical properties can be studied using the  usual tools of standard perturbation theory, which may lead to an alternative solution of the long-standing problem of the dynamics in the GLC gauge. The use of GLC gauge directly in standard perturbation theory will be the subject of future studies.

In summary, we have presented a general framework to compute second-order cosmological observables,  within the light-cone perturbation theory (the GLC gauge). We have validated our method by recovering the known source terms in the second-order angular distance–redshift relation and identified new observer contributions, which renders the result free of divergences. 
This new perturbation theory can conveniently pave the way for the computation of other cosmological observables on the light-cone up to second order, as, for example, the galaxy number count \cite{Bertacca:2014dra,Bertacca:2014wga,DiDio:2014lka, DiDio:2015bua,Magi:2022nfy} or the redshift drift \cite{Marcori:2018cwn,Bessa:2023qrr}.


\section*{Acknowledgements}
We are very thankful to Ruth Durrer, Matteo Magi and Fabian Schmidt for useful discussions. 
PB and GM are supported in part by the Istituto Nazionale di Fisica Nucleare (INFN) through the Commissione Scientifica Nazionale 4 (CSN4),  under the Iniziativa Specifica (IS) Theoretical Astroparticle Physics (TAsP) and the IS Quantum Fields in Gravity, Cosmology and Black Holes (FLAG). The work of GM and MRMS  was  supported by the research grant number 2022E2J4RK \quotes{PANTHEON: Perspectives in Astroparticle and
Neutrino THEory with Old and New messengers} under the program PRIN 2022 funded by the Italian Ministero dell’Universit\`a e della Ricerca (MUR) and by the European Union – Next Generation EU. GF acknowledges the COST Action CosmoVerse, CA21136, supported by COST (European Cooperation in
Science and Technology).
GF is also member of the Gruppo Nazionale per la Fisica Matematica (GNFM) of the Istituto Nazionale di Alta Matematica (INdAM). The work of MRMS was supported by the Brazilian National Council for 
Scientific and Technological Development – CNPq, under project grant 447129/2024-4.

\appendix

\section{Free-Falling Observer}
\label{app:A}

Let us now show how both the SG and OSG observers are free-falling.
In particular, the observer defined by the OSG is free-falling despite the fact that $\phi^{(2)}_{\t{OSG}} \neq 0$.

As a first step, let us consider the observer associated with the proper time $\tau$ in a generic gauge. Then, we can prove that the condition\footnote{As a matter of fact, it is enough that $g^{\tau \tau}=\t{const}$.} $g^{\tau \tau}=-1$  non-perturbatively defines a free-falling observer, regardless of the specific form of the metric. Indeed, following \cite{Fanizza:2015zgz}, by taking the 4-velocity $u_\mu = - \pa_\mu \tau = \delta_\mu^\tau$ associated with this observer, we have 
\begin{equation}
    \pa^\nu \tau \nabla_\nu (\pa_\mu \tau) = - g^{\tau\nu}\Gamma^\tau_{\mu \nu} = 
    -\frac{1}{2}g^{\tau\nu}g^{\tau\rho}\pa_\mu g_{\nu \rho} \, .
\end{equation}
Since we are assuming $g^{\tau\tau}=-1$, it follows  that
\begin{equation}
  \partial_\mu g^{\tau\tau}= \pa_\mu (g^{\tau\rho}g^{\tau\nu}g_{\nu \rho}) =0 \, , 
\end{equation}
from which we have
\begin{equation}
    g^{\tau\nu}g^{\tau\rho}\pa_\mu g_{\rho \nu} = -g^{\tau\nu}g_{\rho \nu}\pa_\mu g^{\tau\rho}-g^{\tau\rho}g_{\nu \rho}\pa_\mu g^{\tau \nu} = -2\pa_\mu g^{\tau\tau} =0 \, , 
\end{equation}
thus proving that $\pa^\nu \tau \nabla_\nu (\pa_\mu \tau)=0$.

As a second step, and a non-trivial check of our new perturbation theory, we will now  prove that this is indeed  the case for both the SG and the OSG, by working directly within our new perturbation theory on the light-cone and up to the second perturbative order.  
Let us resume the GLC gauge-fixing conditions to the first and second perturbative order:
\begin{align}
& L^{(1)}=0 \, , \quad\quad\quad\quad\quad\quad\quad\quad \quad\quad\quad\quad L^{(2)}=0\,,\notag \\[1ex]
& v^{(1)} =0 \, ,\quad\quad\quad\quad\quad\quad\quad\quad \quad\quad\quad\quad v^{(2)} =0\, , \notag \\[1ex]
& \hat{v}^{(1)} =0 \, ,\quad\quad\quad\quad\quad\quad\quad\quad \quad\quad\quad\quad  \hat{v}^{(2)} =0\, , \notag \\[1ex]
& N^{(1)}+2aM^{(1)} =0 \, ,\quad\quad \quad\quad\quad\quad\quad N^{(2)}- \frac{1}{4}
(N^{(1)})^2 +2aM^{(2)}- U^2_{(1)} =0\, 
\label{eq:fixing-GLC-gauge-orde1-2}
\end{align}
and the SG non-perturbative ones
\begin{align}
& a^2 L^{(n)} + N^{(n)}  + 2a M^{(n)}  = 0 \, , \nex 
& N^{(n)} =-a M^{(n)} \, , \nex 
&v^{(n)}  = -\frac{1}{a} u^{(n)} \, , \nex 
&\hat{v}^{(n)}  = -\frac{1}{a} \hat{u}^{(n)} \, . 
\label{eq:SG-lightcone-app}
\end{align}
Then, inverting the light-cone metric \eqref{eq:metricGLC}, we can obtain $g^{\tau \tau}$ at first and second order in perturbation theory before fixing any gauge, namely
\begin{align}
    g^{\tau \tau}_{(1)} &= -a^2 L^{(1)}-2aM^{(1)}-N^{(1)}
    \, , \nex
    g^{\tau \tau}_{(2)} &= -a^2 L^{(2)}-2aM^{(2)}-N^{(2)}- (a^2L^{(1)})^2-4a^3L^{(1)}M^{(1)} -3a^2 (M^{(1)})^2 \nex
    & \quad -2a^2 L^{(1)}N^{(1)}   -2aM^{(1)}N^{(1)} +\bar{\ga}^{ab}U^{(1)}_a U^{(1)}_b + a^2 \bar{\ga}^{ab}V^{(1)}_a V^{(1)}_b +2a\bar{\ga}^{ab}V^{(1)}_a U^{(1)}_b \, .
\end{align}
Then, by fixing the GLC gauge we have $L^{(n)}=0$, $v^{(n)}=0$ and $\hat{v}^{(n)}=0$, so we are left with 
\begin{align}
 g^{\tau \tau}_{(1, \t{GLC})} &= -2aM^{(1)} - N^{(1)}=0 \, , \nex
 g^{\tau \tau}_{(2, \t{GLC})} &= -2aM^{(2)} -\frac{1}{4}(N^{(1)})^2+2aM^{(2)}-U^2_{(1)}-a\big [ 3a(M^{(1)})^2 +2M^{(1)}N^{(1)} \big ]+U^2_{(1)}\nex
 &  = -\frac{1}{4}(2aM^{(1)})^2 -a \big [  3a(M^{(1)})^2 - 4a (M^{(1)})^2  \big ]=0 \,  ,
\end{align}
as expected. On the other hand, fixing the light-cone counterpart of the SG, we have 
\begin{align}
 g^{\tau \tau}_{(1, \t{SG})} &= -a^2 L^{(1)}-2aM^{(1)}-N^{(1)}
 =0 \, ,\nex
 g^{\tau \tau}_{(2, \t{SG})} &= -(N^{(1)}+2aM^{(1)})^2 + 4aM^{(1)}(N^{(1)}+2aM^{(1)}) -3a^2(M^{(1)})^2  \nex
 & \quad + 2N^{(1)}(N^{(1)}+2aM^{(1)}) -2aM^{(1)}N^{(1)}  \nex
 & = (N^{(1)})^2 + a^2 (M^{(1)})^2 +2aM^{(1)}N^{(1)}=0 \, ,
\end{align}
thus confirming that the observer defined by the OSG is free-falling as well as the one defined by the SG. 
  
\section{Table of Notation}
\label{app:table}

\begin{table}[H]
\scriptsize
\renewcommand{\arraystretch}{1.4}
\setlength{\arraycolsep}{13pt}
\[
\begin{array}{ccc}
\toprule
    \text{Symbols} & \text{Definition of the symbol}  & \text{Equation} \\
 \midrule  
y^{\mu}=(\eta, r,\theta^{a}) & \text{Conformal spherical coordinates} & \eqref{eq:FLRW-metric-perturb}\\
        \bar{g}_{\mu \nu}, g^{(1)}_{\mu \nu}, g^{(2)}_{\mu \nu} & \text{Background and perturbed standard metrics} & \eqref{eq:FLRW-metric-perturb}  \\
        a, \bar{\ga}^{\text{FLRW}}_{ab} & \text{Scale factor \& induced metric on $\mathbb{S}^2$ } & \eqref{eq:FLRW-metric-perturb}, \eqref{eq:gamma-ab-flrw}\\
      \phi, \mathcal{B}_i, \mathcal{C}_{ij} & \text{Standard metric perturbations} & \eqref{eq:FLRW-metric-perturb} \\
   B, B_i, \psi, E, F_i, h_{ij} &\text{Decomposition of $\mc{B}_i$ and $h_{ij}$} & \eqref{eq:FLRW-metric-perturbations}, \eqref{eq:Cij-decomposed}\\
   D_{ij} & \text{Traceless  derivative operator} & \eqref{eq:Dij-def}\\
   x^\mu = (\tau, w , \tilde{\theta}^a) & \text{GLC coordinates}& \eqref{eq:spherical-to-glc}\\
   \bar{f}_{\mu \nu}, f^{(1)}_{\mu \nu}, f^{(2)}_{\mu \nu} & \text{Background and perturbed light-cone metrics} & \eqref{eq:fbar}, \eqref{eq:metricGLC}\\
   \bar{\ga}_{ab}=r^2 q_{ab} & \text{Metric on $\mathbb{S}^2$} & \eqref{eq:gamma-ab-glc}\\
     L, M, N, V_a, U_a, \ga_{ab} &\text{Light-cone perturbations} & \eqref{eq:metricGLC} \\
     \ka_{ab}, \varepsilon_{ab} & \text{Volume form on $\mathbb{S}^2$ \& Levi-Civita symbol} & \eqref{eq:volume-form}\\
     v, \hat{v}, u,  \hat{u}, \nu, \mu, \hat{\mu} &\text{Decomposition of $V_a$, $U_a$ and $\ga_{ab}$} & \eqref{eq:SPS-vec-tensor}\\
     D_{ab}, \tilde{D}_{ab} & \text{Traceless  derivative operator on $\mathbb{S}^2$ \& its dual }& \eqref{eq:D-ab-def}\\
     (\ka^\eta, \ka^r, \ka^a) & \text{Standard  gauge modes}& \eqref{eq:lie-deriv-gauge-transf}, \eqref{eq:gauge-trans-coord}\\
      \Phi, \Psi
      &\text{Gauge-invariant  Bardeen potentials} & \eqref{eq:bardeen-1st}, \eqref{eq:bardeen-2nd}\\
      (\xi^\tau, \xi^w, \xi^a), \chi, \hat{\chi} & \text{Light-cone  gauge modes \& decomposition of  $\xi^a$} & \eqref{eq:gauge-modes-glc-def}, \eqref{eq:xi-a-decomposed}\\
      \Upsilon, \mc{U}^a, g_{ab} & \text{Functions of the GLC metric} & \eqref{eq:GLCmetric-start}\\
      g_{(4)} & \text{Determinant of the GLC metric} & \eqref{eq:properties-glc}\\
      u^\mu & \text{4-velocity of a free-falling observer in GLC gauge} & \eqref{eq:4-velocity-GLC}\\
      k^\mu & \text{Photon 4-momentum in GLC gauge} & \eqref{eq:normal-vector-glc-light-cone}\\
      w_0, \chi_0, \hat{\chi}_0 & \text{Gauge freedoms at the observer in GLC gauge} & \eqref{eq:xi-1st-order}\\
      \zeta^A, J^A_B & \text{Geodesic deviation vector \& Jacobi map} & \eqref{eq:jacobi-map}\\
      \ms{V}, \ms{N}, \ms{M}, \hat{\ms{M}}, \ms{U}, \hat{\ms{U}}&\text{Gauge-invariant light-cone perturbations} & \eqref{eq:gauge-inv-quantities-1st-order}, \eqref{eq:glc-gauge-inv-2nd-order}\\
      P, Q & \text{Velocity potential \& Shapiro time-delay in PG} & \eqref{eq:P-Q-defined}\\
      \Psi^\text{I}, \Psi^\text{A} & \text{Isotropic \& anisotropic PG potentials} & \eqref{eq:IA-potentials-defined}\\
      \tau_z, \tau^{(1)}_z, \tau^{(2)}_z & \text{Source's proper time at  observed $z$ \& perturbations} & \eqref{eq:time-source-expanded}, \eqref{eq:time-z-perturbations}\\
      \bar{\ga}, \ga^{(1)}, \ga^{(2)} & \text{Determinant of $\bar{\ga}_{ab}$, $\ga^{(1)}_{ab}$, $\ga^{(2)}_{ab}$} & \eqref{eq:det-perturbed}\\
      d_\text{A}(z) & \text{Angular distance–redshift relation} & \eqref{eq:dA(z)-split}, \eqref{eq:perturbations-dA-z}\\
      v_{||}, v_{a \perp} & \text{Radial \& orthogonal peculiar velocity} & \eqref{eq:pecvel-1st-defined}, \eqref{eq:vperp-v2par}\\
      d_{\text{pos}}(z) & \text{Doppler terms of $d_\text{A}(z)$} & \eqref{eq:dpos}\\
      d_{\text{mixed}}(z) & \text{Doppler $\times$ (SW/ISW/lensing) terms of $d_\text{A}(z)$} & \eqref{eq:dmixedI}, \eqref{eq:dmixedA}\\
      d_{\text{path}}(z) & \text{Purely SW/ISW/lensing terms of $d_\text{A}(z)$} & \eqref{eq:pathASa}, \eqref{eq:dpathA}\\
      d_{P,\text{pos}}(z), d_{P,\text{mixed}}(z)  & \text{New terms of $d_\text{A}(z)$ with $P_\text{o}$} & \eqref{eq:dPpos}-\eqref{eq:dPmixedA}\\
      d_{v,\text{int}}(z)  & \text{New terms of $d_\text{A}(z)$ with $v_{||\text{o}}$} & \eqref{eq:dvint}\\
      
\bottomrule
\end{array}
\]
\vspace{-15pt}
    \caption{List of the main symbols used in this paper. Here, $\mathbb{S}^2$ is the 2-dimensional sphere, PG stands for \quotes{Poisson Gauge}, $P\o$ is an observer monopole term and $v_{||\text{o}}$ is the radial peculiar velocity at the observer position.
    }
    \label{tab:notation}
\end{table}

\bibliographystyle{JHEP}
\bibliography{Ref-GLCPert-order2-GLC}

\end{document}